\documentclass[a4paper,11pt]{article}
\pdfoutput=1 

\usepackage{jcappub} 
\usepackage[T1]{fontenc} 
\usepackage{xfrac}
\usepackage{mathrsfs}
\usepackage{amsmath,amssymb}
\usepackage{mathtools}
\usepackage{enumerate}
\usepackage[dvipsnames]{xcolor}
\usepackage{aas_macros}
\usepackage{xspace}

\usepackage[T1]{fontenc} 
\usepackage{natbib}
\definecolor{internationalkleinblue}{rgb}{0.0, 0.18, 0.65}
\hypersetup{colorlinks=true, urlcolor=internationalkleinblue, linkcolor=internationalkleinblue, citecolor=internationalkleinblue}

\newcommand\Tstrut{\rule{0pt}{3ex}}   
\newcommand{\txtp}{3$\times$2pt\xspace}

\title{Galaxy bias in the era of LSST: perturbative bias expansions}

\author[a]{Andrina Nicola}
\author[b, c]{Boryana Hadzhiyska}
\author[d]{Nathan Findlay}
\author[e]{Carlos Garc\'ia-Garc\'ia}
\author[e]{David Alonso}
\author[f]{An\v{z}e Slosar}
\author[g]{Zhiyuan Guo}
\author[h, i]{Nickolas Kokron}

\author[j, k]{Ra\'ul Angulo}
\author[l, m]{Alejandro Aviles}
\author[n]{Jonathan Blazek}
\author[o, p]{Jo Dunkley}
\author[q]{Bhuvnesh Jain}
\author[j]{Marcos Pellejero}
\author[r, s]{James Sullivan}
\author[g]{Christopher W. Walter}
\author[e]{Matteo Zennaro}
\author{The LSST Dark Energy Science Collaboration}

\affiliation[a]{Argelander Institut f\"ur Astronomie, Universit\"at Bonn, Auf dem H\"ugel 71, 53121 Bonn, Germany}
\affiliation[b]{Miller Institute for Basic Research in Science, University of California, Berkeley, CA, 94720, USA}
\affiliation[c]{Physics Division, Lawrence Berkeley National Laboratory, Berkeley, CA 94720}
\affiliation[d]{Institute of Cosmology and Gravitation, University of Portsmouth, Burnaby Road, Portsmouth, PO1 3FX, United Kingdom}
\affiliation[e]{Department of Physics, University of Oxford, Denys Wilkinson Building, Keble Road, Oxford OX1 3RH, United Kingdom}
\affiliation[f]{Brookhaven National Laboratory, Physics Department, Upton, NY 11973, USA}
\affiliation[g]{Department of Physics, Duke University, Durham NC 27708, USA}
\affiliation[h]{Kavli Institute for Particle Astrophysics and Cosmology and Department of Physics, Stanford University, Stanford, CA, USA}
\affiliation[i]{Kavli Institute for Particle Astrophysics and Cosmology, SLAC National Accelerator Laboratory, Menlo Park, CA, USA}
\affiliation[j]{Donostia International Physics Center (DIPC), Paseo Manuel de Lardizabal, 4, 20018 Donostia-San Sebastián, Spain}
\affiliation[k]{IKERBASQUE, Basque Foundation for Science, 48013, Bilbao, Spain}
\affiliation[l]{Departamento de F\'isica, Instituto Nacional de Investigaciones Nucleares,
Apartado Postal 18-1027, Col. Escand\'on, Ciudad de M\'exico, 11801, M\'exico}
\affiliation[m]{Consejo Nacional de Ciencia y Tecnolog\'ia, Av. Insurgentes Sur 1582,
Colonia Cr\'edito Constructor, Del. Benito Ju\'arez, 03940, Ciudad de M\'exico, M\'exico}
\affiliation[n]{Department of Physics, Northeastern University, Boston, MA, 02115, USA}
\affiliation[o]{Department of Physics, Jadwin Hall, Princeton University, Princeton, NJ 08544, USA}
\affiliation[p]{Department of Astrophysical Sciences, Peyton Hall, Princeton University, Princeton, NJ 08544, USA}
\affiliation[q]{Center for Particle Cosmology, Department of Physics and Astronomy,
University of Pennsylvania, Philadelphia, PA 19104, USA}
\affiliation[r]{Department of Astronomy, University of California, Berkeley, CA 94720, USA}
\affiliation[s]{Berkeley Center for Cosmological Physics, University of California, Berkeley, CA 94720, USA}

\emailAdd{anicola@uni-bonn.de}

\abstract{Upcoming imaging surveys will allow for high signal-to-noise measurements of galaxy clustering at small scales. In this work, we present the results of the Rubin Observatory Legacy Survey of Space and Time (LSST) bias challenge, the goal of which is to compare the performance of different nonlinear galaxy bias models in the context of LSST Year 10 (Y10) data. Specifically, we compare two perturbative approaches, Lagrangian perturbation theory (LPT) and Eulerian perturbation theory (EPT) to two variants of Hybrid Effective Field Theory (HEFT), with our fiducial implementation of these models including terms up to second order in the bias expansion as well as nonlocal bias and deviations from Poissonian stochasticity. We consider a variety of different simulated galaxy samples and test the performance of the bias models in a tomographic joint analysis of LSST-Y10-like galaxy clustering, galaxy-galaxy-lensing and cosmic shear. We find both HEFT methods as well as LPT and EPT combined with non-perturbative predictions for the matter power spectrum to yield unbiased constraints on cosmological parameters up to at least a maximal scale of $k_{\mathrm{max}}=0.4 \; \mathrm{Mpc}^{-1}$ for all samples considered, even in the presence of assembly bias. While we find that we can reduce the complexity of the bias model for HEFT without compromising fit accuracy, this is not generally the case for the perturbative models. We find significant detections of non-Poissonian stochasticity in all cases considered, and our analysis shows evidence that small-scale galaxy clustering predominantly improves constraints on galaxy bias rather than cosmological parameters. These results therefore suggest that the systematic uncertainties associated with current nonlinear bias models are likely to be subdominant compared to other sources of error for tomographic analyses of upcoming photometric surveys, which bodes well for future galaxy clustering analyses using these high signal-to-noise data.}

\begin{document}
\maketitle
\flushbottom

\section{Introduction} \label{sec:intro}

The last three decades have seen the emergence of the $\Lambda$CDM cosmological model as the concordance model preferred by a number of different cosmological probes. Current and future surveys across all wavelengths will allow us to put this model to its most stringent test to date. In the optical wavelength range, these include the Dark Energy Survey (DES)\footnote{\url{https://www.darkenergysurvey.org/}.}, the Hyper-Suprime Cam Survey (HSC)\footnote{\url{https://hsc.mtk.nao.ac.jp/ssp/0}.}, the Kilo Degree Survey (KiDS)\footnote{\url{https://kids.strw.leidenuniv.nl/}.}, the Dark Energy Camera Legacy Survey (DECaLS)\footnote{\url{https://www.legacysurvey.org/decamls/}.}, the Baryon Oscillation Spectroscopic Survey (BOSS)\footnote{\url{https://www.sdss4.org/surveys/boss/}.}, the Dark Energy Spectroscopic Instrument (DESI)\footnote{\url{https://www.desi.lbl.gov/}.}, the Rubin Observatory Legacy Survey of Space and Time (LSST)\footnote{\url{https://www.lsst.org/}.}, Euclid\footnote{\url{https://www.euclid-ec.org/}.}, and the Roman Space Telescope\footnote{\url{https://roman.gsfc.nasa.gov/}.}. 

Alongside weak gravitational lensing, galaxy clustering is one of the main probes observable with these surveys. This powerful cosmological probe offers the promise to deliver tight constraints on modifications of $\Lambda$CDM, such as neutrino masses \cite{Lattanzi:2017} and primordial non-Gaussianity \cite{Dalal:2008}, as well as the physics of galaxy formation. Theoretical modeling of galaxy clustering on small scales is hampered for two main reasons: First, on small scales the clustering of Dark Matter (DM) becomes non-linear. Second, on these scales, the relation between galaxy tracers and the DM field also becomes non-linear and mildly non-local. In the absence of baryons, the clustering of dark matter in the mildly nonlinear regime can be modeled either using analytic perturbative approaches or using N-body simulations (see e.g. Refs.~\cite{Carrasco:2014, Smith:2003, Knabenhans:2019}). The relation between galaxies and dark matter on small scales depends on the physics of galaxy formation, which involves a variety of different processes and spans several orders of magnitude in scale. These processes are impossible to model ab-initio on a cosmological scale, even using the highest-accuracy hydrodynamic simulations (see e.g. Ref.~\cite{Springel:2018}). The relation between galaxies, or any biased tracer, and the underlying DM distribution therefore presents the largest theoretical systematic uncertainty in galaxy clustering analyses. Several approaches have been developed to model these effects, and they can be subdivided in two different categories: (i) perturbative models, and (ii) phenomenological models. The former models use a perturbative expansion to jointly model the non-linear evolution of the DM distribution and its relation to the distribution of biased tracers. This expansion can be either performed in the initial conditions or at late times, leading to two distinct frameworks within which to model tracer bias, Lagrangian or Eulerian perturbation theory (PT) (see e.g. Refs.~\cite{Bernardeau:2002, McDonald:2009, Desjacques:2018, Matsubara:2008, Carlson:2013, Vlah:2016}). The effective field theory of Large-Scale Structure (EFToLSS, hereafter we use EFT for short) presents a closely related approach that treats cosmological fields at mildly non-linear scales as effective fields emerging from a more complete theory describing small-scale structure formation \cite{Carrasco:2012, Baumann:2012}. A number of studies have analyzed the reach of these methods and have found them to be accurate up to maximal wave numbers of $k_{\mathrm{max}}\sim 0.1 - 0.3 \: h \; \mathrm{Mpc}^{-1}$ at redshift $z=0$ (see e.g. Ref.~\cite{Munari:2017, Modi:2020, Eggemeier:2020, Fonseca:2020, Pandey:2020, Goldstein:2022}). Recently, Ref.~\cite{Modi:2020} proposed a new method aimed to improve upon this reach by developing a hybrid bias model that combines the accuracy of N-body simulations with the theoretical underpinning of Lagrangian perturbation theory, called Hybrid Effective Field Theory (HEFT) hereafter. In Ref.~\cite{Modi:2020} it was shown that this model allows for an accurate fit to N-body data up to $k_{\mathrm{max}}\sim 0.6 \: h \; \mathrm{Mpc}^{-1}$ at redshift $z=0$. In contrast to these methods, phenomenological models of galaxy bias typically rely on the Halo Model \cite{Ma:2000, Peacock:2000, Seljak:2000} coupled with a Halo Occupation Distribution (HOD). These models are built on the assumption that all matter in the Universe exists in the form of halos and that galaxies populate these halos with statistics solely determined by halo mass. The advantage of these models is that they are, despite their conceptual simplicity, surprisingly successful at explaining clustering with decent accuracy well into the non-linear regime, where perturbative approaches falter. Their main disadvantage is that they rely on a number of heuristic assumptions on a qualitative level, and thus cannot strictly be shown to provide a complete description of clustering. Additionally, predictions based on the Halo Model typically show inaccuracies in the transition region between the 1- and 2-halo terms and do not account for smearing of the Baryonic Acoustic Oscillation feature (see e.g. Refs.~\cite{Mead:2015, Chaves:2023}).

All of the methods outlined above have successfully been applied to data (see e.g. Ref.~\cite{Beutler:2017, dAmico:2020, Ivanov:2020, White:2022, Zhou:2021, Pandey:2020}), but they tend to be computationally intensive, thus making parameter inference an expensive part of cosmological analyses. Therefore, several recent works have developed hybrid methods that couple these bias models with machine learning methods to build emulators that can generate fast predictions for statistics involving biased tracers (see e.g. Refs.~\cite{Kokron:2021, Zennaro:2021, Hearin:2022, Eggemeier:2023, DeRose:2023, Pellejero:2022a, Pellejero:2023b}).

Combined in a so-called ``\txtp'' analysis, galaxy clustering and weak lensing form a key component of the cosmological analysis planned by upcoming photometric galaxy redshift surveys such as Euclid and LSST. A large amount of cosmological and astrophysical information will be contained in galaxy clustering at small spatial scales. In order to ensure robust and unbiased constraints from these data, it is crucial to assess the performance of different bias models in this high signal-to-noise regime.

In this work, we aim to perform a consistent comparison of non-linear galaxy bias models and assess their performance in \txtp analyses including high-precision galaxy clustering data from Stage IV surveys, using as an example the LSST survey. To this end, we use the \textsc{AbacusSummit} simulations and generate simulated data vectors and corresponding covariances, loosely matching LSST Year 10 (Y10) data \cite{DESC:2018}. We then analyze these data using a number of current non-linear galaxy bias models. Specifically, we employ Eulerian (or Standard) Perturbation Theory, Lagrangian Perturbation Theory, and two implementations of HEFT, \texttt{anzu} and \texttt{BACCO}. We fit all of these models to the simulated data, and assess the performance of each model based on the accuracy of the returned cosmological parameter constraints and the goodness-of-fit. Our results have implications for photometric surveys beyond LSST, but do not directly apply to spectroscopic surveys such as DESI, as we do not model a number of effects important in this regime, such as redshift space distortions (RSDs).

There are three over-arching questions that we would like to answer in this work. Which model and approach offers the most robust and accurate constraints on fundamental cosmological parameters given the high-precision of forthcoming photometric Stage IV surveys? How deep into the non-linear regime can we go using the best-performing method? How much do constraints on cosmological parameters improve as we push to increasingly smaller scales?

The last question is particularly interesting, because it will guide us in further theoretical developments. While linear scales retain the most amount of memory regarding the primordial fields and their subsequent evolution, we expect a relative loss of cosmological sensitivity on non-linear scales\footnote{Ref.~\cite{Modi:2017} for example, show that the halo- and matter fields start to decorrelate once one-loop corrections to the power spectrum become significant.}. Despite this fact, the number of observable modes increases significantly at smaller scales, and it is thus interesting to investigate the impact of these competing effects. Previous galaxy clustering analyses using HODs have found significant improvements in cosmological constraining power when increasing the minimal scale included in the analysis (see e.g. Refs.~\cite{Krause:2017, Zhai:2019, Lange:2022}), while analyses based on PT have found smaller effects, tied to degeneracies between cosmological and bias parameters (see e.g. Ref.~\cite{Chen:2022}). Here, we aim to investigate these questions also in the light of upcoming Stage IV photometric surveys.

This manuscript is structured as follows. In Sec.~\ref{sec:pt}, we introduce perturbative bias models, and in Sec.~\ref{sec:obs} we describe the observables considered. Section \ref{sec:simulations} gives an overview of the simulations employed, while Sec.~\ref{sec:methods} describes the methodology used in our analysis. We
present our results in Sec.~\ref{sec:results} and conclude in Sec.~\ref{sec:conclusions}. Implementation details are deferred to the Appendices.

\section{Perturbative bias models}\label{sec:pt}
The basic premise behind perturbative approaches to describe galaxy biasing is acknowledging the presence of complex physical, non-gravitational processes behind the formation of galaxies. These processes are non-local, and in general involve all the matter in a region around each galaxy of size $R_g$ (e.g. the Lagrangian size of the parent halo). On scales larger than $R_g$ however, galaxy formation can be described as an effectively local process, thus removing the need to describe these physical processes in detail. In this limit, one can invoke the Equivalence Principle which, in its non-relativistic limit, implies that the only measurable gravitational local quantities in a freely-falling frame are the second derivatives of the gravitational potential $\partial_i\partial_j\Phi$ (see e.g. Ref.~\cite{McDonald:2009}). In other words, on these scales the overdensity of galaxies $\delta_g({\bf x})$ can be described by a general function $F$ of these second derivatives:
\begin{equation}
1+\delta_g=F[\partial_i\partial_j\Phi].\label{eq:bias}
\end{equation}
Perturbative bias models then proceed by expanding $F$ in powers of $\partial_i\partial_j\Phi$. Since $1+\delta_g$ is a scalar quantity, each order in this expansion can only involve scalar combinations of $\partial_i\partial_j\Phi$, which further limits the number of possible unique terms at each order in the expansion. Up to second order in powers of $\partial^2\Phi$, only three terms are allowed: the matter overdensity $\delta_m$ (proportional to the trace $\nabla^2\Phi$), the squared overdensity $\delta_m^2$, and the trace squared of the tidal tensor $s^2\equiv s_{ij}s^{ij}$, where $s_{ij}\equiv\partial_i\partial_j\Phi-\delta_{ij}\nabla^2\Phi/3$.

The approximation of local bias is expected to break down on scales close to or smaller than $R_{g}$ \cite{Desjacques:2018}, and the leading correction to Eq.~\ref{eq:bias} due to non-local processes is given by the Laplacian of the matter density field, i.e. $R_g^2\nabla^2\delta_{m}$ \cite{McDonald:2009, Desjacques:2018}. In Fourier space, we have $\nabla^{2}\delta_{m}(\mathbf{k}) = -k^{2}\delta_{m}(\mathbf{k})$, which is what we include in our model given recent detections of non-local bias for halos (see e.g. \cite{Lazeyras:2019}). We note that a similar expansion of Eq.~\ref{eq:bias} at higher orders in $\partial^2\Phi$ and higher-order derivative operators can be derived when including non-local terms, but we will limit our discussion to the lowest-order contribution described here.

Finally, the details of galaxy formation are sensitive to fluctuations in the initial conditions on scales smaller than $R_g$. This leads to stochasticity in the galaxy bias relation which, at lowest order, can be captured by an additional stochastic field $\varepsilon$ that is uncorrelated on large scales (i.e. it is assumed to have a white power spectrum on $k\ll R_g^{-1}$) \citep{1999ApJ...525..543M,1999ApJ...522...46T,1999ApJ...520...24D}, and does not correlate with any of the perturbative terms described above. 

Under the above assumptions, in this work, we describe the galaxy overdensity $\delta_{g}$ perturbatively up to second order as
\begin{equation}\label{eq:pertbias}
1+\delta_g=1+b_1\delta_m+\frac{b_2}{2!}(\delta_m^2-\langle\delta_m^2\rangle)+\frac{b_{s^2}}{2!}(s^2-\langle s^2\rangle)+\frac{b_{\nabla^2}}{2!}\nabla^2\delta_m+\varepsilon,
\end{equation}
where we have employed the Eulerian bias picture (though one could analogously expand the galaxy field in the Lagrangian picture). In Eq.~\ref{eq:pertbias}, we have removed the variance of the quadratic fields to ensure a mean zero galaxy overdensity. Furthermore, the quantity $b_1$ denotes the linear bias, $b_2$ is the quadratic bias, $b_{s^2}$ is the tidal bias, $b_{\nabla^2}$ denotes the non-local bias, and finally $\varepsilon$ is the stochastic contribution.
Different perturbative bias models perform this expansion at different points in time, and can thus be subdivided into Eulerian and Lagrangian approaches. In Eulerian Perturbation Theory (EPT), the perturbative expansion is performed locally at the time corresponding to the galaxy redshift. In Lagrangian Perturbation Theory (LPT) the bias parameters are defined with respect to the initial density field, and galaxy positions are then traced forward in time following their expected trajectories under gravity. If complete to a given order, Eulerian and Lagrangian bias expansions are equivalent \cite{Fry:1996, Baldauf:2012, Chan:2012, Matsubara:2014, Saito:2014}. In the following, we briefly discuss the traditional implementations of Eulerian and Lagrangian perturbation theory as well as an extension of LPT, named Hybrid Effective Field Theory, which aims to track the galaxy Lagrangian trajectories non-perturbatively. For a more detailed description of galaxy bias, the reader is referred to Ref.~\cite{Desjacques:2018}.

\subsection {Eulerian Perturbation Theory}\label{ssec:pt.ept}
In Eulerian perturbation theory, the equations of structure formation are solved by focusing on a particular point in space and following the fluid's movement through this point in time \cite{Bernardeau:2002}. Keeping all terms up to second order, and allowing for non-local Eulerian bias (in both space and time), the galaxy field at any given redshift $z$ can thus be expressed by Eq. \ref{eq:pertbias}, with all quantities ($\delta_g$, $\delta_m$, $s^2$) evaluated at the current galaxy position and time \cite{McDonald:2009, Abidi:2018}.

The galaxy-galaxy and galaxy-matter power spectra in the Eulerian bias framework are thus given by
\begin{equation}\label{eq:pk_eulerian}
  P_{gg}(k, z) = \sum_{i, j} b_ib_j P_{ij}(k, z) + P_{\mathrm{SN}},\hspace{12pt}P_{gm}(k,z)=\sum_i b_i P_{i\delta_m}(k,z),
\end{equation}
where $i, j \in \{\delta_{m}, \delta_{m}^{2}, s^{2}, \nabla^{2}\delta_{m}\}$, and the set $\boldsymbol{b} =\nobreak \{b_{1},\nobreak b_{2},\nobreak b_{s^{2}},\nobreak b_{\nabla^{2}}\}$ denotes the corresponding bias parameters. As a technical subtlety, we note that in purely perturbative approaches, those bias parameters are the renormalized version of the ``bare'' bias parameters in Equation \ref{eq:pertbias}. In full generality, the power spectrum of first and third order bias terms (proportional to $b_1 b_{\rm 3NL}$) gives rise to terms of the same order as those generated by the auto-correlation of second order fields \cite{Saito:2014}. These power spectra however, are strongly degenerate with other terms and can thus be absorbed into lower-order bias coefficients. In this work, we therefore do not consider bias terms beyond second order but note that these are crucial for fitting e.g. higher-order statistics. $P_\mathrm{SN}$ is the power spectrum of the stochastic term $\varepsilon$ in Eq. \ref{eq:pertbias}, which is assumed to be scale-independent on the scales considered in this analysis. At the lowest order, this term can be thought of as the Poisson noise associated with the discrete nature of galaxy tracers, but it also incorporates a variety of other effects. Physically, these phenomena are described as halo exclusion, but in perturbation theory, they naturally arise as a renormalization of the terms that result in white power spectra on large scales \cite{Baldauf:2013, Kokron:2022}. As pure Poisson noise gives rise to a stochastic power spectrum $P_\mathrm{SN} = \bar{n}_g^{-1}$, we expect this term to be of the same order, but not exactly equal.

The power spectra between the different terms in the bias expansion ($P_{ij}$ in Eq. \ref{eq:pk_eulerian}) can be computed using Eulerian perturbation theory (although see Section \ref{ssec:pt.choices}), and to do so, we use the \texttt{FAST PT}\footnote{The code can be found at \url{https://github.com/JoeMcEwen/FAST-PT}.} package \cite{McEwen:2016,Fang:2017}.

As described in more detail below, in this work we fit the set of bias parameters $\boldsymbol{b} =\nobreak \{b_{1},\nobreak b_{2},\nobreak b_{s^{2}},\nobreak b_{\nabla^{2}}\}$ and the shot noise parameter $P_{\mathrm{SN}}$ to simulated data from \textsc{AbacusSummit}. 

\subsection{Lagrangian Perturbation Theory}\label{ssec:pt.lpt}
In the Lagrangian bias picture \cite{Matsubara:2008,Vlah:2016, Senatore:2015, Porto:2014}, the perturbative expansion of Eq. \ref{eq:pertbias} is applied to the proto-galaxy field from which galaxies form in the initial conditions (i.e. at high redshifts during matter domination). In this case, $\partial_i\partial_j\Phi$ is the Hessian of the linear gravitational potential, $\delta_m$ is the linear Lagrangian density etc., and all fields are evaluated at the initial Lagrangian coordinates ${\bf q}$, and denoted by the subscript $L$ (see e.g. Ref.~\cite{Vlah:2016}).

Once the initial proto-galaxy overdensity field is established, its evolution is determined by the Lagrangian trajectories of the galaxies under gravity. Thus at late times, when galaxies are actually observed, the galaxy overdensity $\delta_g$ at Eulerian coordinates ${\bf x}$ is
\begin{equation}\label{eq:advection}
  1 + \delta_{g}(\mathbf{x}, z) = \int \mathrm{d}^{3}\mathbf{q} \;\delta^3(\mathbf{x}-\mathbf{q}-\boldsymbol{\Psi}(\mathbf{q}, z))\,( 1 + \delta_g^L({\bf q}) ),
\end{equation}
where $\delta^3$ is the 3-dimensional Dirac delta function, $\delta_L^g$ is the Lagrangian-space galaxy overdensity, given by Eq. \ref{eq:pertbias} in terms of the different bias expansion operators in the initial conditions, and $\boldsymbol{\Psi}$ is the Lagrangian displacement field. In the usual parlance of Lagrangian perturbation theory, the final galaxy overdensity is found by ``advecting'' the Lagrangian bias overdensity to the final Eulerian coordinates ${\bf x}$.

As in the Eulerian bias expansion, the galaxy-galaxy and galaxy-matter power spectra are given by Eq. \ref{eq:pk_eulerian}, with the exception that in this case the indices $i,j$ run over an extended set of operators $\{1, \delta_{L}, \delta_{L}^{2}, s_{L}^{2}, \nabla^{2}\delta_{L}\}$, with corresponding bias parameters $\boldsymbol{b} =\nobreak \{b_{0}, \nobreak b^{L}_{1},\nobreak b^{L}_{2},\nobreak b^{L}_{s^{2}},\nobreak b^{L}_{\nabla^{2}}\}$. Here, $b_0\equiv1$ is not a free parameter, and $P_{11}$ denotes the non-linear matter power spectrum. Finally, $P_\mathrm{SN}$ denotes a stochastic term as discussed in the previous section. The reason for the additional term is the fact that, in the Lagrangian picture, the advection of completely homogeneous density field simply yields the inhomogeneous matter density in Eulerian space. This does not change the number of free parameters of the model (since $b_0$ is fixed), but it implies that the Lagrangian and Eulerian bias parameters are not identical to one another (hence the $L$ superscripts above).

To calculate the displacement field $\boldsymbol{\Psi}$ and thus the power spectra of the advected operators (i.e. the $P_{ij}(k)$ in Eq. \ref{eq:pk_eulerian}) we can use Lagrangian perturbation theory. The details of this calculation can be found in e.g. Ref.~\cite{Matsubara:2008a}. In this work, we compute these LPT power spectra using \texttt{velocileptors}\footnote{The code can be found at \url{https://github.com/sfschen/velocileptors}.}, which is described in Ref.~\cite{Chen:2020}.

\subsection{HEFT}\label{ssec:pt.heft}
The physical processes underlying galaxy formation are complex, and thus formulating a non-perturbative bias model based on first principles is commensurately difficult. However, non-linear evolution under gravity is a simpler problem that can be solved to high accuracy numerically via $N$-body simulations. This fact may be used to formulate a hybrid bias expansion, where the relation between galaxy overdensity and $\partial_{i}\partial_{j}\Phi$ is given perturbatively at early times, and the subsequent evolution under gravity (i.e. solving for the Lagrangian displacement $\boldsymbol{\Psi}$) is carried out numerically via simulations. First proposed in Ref.~\cite{Modi:2020} and subsequently explored by Refs.~\cite{Kokron:2021, Zehavi:2011, Hadzhiyska:2021}, this approach allows for higher accuracy predictions while keeping the physical intuition of LPT.

Building on Ref.~\cite{Modi:2020}, two separate works \cite{Kokron:2021, Zennaro:2021} employed suites of $N$-body simulations to build emulators for the different power spectra in Eq.~\ref{eq:pk_eulerian}, corresponding to the advected fields $\{1,\delta_L,\delta_L^2,s_L^2,\nabla^2\delta_L\}$. These emulators thus allow for the computation of HEFT predictions for a wide range of cosmological models. We describe these briefly below, but refer the reader to Refs.~\cite{Kokron:2021, Zennaro:2021} for further details.
\begin{itemize}
  \item {\bf anzu}: In Ref.~\cite{Kokron:2021}, the authors use the \texttt{Aemulus} suite of $N$-body simulations to compute predictions for the LPT basis spectra. This simulation suite has been designed for emulating cosmological quantities, and covers the parameter space $\vartheta = \{\Omega_{b} h^{2}, \Omega_{c} h^{2},\sigma_{8}, H_{0}, n_{s}, N_{\mathrm{eff}} , w\}$ within priors set by a combination of current observational constraints. Using the base power spectra, $P_{ij}(k, z)$, computed from \texttt{Aemulus} for a broad range of cosmological models, \texttt{anzu} employs polynomial chaos expansions \cite{Wiener:1938} to emulate these quantities, and is publicly available on \texttt{github}\footnote{The code can be found at \url{https://github.com/kokron/anzu}.} (see Ref.~\cite{DeRose:2023} for an updated version of this emulator).
  \item {\bf BACCO}: The analysis presented in Ref.~\cite{Zennaro:2021} uses a similar approach: The authors use the \texttt{BACCO} suite of numerical simulations coupled with cosmology rescaling \cite{Angulo:2021} to create a library of LPT base power spectra that cover the parameter space $\vartheta = \{\Omega_{m}, \Omega_{b},\sigma_{8}, n_{s}, h, M_{\nu} , w_{0}, w_{a}\}$. Using these power spectra, the authors construct an emulator using a simple Neural Network\footnote{The code can be found at \url{https://bacco.dipc.org/emulator.html}.}.
\end{itemize}

Various works \cite{Modi:2020, Hadzhiyska:2021, Kokron:2021, Zennaro:2021} have found that this hybrid approach is able to reproduce the galaxy-galaxy and galaxy-matter power spectra in real space down to significantly smaller scales than the purely perturbative approaches. In general, precision of a few per cent can be achieved up to $k_\mathrm{max} \sim 0.6\,h \mathrm{Mpc}^{-1}$ for redshifts $0 \lesssim z \lesssim 1$.

\subsection{Relations between bias models}\label{ssec:pt.bias-consistency}
As galaxies found in the evolved, Eulerian field can be traced back to proto-galaxies in Lagrangian space at early times, we can relate Lagrangian bias parameters to their Eulerian counterparts \cite{Desjacques:2018}. Assuming coevolution of the galaxy and the matter distribution, which is equivalent to galaxy number conservation and vanishing velocity bias, allows us to derive simple relations between bias parameters defined at early times to those defined at all later times. Physically, this is due to the fact that the density ratio of conserved, comoving fluids is unchanged under gravity owing to the equivalence principle. Expanding the galaxy overdensity in Lagrangian space up to second order neglecting non-local terms, and using the continuity equation, leads to a relation between Lagrangian and Eulerian bias parameters given by \cite{Desjacques:2018}
\begin{align}
  b_{1} &= 1 + b^{L}_{1}, \nonumber \\
  b_{2} &= \frac{8}{21}b^{L}_{1} + b^{L}_{2}, \label{eq:bias-consistency}\\
  b_{s^{2}} &= -\frac{4}{7}b^{L}_{1} + b^{L}_{s^{2}}. \nonumber
\end{align}
Note that we have adjusted the prefactors to match the bias definition in Eq.~\ref{eq:pertbias}. A popular toy model for galaxy bias is the so-called local-in-matter-density (LIMD) Lagrangian bias. In this model, we assume the galaxy overdensity at early times to be solely a function of the local matter density, which amounts to setting $b^{L}_{s^{2}}=b^{L}_{\nabla^{2}}=0$ in Eq.~\ref{eq:pertbias}. In this special case Eq.~\ref{eq:bias-consistency} implies a relation between Eulerian bias parameters given by $b_{s^{2}} = -\frac{4}{7}(b_{1} - 1)$ \cite{Saito:2014}. This shows that gravitational evolution leads to a bias with respect to the squared tidal field at late times, even in absence of such a bias at early times, i.e. LIMD at early times is inconsistent with LIMD at late times. In general, we do not expect these relations to hold exactly, as galaxy evolution is a complex process determined by forces other than gravity such as momentum transfer due to baryonic feedback and radiation pressure \cite{Desjacques:2018}. In addition, several works have investigated empirical relations between bias parameters, beyond coevolution. Refs.~\cite{2016JCAP...02..018L,2018JCAP...09..008L} for example have found strong correlations between bias parameters for halos, and Refs.~\cite{2021JCAP...08..029B, Zennaro:2022} have found similar results for galaxies, albeit with a larger intrinsic scatter and slightly different relations between bias parameters. In Sec.~\ref{ssec:res.coev} we will test the validity of a subset of these relations for the galaxy samples considered in this work.

Finally, we note that gravitational co-evolution will generate third-order bias terms in Eulerian space from a second-order bias expansion in Lagrangian space. The models described in Sections \ref{ssec:pt.ept} and \ref{ssec:pt.lpt} are therefore not fully equivalent. However, in this work we have chosen to sacrifice full theoretical consistency for consistency in the number of bias parameters used for each model. We will further discuss this choice and how it might affect our results in Sec.~\ref{ssec:res.coev}.

\subsection{Implementation choices}\label{ssec:pt.choices}
When quantifying the validity of the different bias expansions introduced above, we will make use of a few implementation-specific choices.

\paragraph{Bias evolution:}
The perturbative bias expansions explored here, in their different incarnations, determine the scale dependence of the different terms contributing to the final galaxy power spectra, but do not impose any restrictions on the redshift dependence of the bias parameters. As described in more detail below, our analysis is based on angular cross-correlations between galaxies in different tomographic redshift bins. Since each tomographic bin in principle corresponds to a distinct galaxy sample, we assume different and independent bias parameters in each bin. We consider relatively broad redshift bins, and since we expect galaxy properties and thus galaxy bias to evolve with redshift, our model must account for this. In our fiducial implementation of all perturbation theory models, we therefore allow for a redshift evolution in the lowest-order (linear) bias parameters, $b_{1}$ and $b^{L}_{1}$. We assume a linear bias evolution with redshift of the form (e.g. for Eulerian bias):
\begin{equation}
  b_{1}(z) = b_{1} + b_{1, p}(z-\bar{z}),
\end{equation}
where $\bar{z}$ denotes the mean redshift of each bin, and we use a similar expression for $b^{L}_{1}$\footnote{We consider a linear relation between bias and redshift as this corresponds to the lowest-order Taylor expansion of a possibly more complex relation, but can be generalized to higher orders.}. By default all other bias parameters are assumed constant within each redshift bin, although we will also investigate the impact of allowing for a redshift-evolution of $b_{2}$ and $b^{L}_{2}$ in Sec.~\ref{sec:results}.

\paragraph{Decorated PT:}
As stated in Sections \ref{ssec:pt.ept} and \ref{ssec:pt.lpt}, in the EPT and LPT frameworks, the power spectra of the bias expansion terms ($P_{ij}$ in Eq. \ref{eq:pk_eulerian}), may be computed to a given order in Eulerian or Lagrangian perturbation theory. This perturbative approach can be potentially improved by replacing the $P_{\delta_m\delta_m}(k,z)$ terms by its non-perturbative prediction calculated e.g. via the \textsc{halofit} fitting function \cite{Smith:2003, Takahashi:2012}, effectively re-summing all PT terms contributing to it. This selective resummation approach has been found to improve the quality of the fit while remaining largely unbiased \cite{Pandey:2020, Pandey:2021, Krolewski:2021}. When studying the EPT and LPT frameworks, we therefore consider two different approaches:
\begin{description}
  \item[\texttt{PTPk} approach:] We use Eulerian or Lagrangian perturbation theory at next-to-leading order to compute all terms in Eq.~\ref{eq:pk_eulerian}. The power spectrum of the non-local term and the matter density is approximated as
  \begin{equation}
    P_{\delta_m,\nabla^2\delta_m}(k,z) \rightarrow -k^2P_{1-\rm loop}(k,z),
  \end{equation}
  where $P_{1-\rm loop}$ is the 1-loop matter power spectrum and we have used that in Fourier space $\nabla^{2}\delta_{m}(\mathbf{k}) = -k^{2}\delta_{m}(\mathbf{k})$\footnote{We note that the functional form of this expression is equivalent to the counterterms present in EFT approaches (see e.g. Refs.~\cite{Carrasco:2012, Baumann:2012}). We will further discuss this similarity and its impact on our results in Appendix \ref{sec:ap.bias_model_cons}.}. All other terms in the expansion of $P_{gm}$ and $P_{gg}$ will be calculated using EPT or LPT at next-to-leading order\footnote{In particular, the matter power spectrum is modeled as $P_{mm} = P_{1-\rm loop}$.}.
  \item[\texttt{NLPk} approach:] In a second approach, we use the full non-linear matter power spectrum $P_{\rm NL}(k,z)$ from \textsc{halofit} to compute the term multiplying $b_1$ in $P_{gm}$ and $b_1^2$ in $P_{gg}$ in the EPT case. For LPT, we make the following substitutions in the case of $P_{gm}$ and $P_{gg}$ respectively:
  \begin{align}
    &P_{11}(k,z)+b_1^LP_{1\delta_m}(k,z) \rightarrow (1+b_1^L)P_{\rm NL}(k,z)\\
    &P_{11}(k,z)+2b_1^LP_{1\delta_m}(k,z)+(b_1^L)^2P_{\delta_m\delta_m}(k,z) \rightarrow (1+b_1^L)^2P_{\rm NL}(k,z).
  \end{align}
  We calculate all other terms in the expansion using EPT or LPT at next-to-leading order. As above, the cross-power spectrum of the non-local term and the matter density is approximated as
  \begin{equation}
    P_{\delta_m,\nabla^2\delta_m}(k,z) \rightarrow -k^2P_{\rm NL}(k,z).
  \end{equation}
  As opposed to the other bias models considered, where we keep all auto- and cross-correlations involving $\nabla^{2}\delta_{m}$, in this case we only keep the power spectra multiplying the matter density in order to not mix \textsc{halofit} and PT predictions for the nonlocal bias terms. In the EPT case, we thus only keep the term involving $b_{1}$ and $\nabla^2\delta_m$, while for LPT we make the substitution $P_{1,\nabla^2\delta_m}(k,z)+b_1^LP_{\delta_m,\nabla^2\delta_m}(k,z) \rightarrow -(1+b_1^L)k^2P_{\rm NL}(k,z)$.
\end{description}

\section{Observables}\label{sec:obs}
The key cosmological constraints from LSST and other future photometric surveys will be obtained from a so-called \txtp analysis, i.e. a joint analysis of galaxy clustering, galaxy-galaxy lensing and cosmic shear. In this work, we assess the performance of different galaxy bias models in a Fourier-space-based \txtp analysis, closely matching the specifications for LSST Y10. In full generality, the spherical harmonic power spectrum between probes $a$ and $b$, and redshift bins $i, j$ can be computed using the Limber approximation\footnote{We note that the Limber approximation has been shown to lead to biased results at low multipoles (see e.g. Ref.~\cite{Fang:2020}), but we do not expect this to affect our conclusions as these scales are still well within the linear regime and we choose a conservative minimal multipole of $\ell_{\mathrm{min}}=32$.} \cite{Limber:1953, Kaiser:1992, Kaiser:1998} as:
\begin{equation}\label{eq:cell_limber}
C^{a_{i} b_{j}}_\ell = \int \mathrm{d}z\,\frac{c}{H(z)}\frac{q^{a_{i}}(\chi(z))\;q^{b_{j}}(\chi(z))}{\chi^2(z)} \,P_{ab}\left(z,k=\frac{\ell+1/2}{\chi(z)}\right),
\end{equation}
where $c$ is the speed of light, $H(z)$ denotes the Hubble parameter, and $\chi(z)$ is the comoving distance. In addition, $q^{a_{i}}(\chi(z))$ denotes the window function for probe $a$ and bin $i$, and $P_{ab}(z,k)$ is the three-dimensional power spectrum between probes $a$ and $b$. For galaxy tracers, the window function is given by
\begin{equation}
q^{\delta_{g,i}}(\chi(z)) = \frac{H(z)}{c}p^{i}(z), 
\end{equation}
where $p^{i}(z)$ denotes the normalized redshift distribution of the considered sample. The window function for lensing tracers is
\begin{equation}
q^{\gamma_{i}}(\chi(z)) = \frac{3}{2}\frac{\Omega_{m}H_{0}^{2}}{c^{2}}\frac{\chi(z)}{a} \int_{\chi(z)}^{\chi_{h}}\mathrm{d}z' p^{i}(z')\frac{\chi(z')-\chi(z)}{\chi(z')},
\end{equation}
where $\Omega_{m}$ denotes the fractional matter density today, $H_{0}$ is the current expansion rate, $a$ denotes the scale factor, and $\chi_{h}$ is the comoving distance to the horizon. Furthermore, we have $P_{\gamma\gamma}(z,k)=P_{\delta_{m}\delta_{m}}(z,k)$.

Additionally, we also consider a combination of the \txtp data vector with CMB lensing, and the window function in this case is given by:
\begin{equation}
    q^{\kappa_{\mathrm{CMB}}}(\chi(z))=\frac{3}{2}\frac{\Omega_{m}H^{2}_{0}}{c^{2}}\frac{\chi(z)}{a}\frac{\chi(z_{*})-\chi(z)}{\chi(z_{*})},
\end{equation}
where $z_{*}$ denotes the redshift to the surface of last scattering. 

In this work, we model the uncertainties associated to this data vector analytically. The \txtp covariance matrix generally consists of three different parts: a Gaussian contribution, a non-Gaussian contribution and a contribution due to super-sample covariance (SSC) (see e.g. Ref.~\cite{Krause:2017}). Here we make the simplifying assumption that the data covariance is Gaussian, and thus neglect non-Gaussian and SSC contributions. Including those contributions will have two main effects on the covariance: (i) increasing the size of the error bars, and (ii) correlating different $\ell$-modes. While we expect the first effect to lead to our analysis being conservative since our smaller error bars lead to tighter requirements, the second might affect goodness-of-fit tests in a nonlinear way. We defer an investigation including non-Gaussian covariances to future work. Under the assumption of Gaussianity, we compute the covariance between the spherical harmonic power spectra $C^{ij}_{\ell}$ and $C^{i'j'}_{\ell'}$ using
\begin{equation}
\begin{aligned}
  \mathrm{Cov}_{G}(C^{ij}_{\ell}, C^{i'j'}_{\ell'}) = \langle \Delta C^{ij}_{\ell} \Delta C^{i'j'}_{\ell'} \rangle = \frac{\delta_{\ell \ell'}}{(2\ell+1)\Delta\ell f_{\mathrm{sky}}}\left[(C^{ii'}_{\ell} + N^{ii'}_{\ell})(C^{jj'}_{\ell} + N^{jj'}_{\ell}) \right. \\
  \left. +(C^{ij'}_{\ell} + N^{ij'}_{\ell})(C^{i'j}_{\ell} + N^{i'j}_{\ell})\right],
\end{aligned}
\end{equation}
where $C^{ij}_{\ell}$ denotes the signal part of the spherical harmonic power spectrum, $N^{ij}_{\ell}$ is the corresponding noise, $f_{\mathrm{sky}}$ denotes the fraction of sky covered by the experiment, and $\Delta \ell$ is the width of the $\ell$-bin used for power spectrum estimation. Usually, we set $N^{ij}_{\ell} = \delta_{ij} N^{ii}$, but here we keep a more generic expression in order to cater for potentially scale-dependent noise correlated across different probes. 

In addition, in Refs.~\cite{Kokron:2021, Zennaro:2021}, it was shown that the three-dimensional power spectra obtained from the HEFT emulators exhibit relative errors of around $1\%$ when compared to the power spectra measured directly from the $N$-body simulations. Both works account for this systematic uncertainty through an additional term in their covariance matrix. We follow these analyses assuming full correlation of theoretical errors of any two power spectra, and add a systematic error floor to our covariance matrix given by
\begin{equation}
\mathrm{Cov}_{\mathrm{sys}}(C^{ij}_{\ell}, C^{i'j'}_{\ell'}) = f_{\mathrm{sys}}^{2}C^{ij}_{\ell}C^{i'j'}_{\ell'}\delta_{\ell \ell'},
\end{equation}
where $f_{\mathrm{sys}}$ is the fractional error of the theoretical model, which we set to $f_{\mathrm{sys}}=0.01$ as per Ref.~\cite{Kokron:2021}. The total covariance matrix is thus given by 
\begin{equation}
\mathrm{Cov}(C^{ij}_{\ell}, C^{i'j'}_{\ell'}) = \mathrm{Cov}_{G}(C^{ij}_{\ell}, C^{i'j'}_{\ell'})+\mathrm{Cov}_{\mathrm{sys}}(C^{ij}_{\ell}, C^{i'j'}_{\ell'}).
\label{eq:cov_G+sys}
\end{equation}

\section{Simulations}\label{sec:simulations}
We use the \textsc{AbacusSummit} suite of high-performance cosmological $N$-body simulations \citep{Maksimova:2021} to create the simulated galaxy samples used in this analysis. The \textsc{AbacusSummit} suite was designed to meet the simulation requirements of the Dark Energy Spectroscopic Instrument (DESI) survey and was run with the high-accuracy cosmological code \textsc{Abacus} \citep{Garrison:2019, Garrison:2021}. We utilize one of the ``base''-resolution \textsc{AbacusSummit} boxes, \texttt{AbacusSummit\_base\_c000\_ph006}, at the fiducial cosmology: $\Omega_b h^2 = 0.02237$, $\Omega_c h^2 = 0.12$, $h = 0.6736$, $10^9 A_s = 2.0830$, $n_s = 0.9649$. Its box size is $2000 \ {\rm Mpc}/h$, and it contains 6912$^3$ particles with mass $M_{\rm part} = 2.1 \times 10^{9} \ M_\odot/h$.
  
To construct the mock galaxy samples from the $N$-body outputs, we utilize the 10\% particle subsample (i.e., including both the particle \texttt{A} and \texttt{B} subsamples, which constitute 3\% and 7\% of the particles, respectively), which is selected randomly and is consistent across redshift, and the on-the-fly halo catalogues, which are generated using the halo-finding algorithm \textsc{CompaSO} \citep{Hadzhiyska:2022}. Specifically, we apply the \textsc{AbacusHOD} model \citep{Yuan:2021} to the halo catalogue outputs at $z = 0.1, \allowbreak 0.3, \allowbreak 0.5, \allowbreak 0.8, \allowbreak 1.1, \allowbreak 1.4, \allowbreak 1.7, \allowbreak 2.0, \allowbreak 2.5, \allowbreak 3.0$. The \textsc{AbacusHOD} model builds upon the baseline halo occupation distribution (HOD) model by incorporating various generalizations pertaining to halo-scale physics and assembly bias. 

For the fiducial galaxy samples considered in this analysis, we assume that they are well-approximated by the ``baseline HOD'' model with no decorations. The model is akin to the 5-parameter model of \cite{Zheng:2007}, which gives the mean expected number of central and satellite galaxies per halo given halo mass $M$:
\begin{align}
\bar{N}_{\mathrm{cent}}(M) & = \frac{1}{2}\mathrm{erfc} \left[\frac{\log_{10}(M_{\mathrm{min}}/M)}{\sqrt{2}\sigma_M}\right], \label{equ:zheng_hod_cent}\\
    \bar{N}_{\mathrm{sat}}(M) & = \left[\frac{M-\kappa M_{\mathrm{min}}}{M_1}\right]^{\alpha}\bar{N}_{\mathrm{cent}}(M).
    \label{equ:zheng_hod_sat}
\end{align}
Here, $M_{\mathrm{min}}$ characterizes the minimum halo mass to host a central galaxy, $M_1$ is the typical halo mass that hosts one satellite galaxy, $\sigma_M$ describes the steepness of the transition from 0 to 1 in the number of central galaxies, $\alpha$ is the power law index on the number of satellite galaxies, and $\kappa M_\mathrm{min}$ gives the minimum halo mass to host a satellite galaxy. Central galaxies are placed at the halo centre, whereas satellite galaxies are ``painted'' on random particles.

In this work, we consider two main galaxy samples, which are designed to mimic two distinct galaxy sample choices for clustering analyses: (i) a homogeneous sample of Luminous Red Galaxies (LRGs) with a moderate number density (called `red' hereafter), which will constitute our fiducial sample throughout this work, and (ii) a magnitude-limited, high-number density sample (called `maglim' hereafter)\footnote{Similar samples have been considered in the DES Y3 analyses (see e.g. Ref.~\cite{Rodriguez-Monroy:2022}), and we expect such samples to also be used for clustering analyses of LSST data.}. We model the HODs of these samples based on observational constraints using Equations \ref{equ:zheng_hod_cent} and \ref{equ:zheng_hod_sat}. In order to account for evolution of both samples, we assume a redshift dependence of the three HOD masses ($M_{\rm min}$, $M_1$, and $M_0\equiv\kappa M_{\rm }$) of the form \cite{Nicola:2020}
\begin{equation}
  \log_{10} M_x/M_\odot = \mu_x+\mu_{x,p}\left[\frac{1}{1+z}-\frac{1}{1+z_p}\right],
\end{equation}
with $z_p=0.65$. Using this parameterization, we model the HOD of the maglim sample using the results derived for the HSC sample studied in Ref.~\cite{Nicola:2020}. For the red sample on the other hand, we assume redshift-dependent HOD masses consistent with the DESI LRG selection described in Ref.~\cite{Zhou:2021}. The adopted HOD parameters for both samples are given in Tab.\ref{tab:params}.

\begin{table*}
  \caption{Summary of HOD parameters for the two galaxy samples considered in this work. The parametrization of the redshift-dependent HOD follows Ref.~\cite{Nicola:2020}.} \label{tab:params}
  \begin{center}
    \begin{tabular}{ccc}
      \hline\hline 
      Parameter & red & maglim  \\ \hline \Tstrut
      $\mu_{\mathrm{min}}$ & $12.95$ & $11.88$ \\ 
      $\mu_{\mathrm{min}, p}$ & $-2.0$ & $-0.$ \\
      $\mu_{0}$ & $12.3$ & $11.88$ \\
      $\mu_{0, p}$ & $0.0$ & $-0.5$ \\
      $\mu_{1}$ & $14.0$ & $13.08$ \\ 
      $\mu_{1, p}$ & $-1.5$ & $0.9$ \\
      $\alpha$ & $1.32$ & $1.0$ \\
      $\sigma_{\log M}$ & $0.25$ & $0.4$ \\
      \hline\hline 
    \end{tabular}
  \end{center}
\end{table*} 

We complement these samples with an additional data set for which we also model assembly bias effects. A number of studies have investigated the dependence of clustering statistics on quantities other than halo mass, such as halo concentration, environment or spin (see e.g. Refs.~\cite{Wechsler:2002, Gao:2005, Wechsler:2006, Dalal:2008a}), and some of these works have resulted in significant detections of this so-called assembly bias effect. The nonlinear galaxy bias models considered in this work in principle offer the flexibility to also describe tracers affected by these processes. In order to test this, we generate an additional galaxy sample from \textsc{AbacusSummit} via the fast and efficient decorated HOD model, AbacusHOD \cite{2022MNRAS.510.3301Y}, incorporating assembly bias effects due to halo concentration and environment as described and defined in Refs.~\cite{2020MNRAS.493.5506H,Yuan:2021}. In particular, we adopt non-zero values for the concentration and environment assembly bias parameters, \texttt{A\_c,s} and \texttt{B\_c,s}, for both the central and satellite populations (subscript \texttt{c} referring to centrals and \texttt{s} to satellites), given by $\texttt{A\_c} = -0.73$, $\texttt{A\_s} = -0.24$, $\texttt{B\_c} = -0.0093$, and $\texttt{B\_s} = 0.0037$. These parameters modify the central and satellite halo occupations by reranking the haloes at fixed mass based on their intrinsic concentration and environment, and their values are to be close to the best-fit for CMASS data \cite{2016MNRAS.455.1553R} and modify the central and satellite halo occupations (hence, the subscripts) by re-ranking the haloes at fixed mass by their concentration ($\texttt{A\_c}$ and $\texttt{A\_s}$) and their environment ($\texttt{A\_c}$ and $\texttt{A\_s}$). For full definitions of the parameters and how they are used, see Ref.~\cite{2022MNRAS.510.3301Y}.

We would like to stress that all galaxy samples considered in this work are generated using a simplified HOD model, which includes some extensions beyond the ``vanilla'' model described in Ref.~\cite{Zheng:2005}. However, for maximal realism, one could imagine incorporating other effects such as redshift dependence of the selection functions, deviations of the satellite occupations from a Poisson distribution, and a further dependence of the central occupations on assembly bias. These additional effects might significantly affect the clustering and stochasticity of the samples, and thus potentially also our results. We leave an investigation of more complex HOD models to future work.

In addition in this work, we do not consider the additional uncertainties in realistic samples due to galaxy selection effects and photo-z estimation. These uncertainties are specific to photometric surveys; recent analyses from Stage III surveys, e.g. Ref \citep{Pandey:2021}, have found that imperfect correction of ``survey properties'' appear to dominate statistical uncertainties for red samples in DES. Given LSST's statistical precision, these issues may require significant effort. In addition, the sample selection and redshift binning interplays with bias evolution, magnification and other effects not considered here. Finally, we note that the Euclid survey has planned an ambitious tomographic analysis comprising about a dozen redshift bins. The opportunities and challenges posed by such an analysis require separate investigation.

\section{Methods}\label{sec:methods}
\subsection{Generating smooth spectra}\label{ssec:smoothed_spectra}
We use the simulated galaxy samples from \textsc{AbacusSummit} to compute three-dimensional galaxy-galaxy auto- and galaxy-matter cross-power spectra for all the redshifts covered by the simulations. Sample variance in the power spectra extracted from \textsc{AbacusSummit} is significant on large scales compared to the expected error bars, which could lead to biases in our analysis. As described in detail in Appendix \ref{sec:ap.smoothed_spectra}, we suppress this noise by combining the spectra measured from the simulations with theoretical predictions for the linear power spectrum on large scales. This procedure allows us to model the measured power spectra to an accuracy below the sample variance uncertainties of the \textsc{AbacusSummit} simulation box. The reader is referred to Appendix \ref{sec:ap.smoothed_spectra} for a more detailed description of the methodology.

\subsection{Computing simulated data vectors}\label{ssec:computing_data}

\begin{figure*}
\begin{center}
\includegraphics[width=0.49\textwidth]{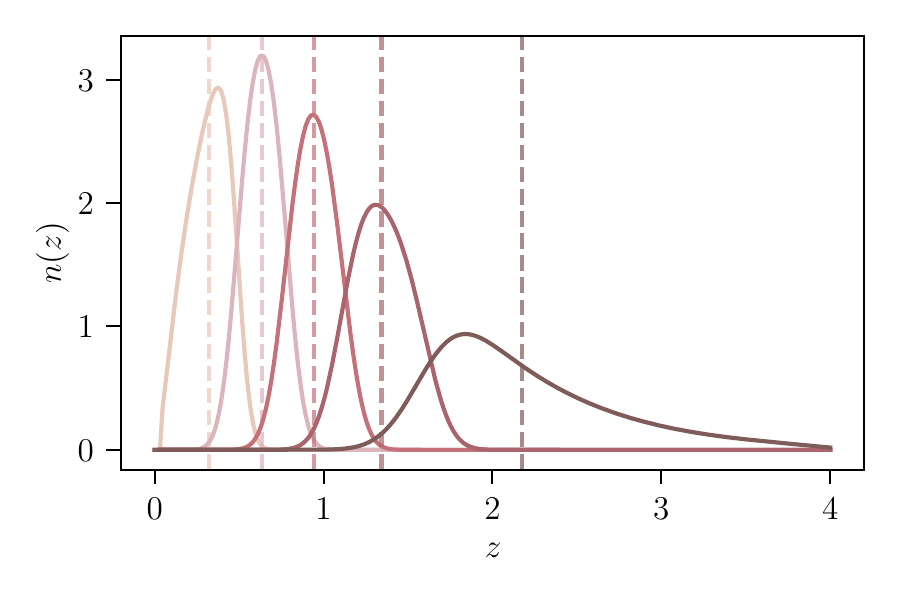}
\includegraphics[width=0.49\textwidth]{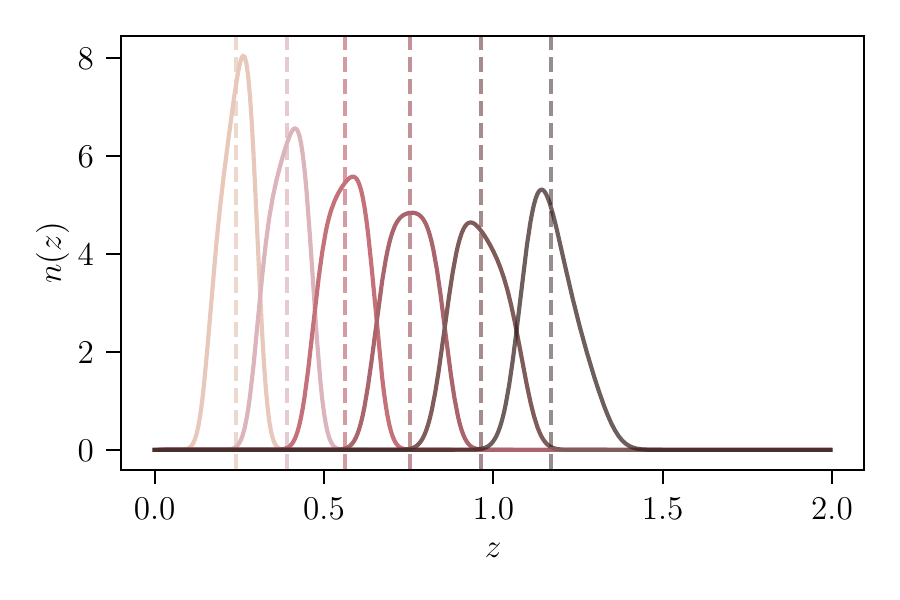}
 \caption{Assumed redshift distributions for the samples considered in this work. \textit{Left panel:} Tomographic redshift bins for both the source sample and the maglim clustering sample. \textit{Right panel:} Tomographic redshift bins for the red clustering sample. The dashed lines denote the mean redshift of each bin. We note that all redshift bins are normalized to unity.}
\label{fig:nzs}
\end{center}
\end{figure*}

In order to compute spherical harmonic power spectra using Eq.~\ref{eq:cell_limber}, we need to specify redshift distributions for the source galaxies as well as the two clustering samples considered in this work. For the source sample, we assume a number density and overall redshift distribution based on the LSST Y10 sample as specified in the LSST DESC Science Requirements document \cite{DESC:2018}. The sample is divided into 5 redshift bins, enforcing an equal number of galaxies in each bin, and assuming Gaussian photometric redshift uncertainties with standard deviation $\sigma_z=0.05(1+z)$. The resulting redshift distributions are shown in the left panel of Fig.~\ref{fig:nzs}, with the inter-bin tails caused by the photo-$z$ uncertainties. For the clustering on the other hand, we assume different distributions for the maglim and the red sample: For the maglim sample we make the assumption that we use the same set of galaxies for lensing and clustering measurements, and thus set the clustering redshift distributions to those assumed for the source sample. For the red sample on the other hand, we assume smaller photometric redshift uncertainties and a lower maximal redshift. The number density and overall redshift distribution is calculated from the luminosity function of red galaxies, following the procedure detailed in Ref.~\cite{2015ApJ...814..145A}. The sample is then divided into 6 redshift bins, evenly spaced in photometric redshift, and assuming a photo-$z$ scatter $\sigma_z=0.02(1+z)$. The resulting tomographic redshift bins are shown in the right panel of Fig.~\ref{fig:nzs}.

Using these redshift distributions, we obtain spherical harmonic power spectra for an LSST-like \txtp analysis. We use the computed three-dimensional power spectra from \textsc{AbacusSummit} and interpolate those in redshift to evaluate Eq.~\ref{eq:cell_limber}. We verified that the number of samples in redshift used in this interpolation was large enough to produce accurate angular power spectra. In a next step, we compute all possible auto- and cross-correlations between probes and tomographic bins, except that we do not include cross-correlations between clustering bins in our analysis. In practice, we perform all projections into angular power spectra using the DESC Core Cosmology Library (\texttt{CCL}\footnote{\url{https://github.com/LSSTDESC/CCL}.}) \cite{Chisari:2019}. This leaves us with a set of 51 or 45 spherical harmonic power spectra for the red or maglim clustering samples respectively, which we evaluate for 22 bandpowers with edges $\ell = \{32,\allowbreak 42,\allowbreak 52,\allowbreak 66,\allowbreak 83,\allowbreak 105,\allowbreak 132,\allowbreak 167,\allowbreak 210,\allowbreak 265,\allowbreak 333,\allowbreak 420,\allowbreak 528,\allowbreak 665,\allowbreak 838,\allowbreak 1054,\allowbreak 1328,\allowbreak 1672,\allowbreak 2104,\allowbreak 2650,\allowbreak 3336,\allowbreak 4200,\allowbreak 5287\}$. When applying a scale cut based on a maximal wavenumber $k_{\mathrm{max}}$ for power spectra involving clustering, we only keep bandpowers for which the mean of the $\ell$-bin satisfies $\ell < k_{\mathrm{max}}\chi(\bar{z})$, where $\chi(\bar{z})$ denotes the comoving distance to the mean redshift of the respective bin. For the cosmic shear power spectra on the other hand, we fix $\ell_{\mathrm{max}}=2000$ for all cases.

As the aim of this work is to assess the performance of different galaxy bias models, we need to prevent biasing our results due to inaccuracies in modeling the matter power spectrum used for cosmic shear predictions. We therefore rescale the \textsc{AbacusSummit} matter power spectrum to match the \textsc{halofit} \cite{Smith:2003, Takahashi:2012} prediction (see e.g. Ref.~\cite{Goldstein:2022} for a similar approach)\footnote{As shown in Ref.~\cite{Hadzhiyska:2023}, the matter power spectrum from \textsc{AbacusSummit} agrees with that from \textsc{halofit} to a level of 2\% for $k<0.5 h$ Mpc$^{-1}$ for the fiducial \textsc{AbacusSummit} cosmology. For other cosmological models however, the discrepancies increase; an example is shown in Ref.~\cite{DeRose:2023}, who find differences of up to 4\% between the \texttt{Aemulus} $\nu$ simulations and \texttt{HMCode2020} \cite{Mead:2021} at $k=0.2 h$ Mpc$^{-1}$. We expect these to be comparable to the differences between \textsc{Abacus} and \textsc{halofit} for a range of cosmological models.}. In Sec.~\ref{sssec:res.samples.halofit}, we explore the effect of not rescaling cosmic shear power spectra. We note that all other power spectra are left unchanged and taken from the \textsc{AbacusSummit} simulations, which means that matter power spectrum mis-modeling might affect our results for $P_{\delta_{g}\delta_{g}}$ and $P_{\delta_{g}\delta_{m}}$. As described in Sec.~\ref{ssec:pt.choices}, we test the sensitivity of LPT and EPT to implementation choices for modeling the matter power spectrum, but leave a more extensive analysis to future work.

We generate covariances for these data as outlined in Sec.~\ref{sec:obs}, modeling LSST Y10 data based loosely on the LSST DESC Science Requirements Document (SRD) \cite{DESC:2018}. We set the fraction of sky covered to $f_{\mathrm{sky}}=0.4$. We model the noise for cosmic shear following Ref.~\cite{Chang:2013}, setting the effective number density of galaxies to $n_{\mathrm{eff}} = 27 \; \mathrm{arcmin}^{-2}$, and the intrinsic ellipticity to $\sigma_{e} = 0.28$. For galaxy clustering on the other hand, we follow an alternative approach: We find the considered samples to exhibit a significant level of non-Poissonian stochasticity. Subtracting an estimate for the Poissonian shot noise of the sample from the simulated three-dimensional power spectra would therefore lead to biased results. We circumvent this issue by considering clustering noise levels consistent with the stochasticity determined by the HODs of the simulated galaxy samples. In practice this means that we do not subtract any Poisson shot noise estimate from the simulated power spectra, but account for it in the theoretical modeling as described in Sections \ref{ssec:pt.ept} and \ref{ssec:pt.lpt}\footnote{We note that this approach amounts to assuming that the noise levels determined by the assumed HODs are consistent with the clustering noise levels expected for LSST Y10 data. Given that we have modeled our samples according to observational results, we believe this to be an acceptable approximation.}.

Finally we consider an extension of our LSST-like \txtp analysis: we additionally model a joint analysis of our fiducial red galaxy sample from LSST with CMB lensing from CMB S4. Practically, we thus extend our fiducial \txtp data vector with all possible cross-correlations with the CMB lensing convergence leading to a set of 62 spherical harmonic power spectra. To model CMB S4 lensing and its associated covariance through Equations \ref{eq:cell_limber} and \ref{eq:cov_G+sys}, we follow the specifications given in the CMB S4 wiki\footnote{The description can be found at \url{https://cmb-s4.uchicago.edu/wiki/index.php/Survey_Performance_Expectations}.} (private communication Toshiya Namikawa and Colin Hill), and use the CMB lensing reconstruction noise computed using the \texttt{cmblensplus}\footnote{The package can be found at \url{https://github.com/toshiyan/cmblensplus}.} code. We assume a common sky coverage of LSST and CMB S4 given by $f_{\mathrm{sky}}=0.4$. 

To fully simulate the expected data vectors, one should draw a noise realization from the covariance matrix and add it to the noiseless prediction for the observables. We choose not to do this: for an ideal model, our result should thus center on the true values of cosmological parameters and the corresponding $\chi^2$ should satisfy $\chi^2=0$. Therefore, any deviation in $\chi^2$ and parameter values indicates systematic errors in the theory modeling (see also Section \ref{ssec:model-performance}).

\subsection{Deriving best-fit parameters and associated uncertainties}\label{ssec:best_fit}

For all bias models and galaxy samples analyzed in this work, we assume a Gaussian likelihood with simulated data and covariance matrix as described above. In order to compare the performance of different bias models and test if the fits to the simulated data return unbiased constraints on cosmological parameters, we would ideally sample the likelihood in an Monte Carlo Markov Chain (MCMC) and thus derive the full posterior of our model parameters. As we are considering a large number of bias models and different implementations, this is computationally expensive (although see Ref.~\cite{2023arXiv230111895H} for potential ways to accelerate this process). We therefore resort to a simplified approach: for all models considered, we run an optimization algorithm to maximize the likelihood and determine the corresponding values of cosmological and bias parameters. Specifically, we use the \texttt{BOBYQA}\footnote{The code can be found at \url{https://numericalalgorithmsgroup.github.io/pybobyqa/build/html/index.html}.} algorithm \cite{Powell:2009, Cartis:2018a, Cartis:2018b} through the \texttt{cobaya}\footnote{The code can be found at \url{https://cobaya.readthedocs.io/en/latest/}.} package as well as Powell's optimization method as implemented in \texttt{scipy} \cite{Powell:1964}. We choose the latter method as the parameter space considered in this analysis exhibits significant degeneracies, and we find that directional optimization as implemented in Powell's method outperforms \texttt{BOBYQA} in most cases. We therefore use Powell's method for most of our fiducial results but always compare to the results obtained using \texttt{BOBYQA} to test the robustness of our conclusions. In addition, we perform a number of further tests to ensure the stability of our results: we rerun each case multiple times to test stability against changing the initial conditions, we also vary the convergence criterion for both methods as well as the number of starting points for \texttt{BOBYQA}, finding consistent results for reasonable settings of these parameters.

In order to characterize biases on cosmological parameters, we additionally need to estimate parameter uncertainties. In this work, we use a Fisher matrix (FM) formalism to forecast uncertainties on cosmological and nonlinear bias parameters. The Fisher matrix allows for propagation of uncertainties on observables (in our case, the spherical harmonic power spectra) to corresponding uncertainties on model parameters. Under the assumption that the dependence of the data covariance matrix $\mathbf{C}$ on the parameters of interest $\theta_{\alpha}$ can be neglected (which is a very good approximation \cite{2019OJAp....2E...3K}), the Fisher matrix $F$ is given by (see e.g. Refs.~\cite{Fisher:1935, Kendall:1979, Tegmark:1997})
\begin{equation}
F_{\alpha \beta} = \frac{\partial \mathbf{D}}{\partial \theta_{\alpha}}\mathbf{C}^{-1}\frac{\partial \mathbf{D}}{\partial \theta_{\beta}},
\end{equation}
where $\mathbf{D}$ denotes the data vector of a given experiment (in our case, a list containing all the power spectra used in the analysis).
The Cram\'er-Rao bound then states that the uncertainty on $\theta_{\alpha}$, marginalized over all other $\theta_{\beta}$ satisfies
\begin{equation}
\Delta \theta_{\alpha} \geq \sqrt{(F^{-1})_{\alpha \alpha}}.
\end{equation}
In this analysis, we consider a fiducial model characterized by two cosmological and a set of six nonlinear bias parameters per clustering redshift bin, i.e. $\boldsymbol \theta = \{\sigma_{8}, \Omega_{c}, b^{p, i}_{1}, \nobreak b^{p, i}_{1p},\nobreak b^{p, i}_{2},\nobreak b^{p, i}_{s^{2}},\nobreak b^{p, i}_{\nabla^{2}},\nobreak P^{p, i}_{\mathrm{SN}}\}$, where $i$ denotes the redshift bin (i.e. $i=1\ldots 5/6$ in our analysis), and $p = L$ in the case of LPT and HEFT. Computing the Fisher matrix requires the assumption of fiducial values for the model parameters, which we set to the best-fit values derived from the minimizer. We compute derivatives of the observables with respect to the model parameters numerically using a five-point stencil with step size $\epsilon = 0.01\theta$, where $\theta$ denotes any parameter considered in our analysis\footnote{For parameters with fiducial value of zero, we set  $\epsilon = 0.01$.}. We test the stability of our results by varying the parameter $\epsilon$ and find our results to be largely insensitive to this choice. 

Fisher matrix analyses are prone to numerical instabilities (see e.g. Refs.~\cite{Euclid:2020, Bhandari:2021, Yahia:2021}). An additional complication in our case is that the posterior is likely to exhibit non-Gaussian features due to degeneracies between model parameters. We therefore test the robustness of our results by comparing our FM constraints to those derived using an MCMC for selected cases, finding reasonable agreement as described in detail in Appendix \ref{sec:ap.MCMC_comp}.

\subsection{Assessing model performance}\label{ssec:model-performance}

For all the bias models considered in this analysis, we assess model performance as a function of the smallest scale (largest wavenumber) $k_{\mathrm{max}}$ considered for galaxy clustering and galaxy-galaxy lensing data. As we are working with angular power spectra, we convert this quantity to a maximal angular multipole for each redshift bin using the approximate relation $\ell_{\mathrm{max}} = k_{\mathrm{max}}\chi(\bar{z})$, where $\bar{z}$ denotes the mean redshift of each bin.

In this work, we assess model performance in several different ways. First, we test if the values of the cosmological parameters recovered from our fits are consistent with their fiducial values within statistical uncertainties as derived from the Fisher matrix analysis. This test does not guarantee good model performance, as even models that do not fit the data can potentially yield unbiased parameter constraints. We therefore additionally assess goodness-of-fit (GOF) of all models considered. As described in Section \ref{ssec:computing_data}, we work with noiseless data vectors. 
 We therefore do not expect the $\chi^{2}$ of the fit to follow a $\chi^{2}$-distribution with mean given by the number of degrees of freedom in the data, but rather be significantly smaller and dominated by the model performance and numerical errors. Based on this, we devise a set of two $\chi^{2}$-tests which we use to validate our results:
 \begin{description}
     \item[Noiseless GOF:] First, we determine the minimal $\chi^{2}$-value of the fit to the noiseless data vector, which we will call $\chi^{2}_{\mathrm{theory}}$. We need to make sure that this quantity is significantly smaller than the $\Delta \chi^{2}$ allowed by statistical uncertainties and the number of degrees-of-freedom of our model. In the presence of noise, we expect the minimal $\chi^{2}$ of the fit to roughly follow a $\chi^{2}$-distribution with number of degrees-of-freedom $dof = n_{\mathrm{data}}-n_{\mathrm{param}}$, where $n_{\mathrm{data}}$ denotes the number of data points, and $n_{\mathrm{param}}$ is the number of parameters in the model\footnote{Strictly, this is only valid for linear models with independent basis functions (see e.g. Ref.~\cite{Andrae:2010}). We will nevertheless use this criterion throughout this analysis, as we are mainly interested in deriving a threshold and not accurate probabilities-to-exceed.}. We call this latter quantity $\chi^{2}_{\mathrm{noise}}$, which in our case is given by $\chi^{2}_{\mathrm{noise}} = n_{C_{\ell}} - n_{\mathrm{param}} - n_{C_{\ell}^{\gamma \gamma}}$, where $ n_{C_{\ell}}$ denotes the total number of spherical harmonics $C_{\ell}$, and $n_{C_{\ell}^{\gamma \gamma}}$ denotes the number of shear-only data points. As described above, we set the matter power spectrum of the simulations to its \textsc{halofit} prediction, and we therefore do not count the shear power spectra as degrees-of-freedom of the model.
     
     In our first test we thus assess that the sum of the expected $\chi^{2}_{\mathrm{noise}}$ and $\chi^{2}_{\mathrm{theory}}$ is reasonably likely given a $\chi^{2}$ distribution with mean $\chi^{2}_{\mathrm{noise}}$. This tests ensures that the differences due to systematic uncertainties in the theory are subdominant to statistical uncertainties. For the noiseless data vector, we also examine the size of the fit residuals.
     
     \item[Noisy GOF:] Secondly, we use our synthetic data vector alongside the covariance matrix to create a noisy realization of the data. We then fit this data using all the bias models discussed and test that the best-fit yields a $\chi^{2}$ consistent with $\chi^{2}_{\mathrm{noise}} + \chi^{2}_{\mathrm{theory}}$. In contrast to earlier, we set $\chi^{2}_{\mathrm{noise}} = n_{C_{\ell}} - n_{\mathrm{param}}$, as we do sample the cosmic shear data as well. We additionally assert that the histogram of the fit residuals is consistent with a Gaussian.
 \end{description} 
In principle, we require the bias models to pass both these tests in order to be deemed acceptable. Given the large number of cases considered in our analysis, and the computational cost of performing the `Noisy GOF' test for each case, in practice we only perform the second test for our fiducial case. As we find largely consistent results between the two tests in this case, we resort to only the `Noiseless GOF' test for the remainder of this work. As is customary, we set a threshold of a $p$-value larger than $p=0.05$ for a model to pass a specific test.

\section{Results}\label{sec:results}

In the following sections, we present the results of our analysis. Unless stated otherwise, all results will be presented for our fiducial, red sample. At the end of the section, we also give a summary of the results obtained for the alternative samples considered.

\subsection{HEFT}\label{ssec:results.heft}
\begin{figure*}
  \begin{center}
    \includegraphics[width=0.45\textwidth]{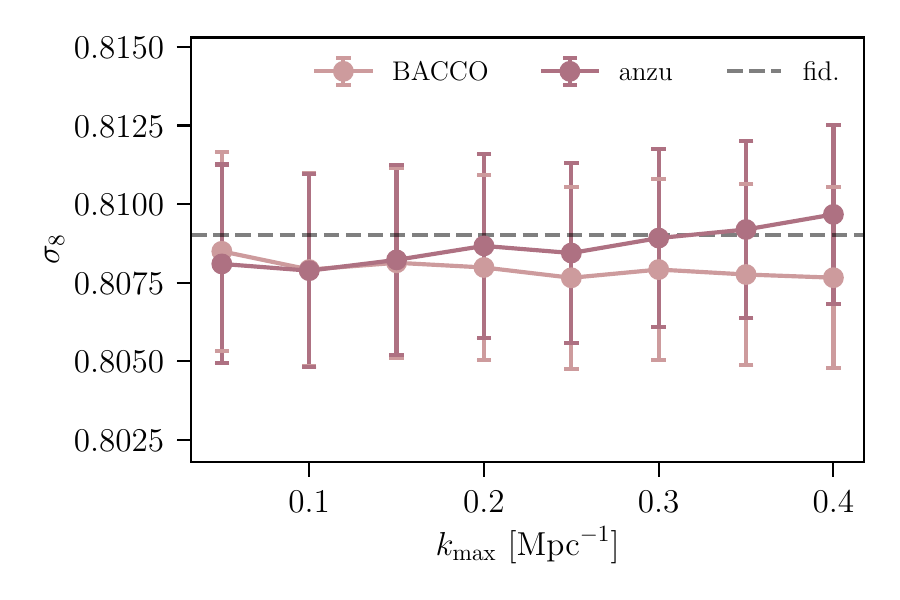}
    \includegraphics[width=0.45\textwidth]{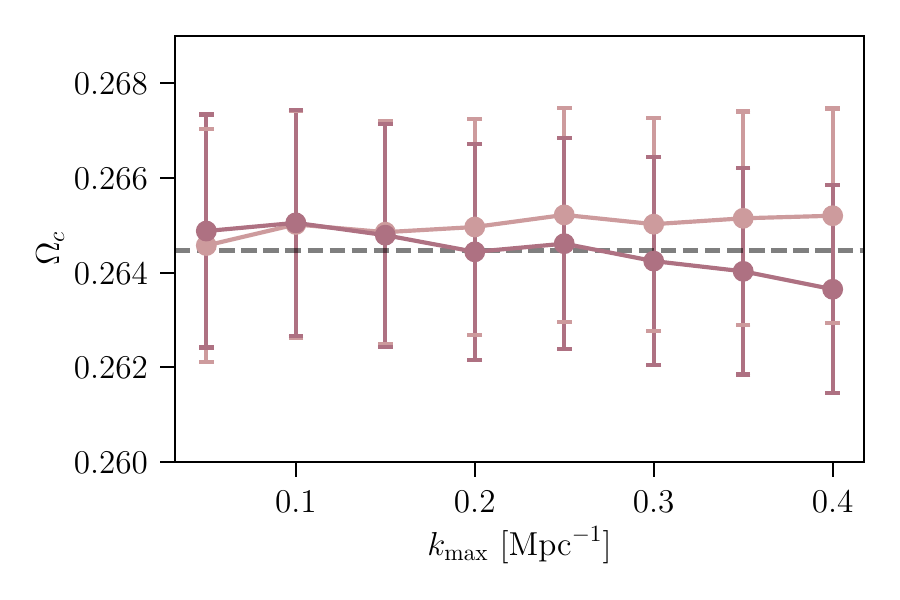}\\
    \includegraphics[width=0.45\textwidth]{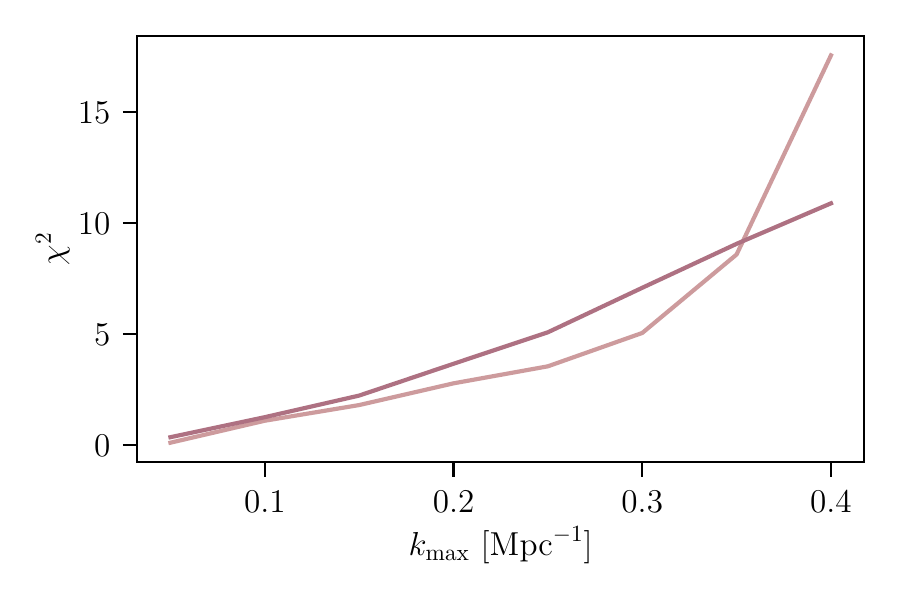}
    \includegraphics[width=0.45\textwidth]{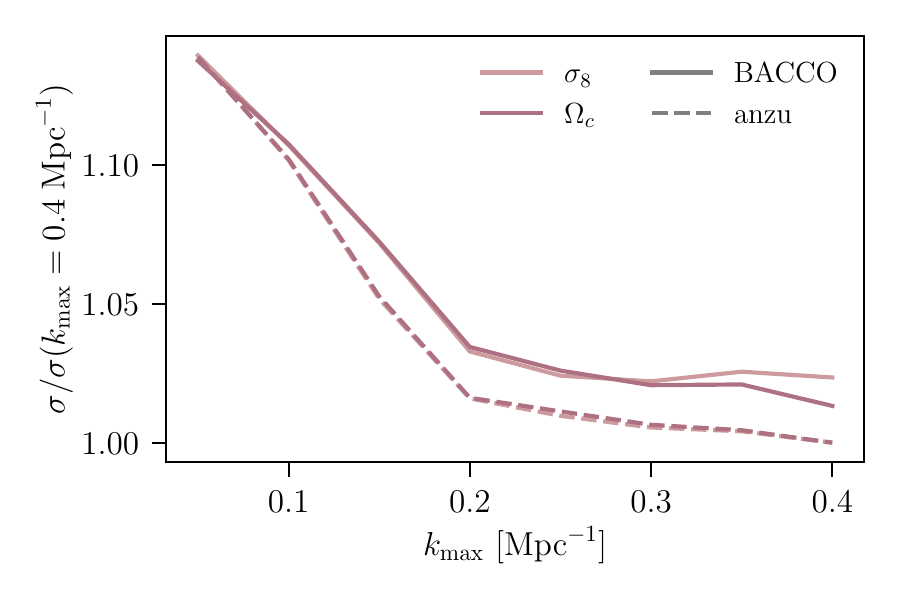}
    \caption{Results of fitting the red galaxy sample with the HEFT implementations considered in this work, \texttt{anzu} and \texttt{BACCO}. \textit{Top panels:} Recovered values of $\sigma_{8}$ and $\Omega_{c}$ as a function of maximal wavenumber, $k_{\mathrm{max}}$. \textit{Bottom panels:} Minimum $\chi^{2}$-value obtained from the fit, and relative uncertainties on $\sigma_{8}$ and $\Omega_{c}$, as a function of $k_{\mathrm{max}}$. The uncertainties are normalized with respect to those obtained for \texttt{anzu} at $k_{\mathrm{max}}=0.4 \; \mathrm{Mpc}^{-1}$.}\label{fig:parameter-fits-heft}
  \end{center}
\end{figure*}

We first investigate the performance of the two HEFT implementations, \texttt{anzu} and \texttt{BACCO}. As discussed in Sec.~\ref{ssec:best_fit}, we minimize the likelihood with respect to two cosmological parameters $\{\Omega_{c}, \sigma_{8}\}$, a well as six bias parameters per redshift bin, $\{b_{1}, b_{1p}, b_{2}, b_{s^{2}}, b_{\nabla^{2}}, P_{\mathrm{SN}}\}$. We keep the remaining cosmological parameters fixed at their fiducial values adopted in \textsc{AbacusSummit}.

The upper panels of Fig.~\ref{fig:parameter-fits-heft} show the values of the cosmological parameters $\sigma_{8}$ and $\Omega_{c}$ recovered from fitting our fiducial red galaxy sample with \texttt{anzu} and \texttt{BACCO}. As can be seen, both HEFT methods are able to recover the true parameter value within $1\sigma$ uncertainty for all values of $k_{\mathrm{max}}$ considered\footnote{For \texttt{BACCO} as well as the PT-based methods discussed in Sec.~\ref{ssec:pt}, we see a sign that the recovered values of $\sigma_{8}$ and $\Omega_{c}$ are systematically under- or over-estimated. These effects are not significant at the precision we are working in, but might be a sign for biases that could become important at higher signal-to-noise. However, this could also be a feature of the particular simulation realization as the bias parameters derived as a function of $k_{\mathrm{max}}$ are highly correlated.}. The bottom left panel of Fig.~\ref{fig:parameter-fits-heft} shows the minimal $\chi^{2}$ obtained for both models, and we can use these values to assess the goodness-of-fit in all cases. The most stringent test for each model is obtained at the maximal $k_{\mathrm{max}}$ considered in our analysis, $k_{\mathrm{max}}=0.4 \; \mathrm{Mpc}^{-1}$ (corresponding to $0.6\,h\mathrm{Mpc}^{-1}$), and in the following we only discuss this case. The results of our goodness-of-fit tests are described in detail in Appendix \ref{ssec:ap.gof.heft}, and here we only give a summary of our main findings. For the noiseless data vector (noiseless GOF), we find that both models, \texttt{anzu} and \texttt{BACCO}, pass our $\chi^{2}$ tests\footnote{We note that for both HEFT methods as well as for our fiducial implementations of LPT and EPT, we find consistent results when we restrict our data vector to only include galaxy clustering and galaxy-galaxy lensing in a so-called 2x2pt analysis.}. In addition, most obtained fit residuals lie within $1\sigma$ of the synthetic data and the relative differences are generally of order $1\%$. We additionally consider two noisy realizations of the data vector (noisy GOF), finding all tests to pass with the exception of the analysis of one of the noisy realizations with \texttt{BACCO}, which we consider to be a statistical fluctuation. In all cases considered, we further find the distribution of fit residuals to be consistent with a Gaussian. 

In order to test for possible systematic modeling uncertainties, we compare the consistency of the bias parameter values obtained for the two HEFT implementations. While a detailed description is deferred to Appendix \ref{sec:ap.bias_model_cons}, we generally find consistency between bias parameters derived using \texttt{anzu} or \texttt{BACCO}. The only exception being $b_{\nabla^{2}}$ for which we find consistently higher values for \texttt{BACCO} than we do for \texttt{anzu}. We attribute these differences to slightly different implementations of the nonlocal power spectra in the two HEFT models, as well as known sensitivities of $b_{\nabla^{2}}$ to small-scale implementation details such as different smoothing scales used when deriving template spectra (see e.g. Ref.~\cite{Desjacques:2018}). 

Finally, we expect that systematics in the modeling might manifest themselves as dependencies of the derived bias parameter values on the maximal wavenumber $k_{\mathrm{max}}$ used in the analysis. In order to test for this, we thus compare the bias parameter values obtained for \texttt{anzu} when varying $k_{\mathrm{max}}$ from $k_{\mathrm{max}}=0.1 \; \mathrm{Mpc}^{-1}$ to $k_{\mathrm{max}}=0.4 \; \mathrm{Mpc}^{-1}$, finding largely consistent results.

Based on these results, the two HEFT implementations considered in this work thus appear promising  for analyzing high-precision data as expected from LSST.

\subsection{LPT/EPT}\label{ssec:pt}
\begin{figure*}
  \begin{center}
    \includegraphics[width=0.45\textwidth]{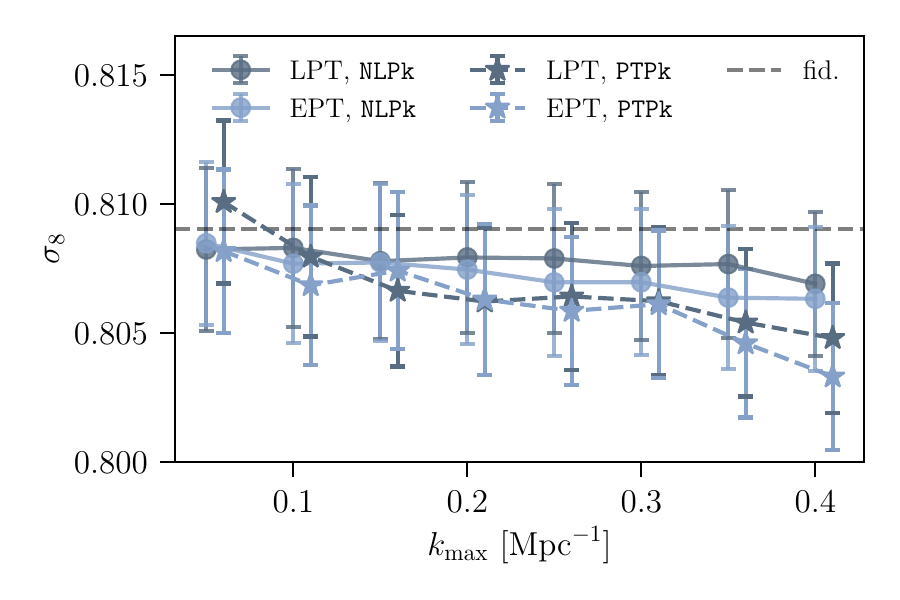}
    \includegraphics[width=0.45\textwidth]{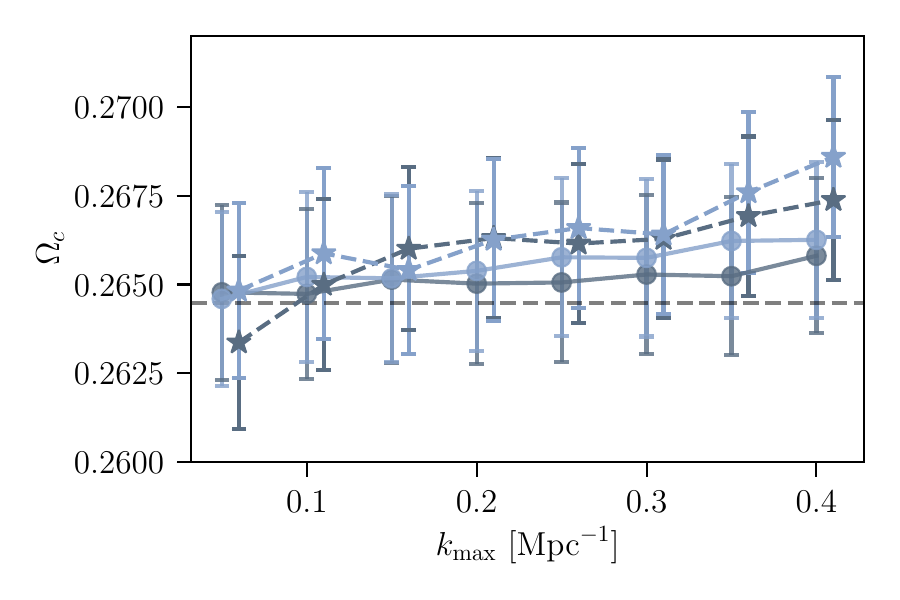}\\
    \includegraphics[width=0.45\textwidth]{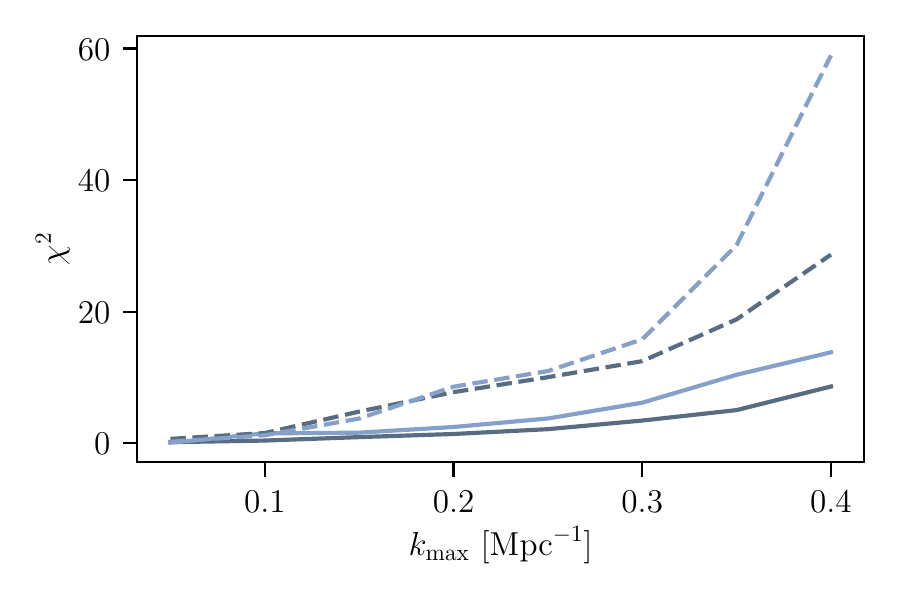}
    \includegraphics[width=0.45\textwidth]{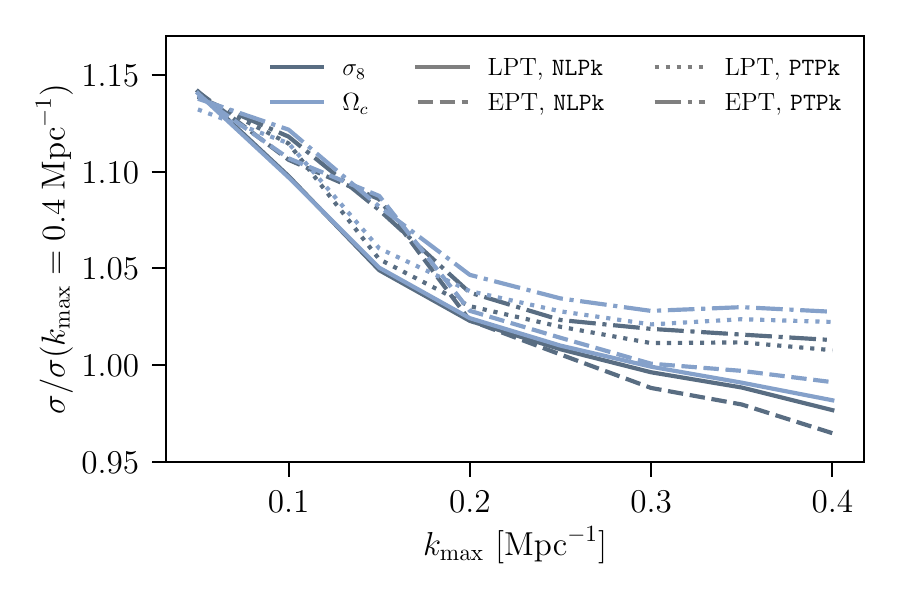}
    \caption{Results of fitting the red sample with the \texttt{NLPk} and \texttt{PTPk} implementations of LPT and EPT. \textit{Top panels:} Recovered values of $\sigma_{8}$ and $\Omega_{c}$ as a function of maximal wavenumber, $k_{\mathrm{max}}$. The data points for the \texttt{PTPk} case have been displaced by $\Delta k_{\mathrm{max}}=0.01 \; \mathrm{Mpc}^{-1}$ for clarity. \textit{Bottom panels:} Minimum $\chi^{2}$-value obtained from the fit, and relative uncertainties on $\sigma_{8}$ and $\Omega_{c}$, as a function of $k_{\mathrm{max}}$. The uncertainties are normalized with respect to those obtained for \texttt{anzu} at $k_{\mathrm{max}}=0.4 \; \mathrm{Mpc}^{-1}$.}\label{fig:parameter-fits-spt}
  \end{center}
\end{figure*}

In addition to the HEFT implementations discussed above, we also assess the performance of two PT-based models, Lagrangian and Eulerian perturbation theory. As described in Sec.~\ref{ssec:pt.choices}, we consider two different ways to model the matter power spectrum $P_{\delta_m\delta_m}(z, k)$ as well as nonlocal contributions in $P_{\delta_{g}\delta_{g}}$ and $P_{\delta_{g}\delta_{m}}$. These are denoted \texttt{NLPk} for the \textsc{halofit}-based model, and \texttt{PTPk} for the fully perturbative case. We remind the reader that this is separate from our rescaling of the matter power spectrum for modeling cosmic shear (as described in Sec.~\ref{ssec:computing_data}), which is performed for both \texttt{NLPk} and \texttt{PTPk}.

The results obtained from fitting the red galaxy sample with the LPT and EPT models are shown in Fig.~\ref{fig:parameter-fits-spt}. Similarly to the results for the two HEFT implementations, we find that for the \texttt{NLPk} models, the recovered values for $\sigma_{8}$ and $\Omega_{c}$ agree with their fiducial values within $1\sigma$ in all cases. For the \texttt{PTPk} models on the other hand, we find the recovered values for the cosmological parameters to start showing biases of around $1\sigma$ for $k_{\mathrm{max}} \gtrsim 0.3 \; \mathrm{Mpc}^{-1}$, increasing to 1.5 to $2\sigma$ for $k_{\mathrm{max}} = 0.4 \; \mathrm{Mpc}^{-1}$. This is borne out by our goodness-of-fit tests, as described in Appendix \ref{ssec:ap.gof.pt}. We find that both \texttt{NLPk} methods pass our tests on the noiseless data. Despite both \texttt{PTPk} models returning biased constraints on cosmological parameters, we find that the fully perturbative implementation of LPT still passes our goodness-of-fit test, while EPT does not. These results suggest that predicting the galaxy power spectra considered in this analysis up to a maximal wavenumber of $k_{\mathrm{max}} = 0.4 \; \mathrm{Mpc}^{-1}$ requires more accurate modeling of the matter power spectrum and nonlocal terms, which we achieve here through our use of the \textsc{halofit} fitting function. These findings are consistent with previous studies, such as e.g. Refs.~\cite{Pandey:2020, Goldstein:2022, Pandey:2021}, which also found significantly improved performance of EPT bias models when using non-perturbative predictions for the matter power spectrum.  

Nevertheless, comparing to previous analyses that roughly find LPT and EPT bias models to only reach up to $k_{\mathrm{max}} \sim 0.3 \; h \; \mathrm{Mpc}^{-1}$ (see e.g. Ref.~\cite{Fonseca:2020}), these results are somewhat surprising. We believe this to be due both to differences in the considered data vector as well as the criterion chosen to assess goodness-of-fit: first, in this work we focus on spherical harmonic power spectra, which constitute a line-of-sight projection of the underlying three-dimensional power spectrum usually studied in the literature. The line-of-sight projection results in the smoothing of power spectrum features and might thus increase the reach of perturbative nonlinear bias models. In contrast to other works, we additionally do not include redshift space distortions (RSDs) in our analysis, as their impact on projected statistics is expected to be small (although potentially not negligible \citep{2015ApJ...814..145A}), which might also lead to better performance of PT-based models. Finally, our results could also be affected by our pragmatic approach to assess the goodness-of-fit of a given model: we only require the fit to return unbiased constraints on $\Omega_{c}$ and $\sigma_{8}$, as well as returning a $\chi^{2}$-value passing the criteria discussed in Sec.~\ref{ssec:model-performance}. In particular, we do not make any requirements on model residuals as was done in previous analyses, and we include a theoretical relative error floor of $1\%$ in all our covariances. We believe it is for these reasons that our results find perturbative bias models to be applicable up to slightly smaller scales than previous analyses. In Appendix \ref{ap:sec:unit} we further investigate this by performing an analogous analysis of three-dimensional power spectrum data in real space from the \textsc{UNIT} simulation \cite{Chuang:2019}.

As above, in a final test we compare the bias parameter values obtained using our fiducial \texttt{NLPk} implementation of LPT and EPT. As discussed in Appendix \ref{sec:ap.bias_model_cons}, we generally find good agreement between both methods, although the EPT approach seems to prefer significantly larger values for $b_{\nabla^{2}}$ and $P_{\rm SN}$ compared to LPT. In addition, the recovered bias parameters for the two \texttt{NLPk} models are generally consistent with those obtained from the HEFT methods. The only exception being $b_{\nabla^{2}}$, for which we obtain significantly lower values for the HEFT implementations than we do for the perturbation-theory-based models, particularly at high redshift. As further discussed in Appendix \ref{sec:ap.bias_model_cons}, this could be a possible sign for larger model inaccuracies in the PT-based methods as compared to HEFT.

\subsection{Bias consistency relations}\label{ssec:res.coev}

\begin{figure*}
\begin{center}
\includegraphics[width=0.45\textwidth]{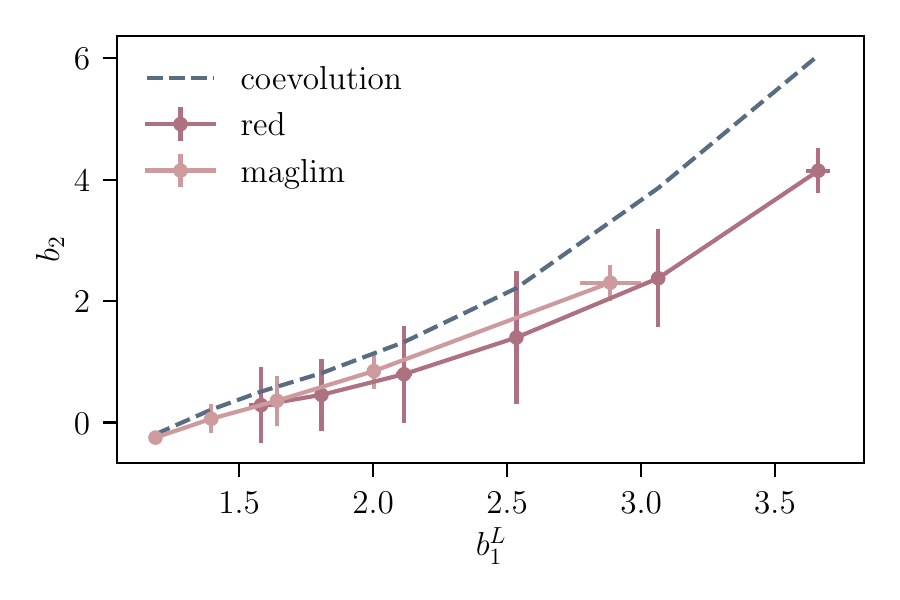}
\includegraphics[width=0.45\textwidth]{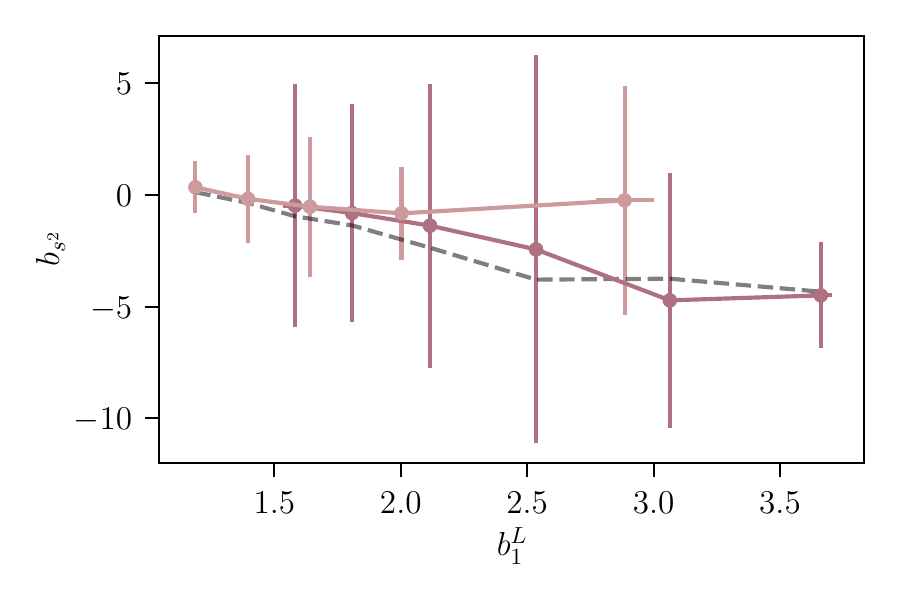}\\
\includegraphics[width=0.45\textwidth]{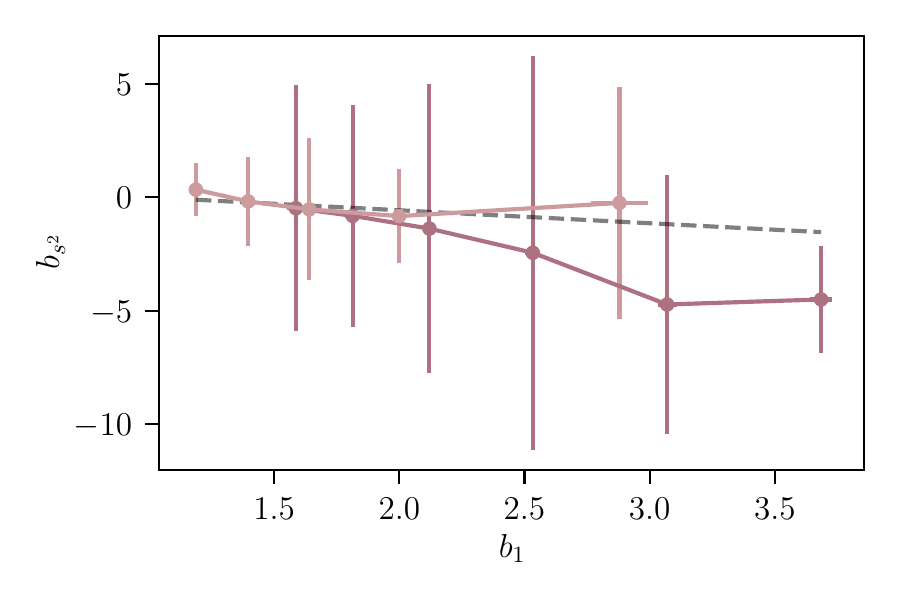}
 \caption{Comparison of the theoretical bias consistency relations given in Eq.~\ref{eq:bias-consistency} to the relations obtained from fitting the \textsc{AbacusSummit} data with the \texttt{NLPk} implementations of EPT and LPT. All cases show results for $k_{\mathrm{max}} = 0.4 \; \mathrm{Mpc}^{-1}$. \textit{Top panels:} Consistency relations between LPT and EPT biases. \textit{Bottom panel:} Consistency relation for EPT biases in case of local-in-matter-density (LIMD) Lagrangian bias.}
\label{fig:bias-consistency}
\end{center}
\end{figure*}

As discussed in Sec.~\ref{ssec:pt.bias-consistency}, we expect Eulerian and Lagrangian bias parameters to exhibit consistency relations under purely gravitational evolution. While we do not anticipate these relations to hold exactly due to non-gravitational processes involved in galaxy formation, they present a complementary means for validating our results, and our bias parameter fitting procedure in particular. We therefore investigate the validity of the coevolution relations given in Eq.~\ref{eq:bias-consistency}. The results using the \texttt{NLPk} models with $k_{\mathrm{max}} = 0.4 \; \mathrm{Mpc}^{-1}$ are illustrated in Fig.~\ref{fig:bias-consistency} both for our fiducial, red sample as well as the maglim sample, which we discuss in more detail below: in the upper panels we show the value of $b_{2}$ and $b_{s^2}$ as a function of $b^{L}_{1}, b^{L}_{2}$ and $b^{L}_{1}, b^{L}_{s^2}$, respectively, while the lower panel shows the relation between $b_{s^2}$ and $b_{1}$. In these figures, we also show the uncertainties associated with the various bias parameters. However, as discussed in more detail in Sections \ref{sec:ap.bias_model_cons} and \ref{sec:ap.MCMC_comp}, we find the Fisher matrix-derived uncertainties on bias parameters to be rather unstable for the full model, while they match their MCMC counterparts if we fix either $b_{s^2}$ or $b_{\nabla^2}$. Where possible we thus always show error bars obtained setting $b_{s^2}=0$ (even though this has no effect on cosmological parameter uncertainties). As this is not possible in this case, we caution the reader to keep these instabilities in mind when interpreting the uncertainties. As can be seen from the figure, the obtained bias values largely follow the theoretically expected trends. However, we do see signs of significant deviations, particularly in the consistency between $b_{2}$ and $b^{L}_{1}, b^{L}_{2}$. Further investigating this, we also compare our results to the empirical relations between Lagrangian bias parameters found in Ref.~\cite{Zennaro:2022} and repeat the analysis for the \textsc{UNIT} simulations, finding similar results in both cases. This suggests that these findings are not driven by our use of \textsc{AbacusSummit}, and that empirical bias relations do not provide a better fit to our data than those derived from coevolution. Another possible reason for these results might be our choice to not consider third-order bias parameters in EPT, which breaks the full correspondence between EPT and LPT (as discussed in Sec.~\ref{ssec:pt.bias-consistency}), and might lead to some of these differences being absorbed by lower-order bias parameters. We leave a further investigation to future work. 

\subsection{Alternative samples}\label{ssec:results.alternative_samps}

We investigate the generality of the results presented so far by applying the nonlinear bias models considered in this work to the analysis of three alternative samples, as described in Sections \ref{sec:simulations} and \ref{sec:methods}: a magnitude-limited galaxy sample, a galaxy sample featuring assembly bias, and a joint analysis of our fiducial data vector with CMB lensing data from a CMB S4-like experiment. For all the following cases, we only show the results for $\sigma_{8}$ as the results for $\Omega_{c}$ are qualitatively similar.

\subsubsection{Magnitude-limited sample}\label{sssec:res.samps.maglim}
\begin{figure*}
\begin{center}
\includegraphics[width=0.45\textwidth]{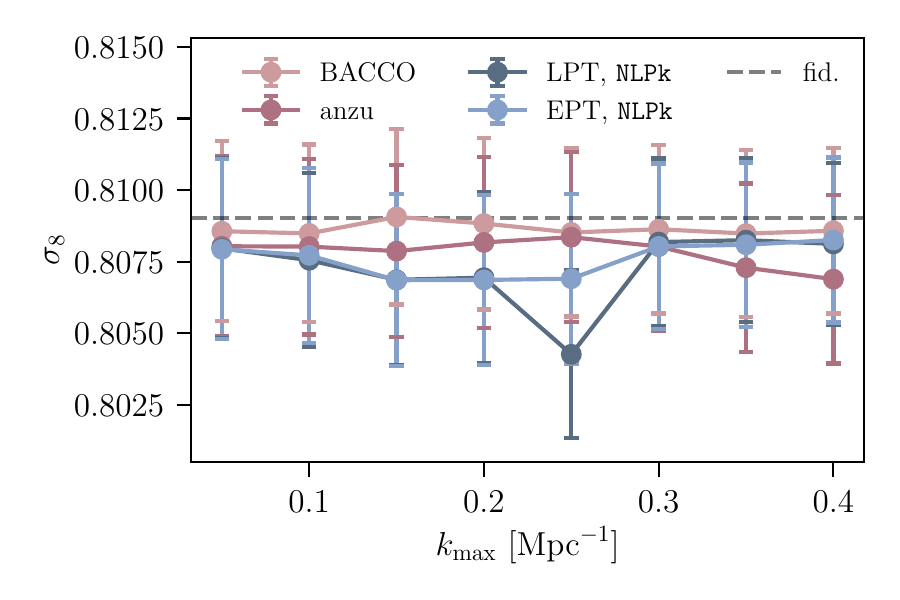}
\includegraphics[width=0.45\textwidth]{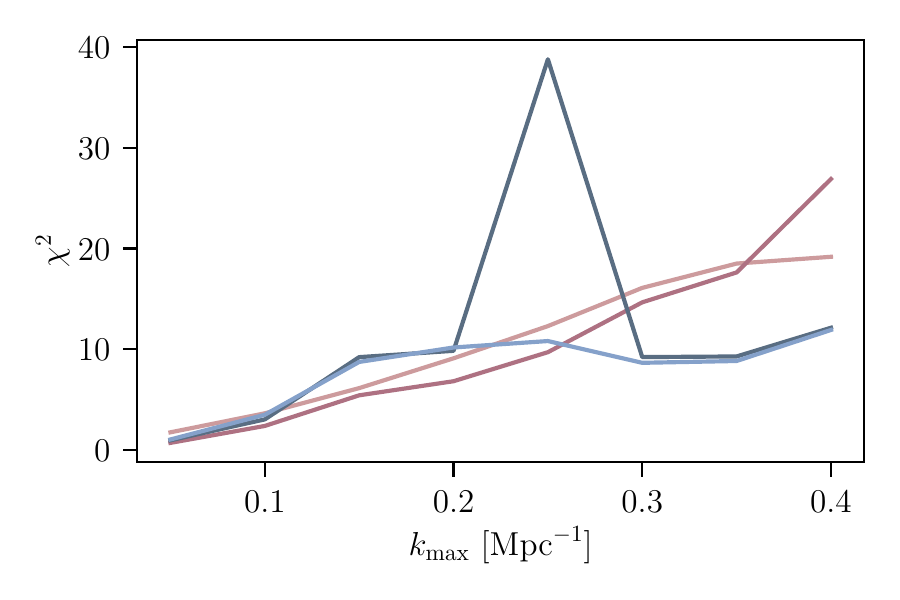}\\
\includegraphics[width=0.45\textwidth]{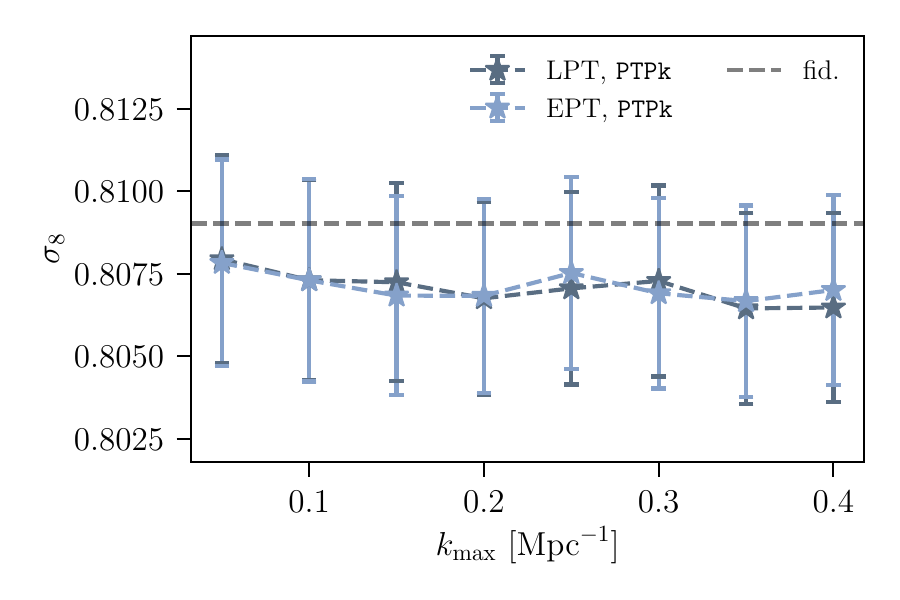}
\includegraphics[width=0.45\textwidth]{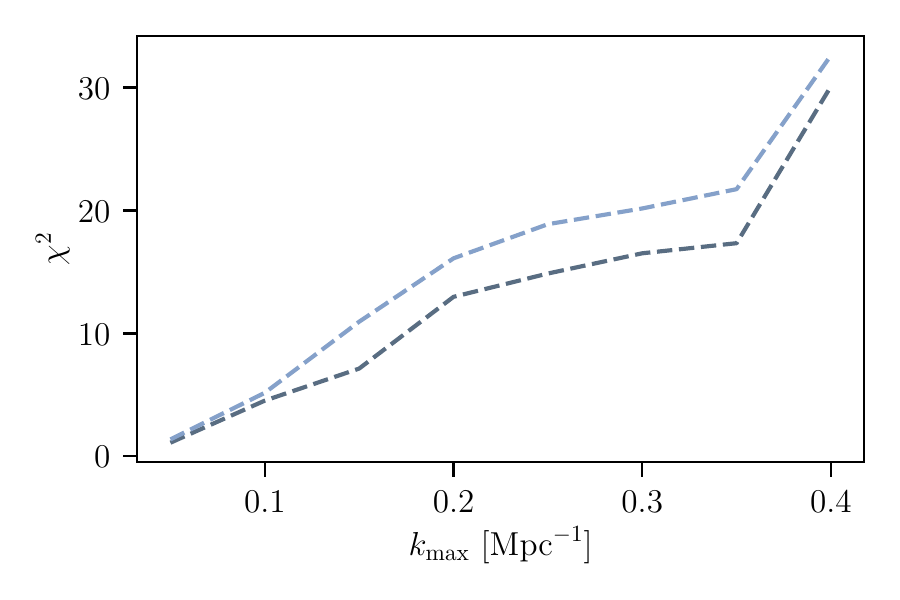}
 \caption{Results of fitting the maglim sample using the nonlinear bias models considered in this work, LPT, EPT, \texttt{anzu} and \texttt{BACCO}. \textit{Top panels:} Recovered values of $\sigma_{8}$ and minimal $\chi^{2}$ obtained from the fit as a function of maximal wavenumber, $k_{\mathrm{max}}$ for HEFT and the \texttt{NLPk} implementation of EPT and LPT. \textit{Bottom panels:} Same as above but using the \texttt{PTPk} implementation of LPT and EPT.}
\label{fig:parameter-fits-heft-hsc}
\end{center}
\end{figure*}

We repeat our analysis using the maglim sample discussed in Sec.~\ref{sec:simulations}, using the same redshift bins for galaxy clustering and weak gravitational lensing. In its fiducial implementation \texttt{BACCO} only covers redshifts $z\leq 1.5$, and we thus need to extend the emulator using LPT at high redshift in order to model the maglim sample. The results obtained for all bias models after implementing this extension are shown in Fig.~\ref{fig:parameter-fits-heft-hsc}. As can be seen from the upper panels we generally recover unbiased results for all our four fiducial bias models\footnote{The only exception is the fit using LPT at $k_{\mathrm{max}}=0.25 \; \mathrm{Mpc}^{-1}$, which is an outlier as compared to the other cases. We suspect this to be due to a parameter degeneracy specific to this case preventing the minimizer to converge to the global minimum. This hypothesis is strengthened by the fact that evaluating the likelihood at $k_{\mathrm{max}}=0.25 \; \mathrm{Mpc}^{-1}$ with the best-fit parameter values derived using data up to $k_{\mathrm{max}}=0.4 \; \mathrm{Mpc}^{-1}$ leads to a significantly lower $\chi^{2}$-value.}. For the maglim sample we only consider five clustering bins, thus leading to an effective number of degrees-of-freedom of $dof = 436$. Following Sec.~\ref{ssec:model-performance}, we thus require $\chi^{2}_{\mathrm{theory}, \mathrm{max}} \leq 49$ to pass our goodness-of-fit test. From Fig.~\ref{fig:parameter-fits-heft-hsc} we can see that all minimal $\chi^{2}$-values derived using our fiducial bias models satisfy this criterion. In the lower panels of Fig.~\ref{fig:parameter-fits-heft-hsc}, we additionally show the recovered parameter values for the \texttt{PTPk} variants of LPT and EPT. As opposed to the results obtained for the red sample, we recover unbiased constraints on $\sigma_{8}$ and $\Omega_{c}$ also for those models. This suggests a higher reach of fully-perturbative bias models for the maglim sample, likely related to the fact that its associated linear bias is significantly smaller than that of the red sample.

\subsubsection{Assembly bias}

\begin{figure*}
\begin{center}
\includegraphics[width=0.45\textwidth]{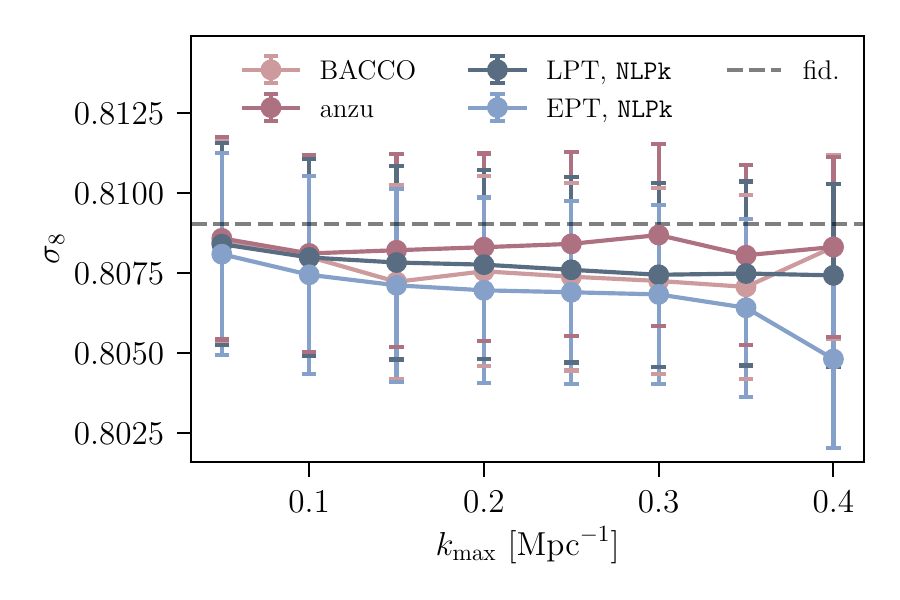}
\includegraphics[width=0.45\textwidth]{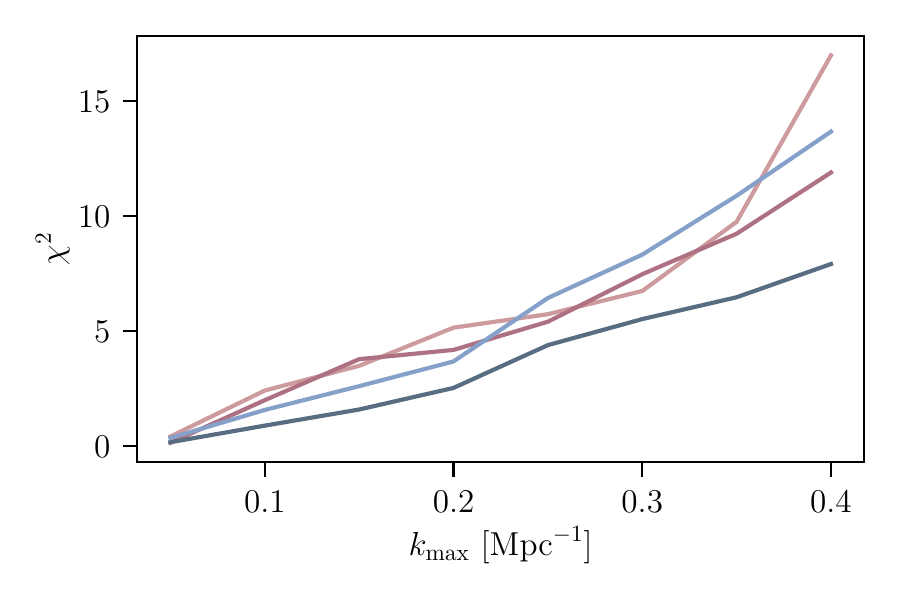}
 \caption{Results of fitting the galaxy sample with assembly bias using the main nonlinear bias models considered in this work, LPT, EPT, \texttt{anzu} and \texttt{BACCO}. Recovered values of $\sigma_{8}$ and minimal $\chi^{2}$ obtained from the fit as a function of maximal wavenumber, $k_{\mathrm{max}}$.}
\label{fig:parameter-fits-heft-AB}
\end{center}
\end{figure*}

Given the current uncertainties on the dependence of galaxy clustering on quantities beyond halo mass, it is important to assess the performance of nonlinear bias models in the presence of these effects. To this end we apply our four fiducial models to the galaxy sample with assembly bias described in Sec.~\ref{sec:simulations}, jointly fitting the combination of galaxy clustering, galaxy-galaxy lensing and cosmic shear. The results for both HEFT methods as well as the \texttt{NLPk} implementation of EPT and LPT are shown in Fig.~\ref{fig:parameter-fits-heft-AB}\footnote{Given the results found for the red sample in Sec.~\ref{ssec:results.heft}, we do not consider the \texttt{PTPk} implementation of EPT and LPT here.}. As can be seen, all four models yield unbiased constraints on cosmological parameters, thus confirming theoretical expectations that these models offer the flexibility to capture the effects of assembly bias as implemented in our simulated data for LSST Y10-like precision (see Ref.~\cite{Kokron:2021} for similar results).

\subsubsection{Including CMB lensing convergence}

\begin{figure*}
\begin{center}
\includegraphics[width=0.45\textwidth]{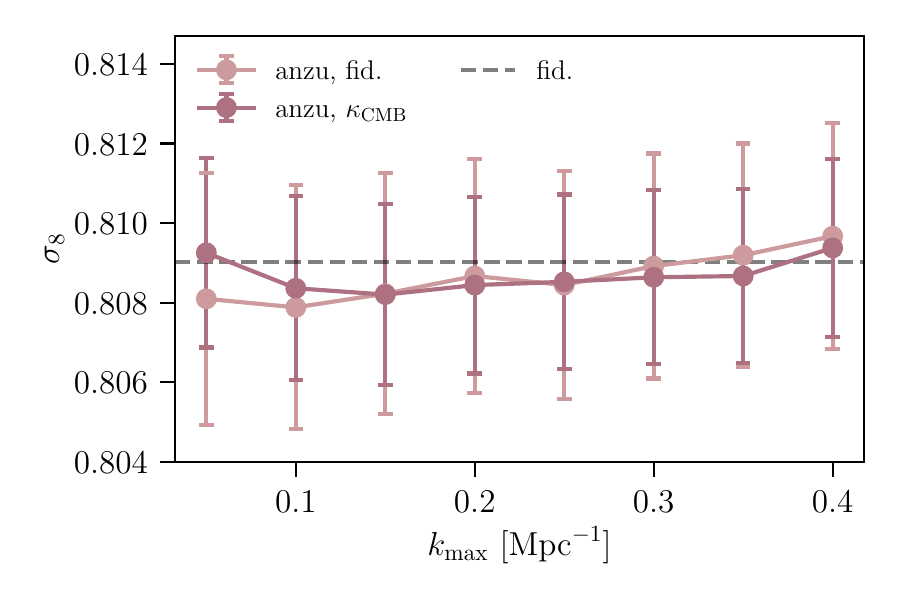}
\includegraphics[width=0.45\textwidth]{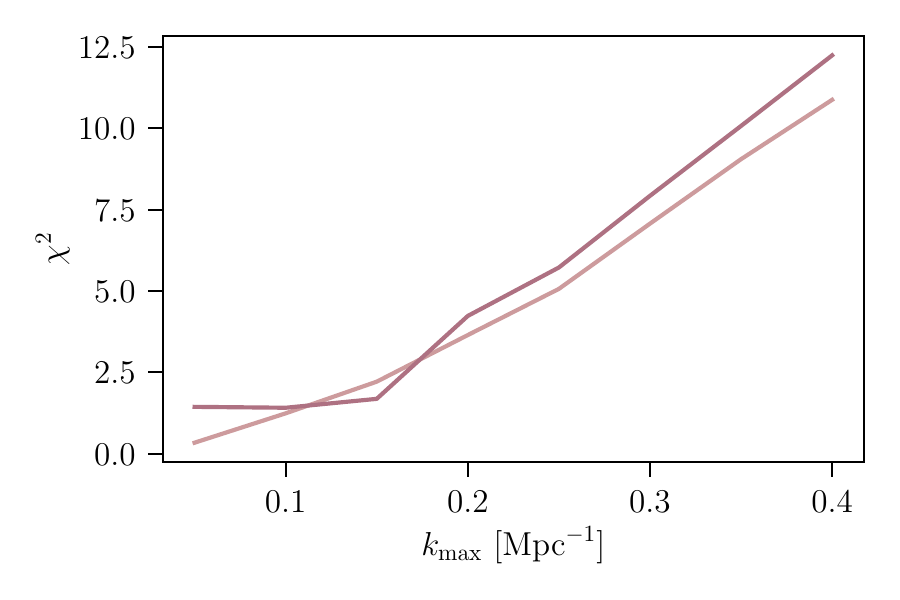}
 \caption{Results of fitting the red galaxy sample combined with CMB S4-like CMB lensing using the four nonlinear bias models considered in this work, as compared to our fiducial results. Recovered values of $\sigma_{8}$ and minimal $\chi^{2}$ obtained from the fit as a function of maximal wavenumber, $k_{\mathrm{max}}$.}
\label{fig:parameter-fits-heft-cmbk}
\end{center}
\end{figure*}

The combination of LSS surveys with CMB measurements has been shown to be a powerful way to constrain deviations from $\Lambda$CDM (see e.g. Refs.~\cite{Mishra-Sharma:2018, Giusarma:2018, Schmittfull:2018, Euclid:2022, Wenzl:2022}), and thus constitutes a key priority for current and future surveys. We therefore additionally test the applicability of current nonlinear bias modeling techniques to the combination of galaxy clustering, galaxy-galaxy lensing, cosmic shear and their cross-correlations with the CMB lensing potential. Specifically, we use the \texttt{anzu} implementation of HEFT to analyze the simulated joint LSST and CMB S4 data vector described in Sec.~\ref{ssec:computing_data}. The inclusion of CMB lensing leads to reduced uncertainties on cosmological parameters by roughly 10\% to 30\% as can be seen from Fig.~\ref{fig:parameter-fits-heft-cmbk}. Nevertheless, we find unbiased recovery of $\sigma_{8}$ and $\Omega_{c}$ for all maximal wavenumbers considered even for this extended data vector, which suggests that these methods meet the accuracy-requirements for future joint analyses of LSST with CMB S4 or SO.

\subsubsection{No \textsc{halofit} rescaling} \label{sssec:res.samples.halofit}

In a final test, we investigate the impact of not rescaling the matter power spectrum to match the \textsc{halofit} prediction. While we do not explicitly show the results, in this case we recover significantly biased constraints on cosmological parameters for all bias models considered. These results suggest that at the precision of LSST Y10, we are sensitive to systematic differences between the \textsc{halofit} fitting function and matter power spectra measured from simulations. This is not related to the problem of characterizing galaxy bias, which is the focus of this paper, and instead is limited to the shear-shear component of the \txtp data vector. A thorough study of the precision with which the matter power spectrum needs to be modeled for LSST (including the impact of baryons) will be the focus of future work.

\subsection{Stochasticity}\label{ssec:results.stochasticity}

\begin{figure*}
\begin{center}
\includegraphics[width=0.6\textwidth]{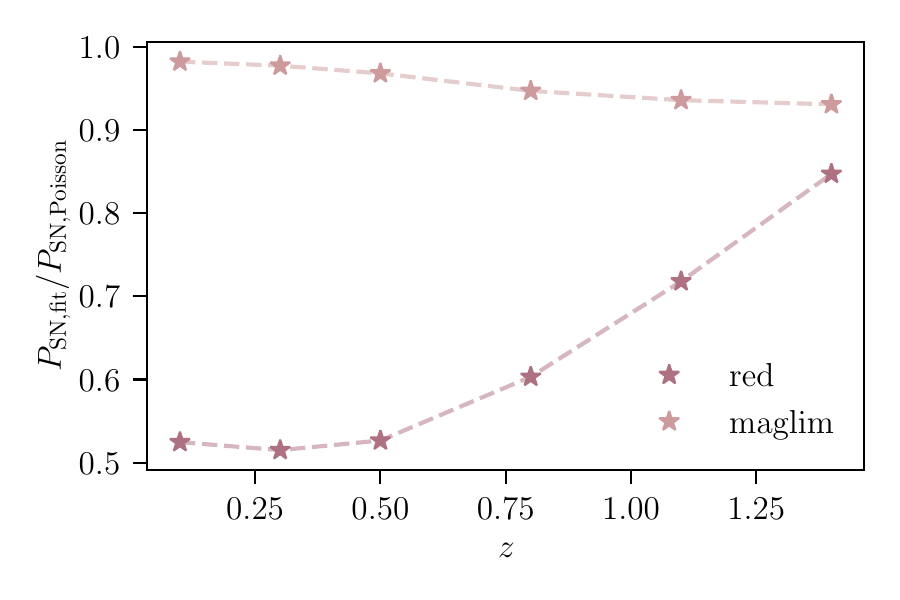}
\caption{Ratio of recovered value of stochasticity power spectrum and its Poissonian expectation as obtained from fitting three-dimensional power spectrum data with the \texttt{anzu} model at discrete redshifts. All results are shown for $k_{\mathrm{max}}=0.4 \; \mathrm{Mpc}^{-1}$.}
\label{fig:sn-consistency}
\end{center}
\end{figure*}

A number of studies have found evidence for non-Poissonian stochasticity in simulated galaxy samples (see e.g. Refs.~\cite{Baldauf:2013, Kokron:2022}). It is therefore interesting to compare the levels of stochasticity obtained in this work with their Poissonian expectations. The results presented so far have been derived using spherical harmonic power spectra, which constitute line-of-sight projections of the associated three-dimensional power spectra. It is therefore difficult to relate the redshift-averaged stochasticities obtained in our analysis to their Poisson counterparts, and we thus use measurements of three-dimensional power spectra to investigate stochasticity. As described in detail in Appendix \ref{sec:ap.pk}, we consider a data vector consisting of $\mathbf{d}=\{P_{gg}(z, k), P_{gm}(z, k), P_{mm}(z, k)\}$ at discrete redshifts, and determine the best-fitting bias parameters. 

The ratio of the derived values of $P_{\rm SN}$ to their Poissonian expectations for $k_{\mathrm{max}}=0.4 \; \mathrm{Mpc}^{-1}$ are shown in Fig.~\ref{fig:sn-consistency} as a function of redshift. As can be seen, for the red sample we detect significantly sub-Poissonian stochasticity, with the largest discrepancies at low redshift. As shown in Ref.~\cite{Kokron:2022}, non-Poissonian stochasticity is due to two different effects: halo exclusion and HOD stochasticity. Halo exclusion denotes the effect that large halos are not a Poisson realization of an underlying density field as the halos cannot overlap \cite{Baldauf:2013}. This effectively decreases the volume available to the halos and thus reduces stochasticity. HOD stochasticity on the other hand denotes the additional stochasticity in galaxy samples caused by the variance in the galaxy-halo-distribution. This effect is always positive and thus acts to increase $P_{\rm SN}$. The stochasticity of a given galaxy sample is thus determined by the combined effects of halo exclusion and HOD stochasticity. Our results therefore suggest that the host halos of the red galaxy sample exhibit significant halo exclusion, which is qualitatively and quantitatively comparable to the findings presented in Ref.~\cite{Kokron:2022} for DESI LRGs. However, our results suggest lower values for $P_{\rm SN}$, especially at low and high redshift. We believe these changes to be due to either a difference in methodology to constrain stochasticity, or differences in the adopted HODs for the DESI LRG sample, or a combination of both\footnote{The differences in HODs might be caused by our assumption of a redshift-evolving HOD model, and comparing secondary halo properties of our sample to those reported in Ref.~\cite{Zhou:2021}, we find the largest differences at low and high redshift, which is also where we see the largest discrepancies with Ref.~\cite{Kokron:2022}.}. 

Performing an analogous analysis for the maglim sample, we find a significantly lower level of sub-Poissonian noise than what is detected for the red sample. Specifically we find a $7\%$ reduction as compared to the Poissonian expectation at most. Comparing the mean halo masses of both samples at redshift $z=0.65$, we find $\bar{M}_{h} = 4.1 \times 10^{13} \; M_{\odot}$ for the red sample, while we find $1.3 \times 10^{13} \; M_{\odot}$ for the maglim sample. In addition, we find significantly higher satellite fractions for the maglim sample. These higher satellite fractions for the maglim sample can be interpreted using the results of e.g. Ref.~\cite{Zehavi:2011} that finds the satellite fraction to increase steeply with decreasing luminosity but also with increasing redness of the sample. The higher satellite fraction of the maglim sample thus suggests that its lower luminosity dominates over its reduced fraction of red galaxies. Combined, these two properties of the maglim sample indicate that it exhibits stochasticity closer to its Poissonian expectation because its host halos show lower halo exclusion and the larger satellite fractions give rise to higher HOD stochasticity. 

As shown in Appendix \ref{sec:ap.consistency}, the analysis in terms of spherical harmonic power spectra leads to similar conclusions, as we find good agreement between bias parameter constraints derived using two- and three-dimensional power spectra.

\subsection{Minimal bias model}
\begin{figure*}
\begin{center}
\includegraphics[width=0.45\textwidth]{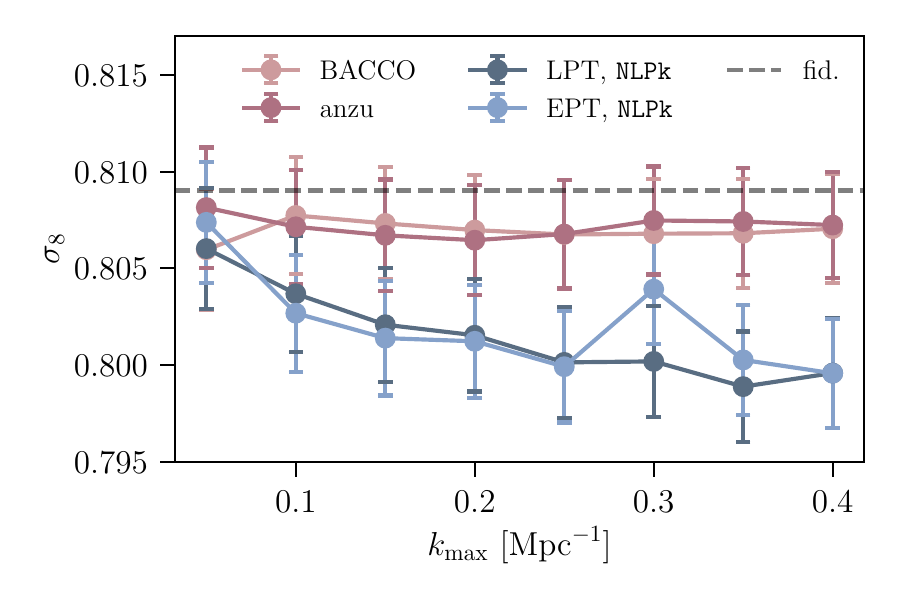}
\includegraphics[width=0.45\textwidth]{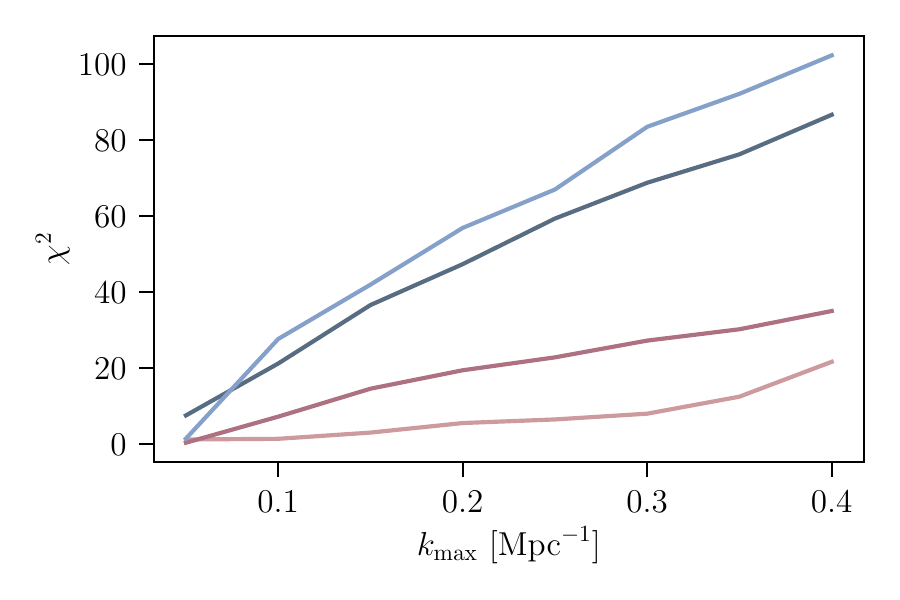}\\
\includegraphics[width=0.45\textwidth]{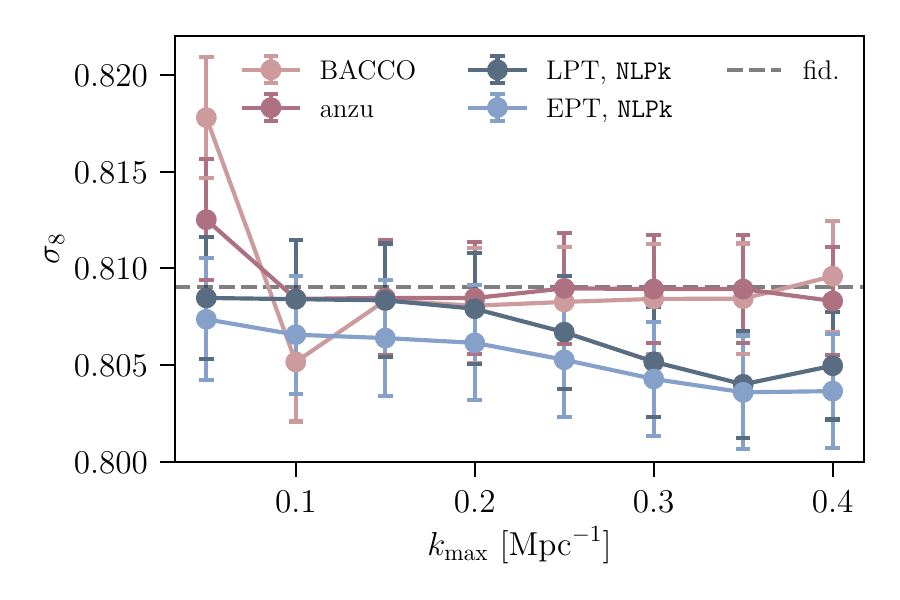}
\includegraphics[width=0.45\textwidth]{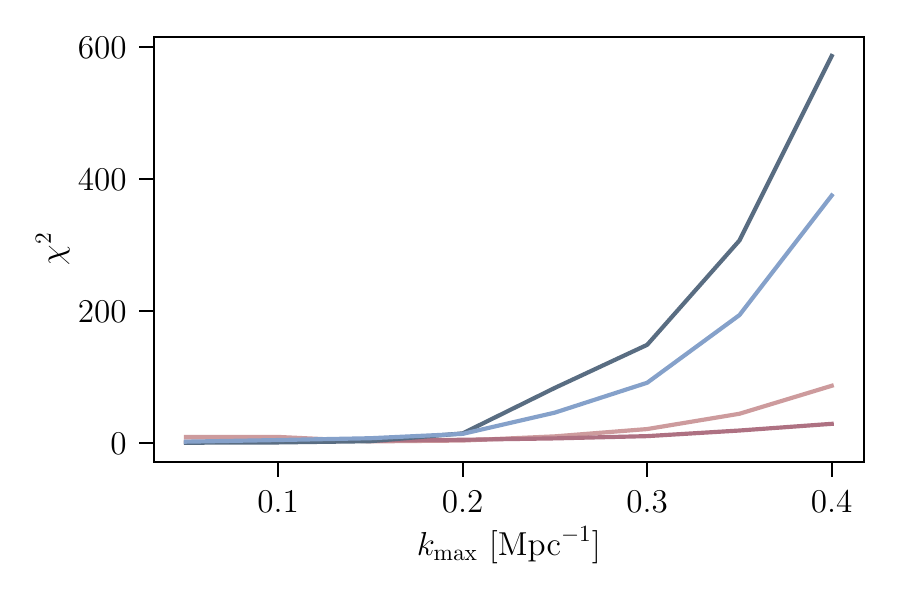}
 \caption{Results of fitting the red galaxy sample with reduced bias models. \textit{Top panels:} Recovered values of $\sigma_{8}$ and minimal $\chi^{2}$ obtained from the fit as a function of maximal wavenumber, $k_{\mathrm{max}}$ when fixing $b_{s^{2}}=0$. \textit{Bottom panels:} The same as in the upper panels but fixing $b_{\nabla^{2}}=0$.}
\label{fig:minimal-model}
\end{center}
\end{figure*}
The results presented so far have been derived considering galaxy bias terms up to quadratic order, resulting in a model with six bias parameters per redshift bin. As shown in Sections \ref{ssec:results.heft}, \ref{ssec:pt} and \ref{ssec:results.alternative_samps}, this model allows us to fit the \textsc{AbacusSummit} data reasonably well, but it is interesting to ask if we can reduce the complexity of the model while maintaining its performance. To this end, we first focus on the red sample and repeat our analysis, setting a number of bias parameters to zero. Specifically, we consider the cases $b_{s^{2}}=0$, $b_{\nabla^{2}}=0$ and $b_{1, p}=0$. This choice corresponds to the most drastic approach to model reduction, and we note that one could also explore alternative methods such as placing tight priors on bias parameters. The values recovered for $\sigma_{8}$ and their corresponding minimal $\chi^{2}$-values when setting $b_{s^{2}}=0$ and $b_{\nabla^{2}}=0$ respectively are shown in Fig.~\ref{fig:minimal-model} for the HEFT methods as well as the \texttt{NLPk} implementation of EPT and LPT. As can be seen, we find that the HEFT methods return unbiased constraints on $\sigma_{8}$ even for these reduced models, while the perturbation-theory-based methods show significant biases, most pronounced at high $k_{\mathrm{max}}$. This is borne out by looking at the $\chi^{2}_{\mathrm{min}}$-values: for $b_{s^{2}}=0$ both HEFT methods pass our goodness-of-fit test at $k_{\mathrm{max}}=0.4\; \mathrm{Mpc}^{-1}$, while the PT methods do not. For $b_{\nabla^{2}}=0$ on the other hand, only \texttt{anzu} recovers a $\chi^{2}$ low enough to pass our test at the largest wavenumber considered. It is interesting to note that while the biases in the PT methods are more pronounced when setting $b_{s^{2}}=0$, the $\chi^{2}_{\mathrm{min}}$-values are significantly worse for the model with $b_{\nabla^{2}}=0$. Additionally, while we do not show the results, we find that all models yield biased constraints on cosmological parameters when we set $b_{1, p}=0$, highlighting the need to account for the evolution of at least the linear bias across each redshift bin at the level of precision achieved by LSST Y10.

We test the stability of these results by comparing the values of the common bias parameters recovered for the three levels of model complexity considered ($\{b_1,b_{1,p},b_2,b_{s^2},b_{\nabla^2}\}$, $\{b_1,b_{1,p},b_2,b_{\nabla^2}\}$, and $\{b_1,b_{1,p},b_2,b_{s^2}\}$) using the \texttt{anzu} implementation of HEFT. Without explicitly showing the results, we find all recovered values to be consistent within uncertainties. This suggests that the values recovered for the linear and quadratic bias, and, perhaps more importantly, for the stochasticity $P_{\rm SN}$, are not driven by inaccuracies in the bias model, but rather represent a genuine physical feature of the sample under consideration.

\begin{figure*}
\begin{center}
\includegraphics[width=0.45\textwidth]{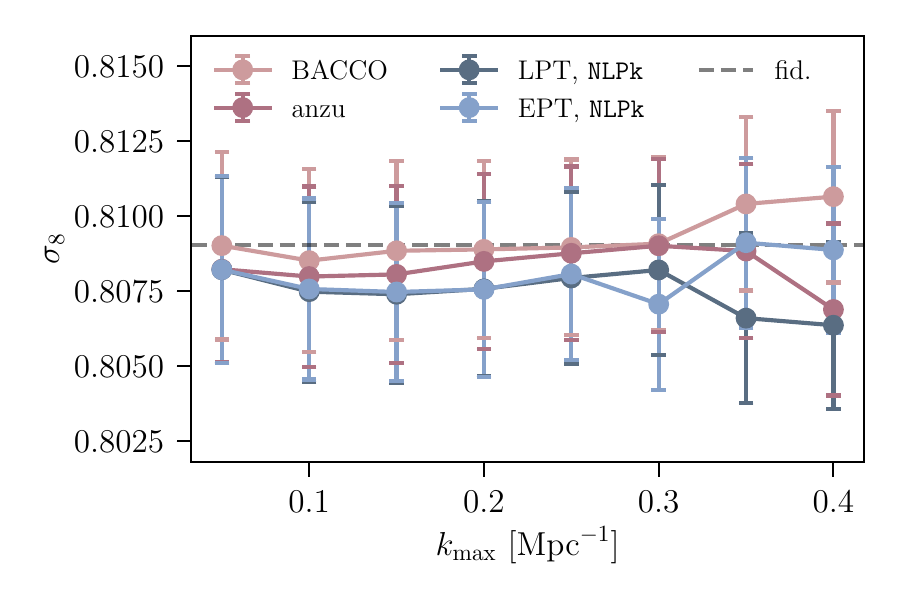}
\includegraphics[width=0.45\textwidth]{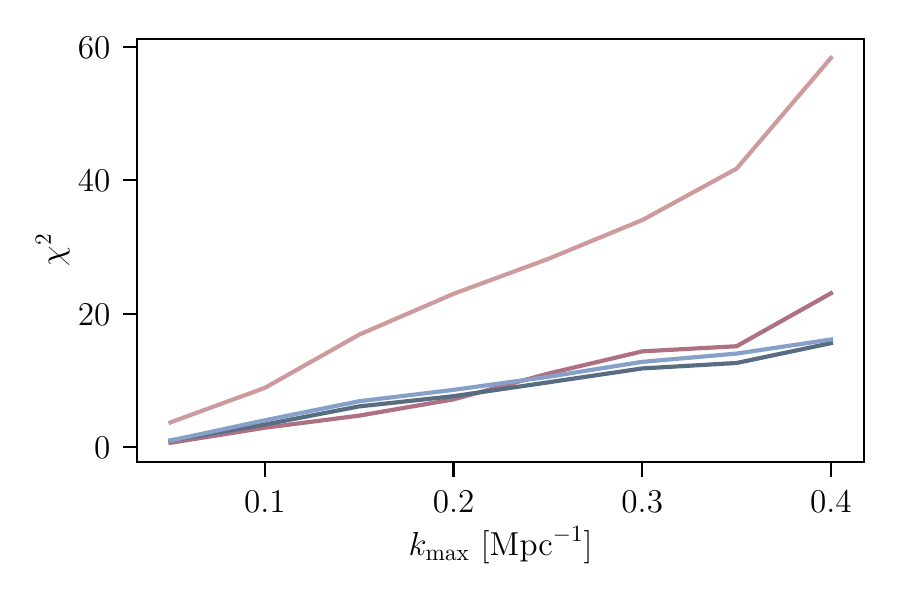}
 \caption{Results of fitting the maglim sample using the nonlinear bias models considered in this work, LPT, EPT, \texttt{anzu}, \texttt{BACCO}, and setting $b_{s^2}=0$ in all cases. The plots show the recovered values of $\sigma_{8}$ and minimal $\chi^{2}$ obtained from the fit as a function of maximal wavenumber, $k_{\mathrm{max}}$.}
\label{fig:parameter-fits-heft-hsc-variants}
\end{center}
\end{figure*}

In order to investigate the sample-dependence of these findings, we repeat our analysis for the maglim and assembly bias samples. The results obtained when setting the tidal bias to zero for the magnitude-limited sample, i.e. $b_{s^{2}}=0$, are shown in Fig.~\ref{fig:parameter-fits-heft-hsc-variants} for the HEFT methods as well as the \texttt{NLPk} implementation of EPT and LPT. As can be seen, all bias models recover unbiased constraints on cosmological parameters, but the \texttt{BACCO} fits lead to $\chi^{2}$-values significantly higher than our threshold\footnote{We note that in contrast to the results presented in Sec.~\ref{sssec:res.samps.maglim}, the \texttt{PTPk} models without tidal bias, $b_{s^{2}}=0$, give biased constraints on cosmological parameters.}. We find similar results when setting the non-local bias to zero, i.e. $b_{\nabla^{2}}=0$, and thus do not show the results explicitly. The only exception is that in this case all models, including \texttt{BACCO}, pass our goodness-of-fit tests. Investigating this further, we repeat our analysis for $b_{s^{2}}=0$ excluding the highest redshift bin for the clustering statistics. In this case, we find unbiased constraints on cosmological parameters as well as minimal $\chi^{2}$-values passing our test criteria for all bias models. The model residuals for \texttt{BACCO} thus appear predominantly driven by the highest redshift bin, and we suspect that the low model performance observed might be caused by inaccuracies in our extension of \texttt{BACCO} to high redshift, but we leave a more detailed analysis to future work. For the assembly bias sample, we find results very similar to those obtained for the red sample, i.e. EPT and LPT yield biased constraints in all cases, while both HEFT implementations show good performance.

These results confirm the expectation that the minimal bias model depends on the data set as well as the actual model. For the red and the assembly bias samples, we find that all PT-based methods (\texttt{NLPk} and \texttt{PTPk}) lead to biases when setting $b_{s^{2}}=0$ or $b_{\nabla^{2}}=0$, while the HEFT methods perform generally well. We attribute this to the fact that a less accurate model will need more parameters to fit a given data set in order to compensate for inaccuracies in the template power spectra. It is interesting that a vanishing tidal bias appears to be more consistent with the data, while we start seeing biases even for the HEFT methods when setting $b_{\nabla^{2}}=0$. In addition, it is worth noting that the scale-dependent modification induced by $b_{\nabla^2}\neq0$ is qualitatively different in real and harmonic space. Thus, these results may be sensitive to our analysis choices, in which scale cuts are imposed in harmonic space. Finally, we find both the HEFT and \texttt{NLPk} methods to yield unbiased results for the maglim sample, thus suggesting that the inaccuracies of PT-based models are most pronounced for more highly biased galaxy samples as compared to their less biased counterparts, and thus their modeling requires a smaller number of bias parameters.

It is worth noting that an alternative approach to finding a minimal bias model would be to make use of the rather tight relations between bias parameters found in the literature (see Sec.~\ref{ssec:pt.bias-consistency} or e.g. Ref.~\cite{Zennaro:2022}) and use them to fix the value of certain bias parameters rather than set them to zero. We leave an investigation thereof to future work.

\subsection{Error as a function of maximal wavenumber}
\begin{figure*}
\begin{center}
\includegraphics[width=0.45\textwidth]{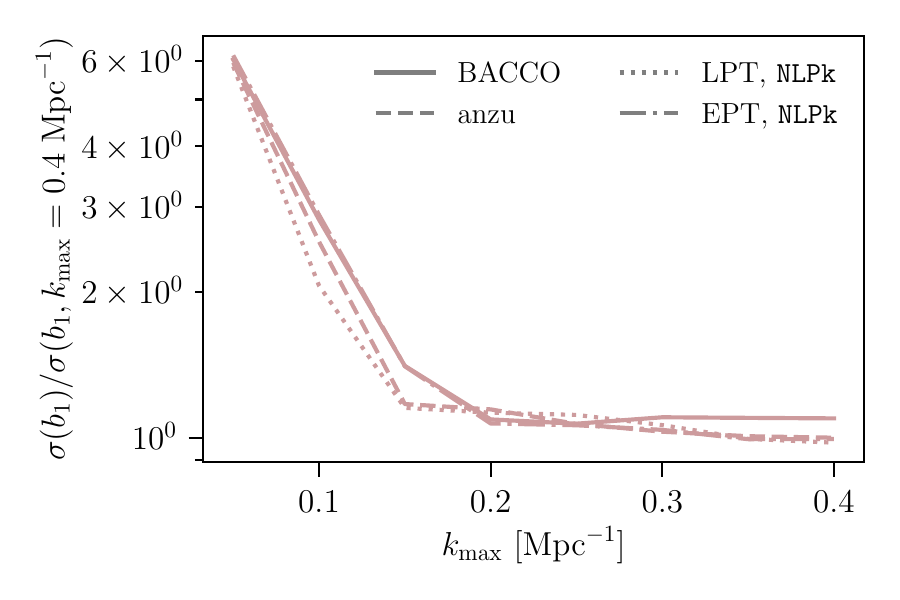}
\includegraphics[width=0.45\textwidth]{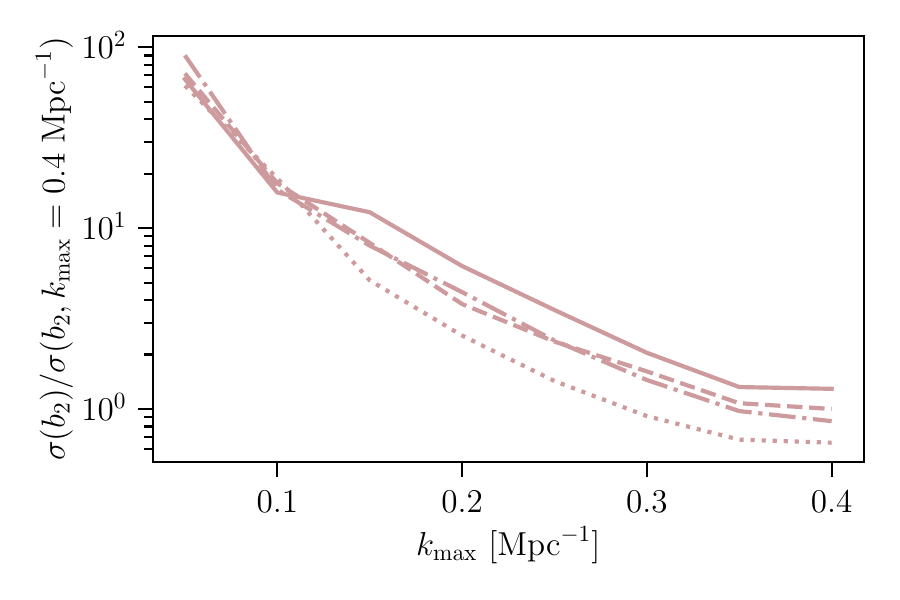}\\
\includegraphics[width=0.45\textwidth]{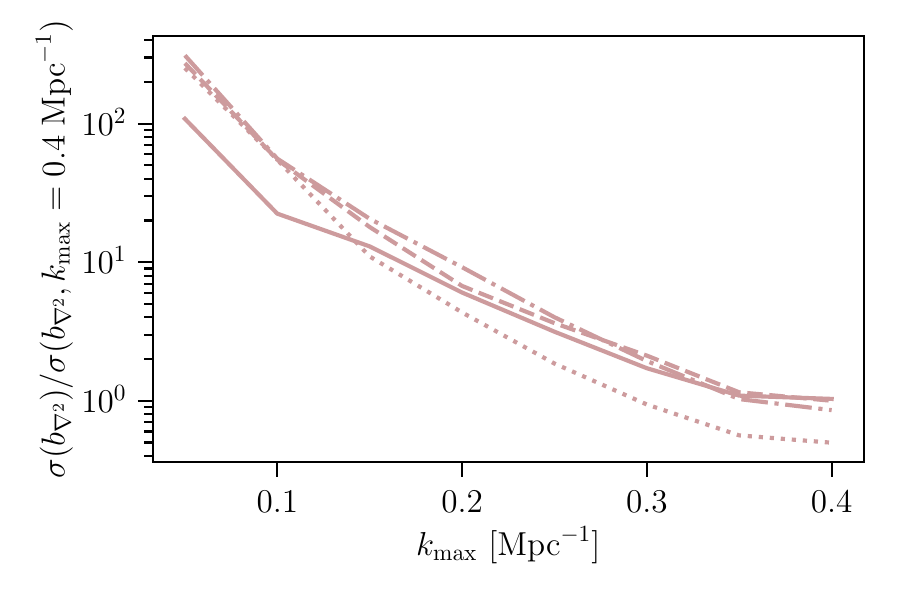}
\includegraphics[width=0.45\textwidth]{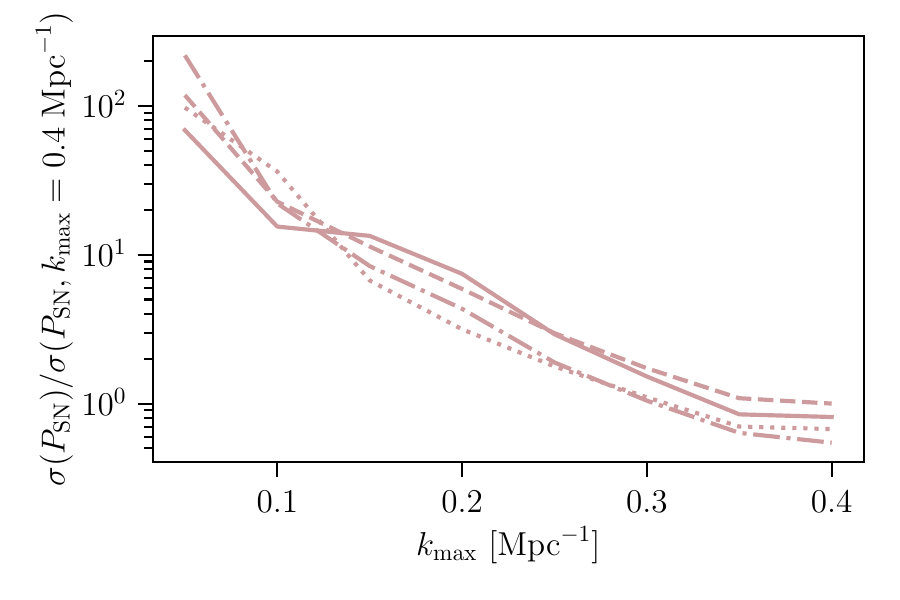}
 \caption{Relative uncertainties on the bias parameters $b_{1}$, $b_{2}$ and $b_{sn}$ for the fourth redshift bin as a function of maximal wavenumber, $k_{\mathrm{max}}$. All uncertainties are normalized with respect to those obtained for \texttt{anzu} at $k_{\mathrm{max}}=0.4 \; \mathrm{Mpc}^{-1}$.}
\label{fig:relative-uncertainties-bias}
\end{center}
\end{figure*}

In order to determine optimal scale cuts for a given analysis involving galaxy clustering data, it is essential to investigate to what extent parameter constraints are tightened by the inclusion of additional small-scale information. In particular it is interesting to investigate if in the setup considered in this analysis, additional small-scale information results in tighter constraints on cosmological parameters, or rather serves to constrain bias parameters more tightly. To this end we compare the uncertainties on cosmological parameters as well as the four bias parameters $b_{1}$, $b_{2}$, $b_{\nabla^{2}}$ and $P_{\rm SN}$ obtained as a function of $k_{\mathrm{max}}$\footnote{We focus on these, as we find $b_{1p}$ to be only weakly constrained in our analysis. We further note that we compute uncertainties setting $b_{s^{2}}=0$, as discussed in Sec.~\ref{ssec:results.heft}.}. The relative uncertainties for $\sigma_{8}$ and $\Omega_{c}$ are shown in the lower right panel of Fig.~\ref{fig:parameter-fits-heft} while Fig.~\ref{fig:relative-uncertainties-bias} illustrates the corresponding relations for the bias parameters. All uncertainties are normalized with respect to those obtained for \texttt{anzu} at $k_{\mathrm{max}}=0.4 \; \mathrm{Mpc}^{-1}$, and as can be seen from Fig.~\ref{fig:parameter-fits-heft}, the marginalized 1$\sigma$ errors on cosmological parameters decrease only by roughly $15\%$ as we increase $k_{\mathrm{max}}$ from $0.05 \; \mathrm{Mpc}^{-1}$ to $0.4 \; \mathrm{Mpc}^{-1}$. These gains are small given the significant increase in the amount of small-scale information included in the fits. Looking at Fig.~\ref{fig:relative-uncertainties-bias} on the other hand, we see that the bias parameter constraints tighten significantly, by up to 2 orders of magnitude, for the same increase in $k_{\mathrm{max}}$, in particular for $b_{\nabla^{2}}$ and $P_{\rm SN}$. These results suggest that in a joint analysis of galaxy clustering, galaxy-galaxy lensing and weak lensing with LSST Y10-like specifications, the information contained in small-scale galaxy clustering mainly serves to constrain galaxy bias parameters as compared to cosmological parameters.

\begin{figure*}
\begin{center}
\includegraphics[width=0.6\textwidth]{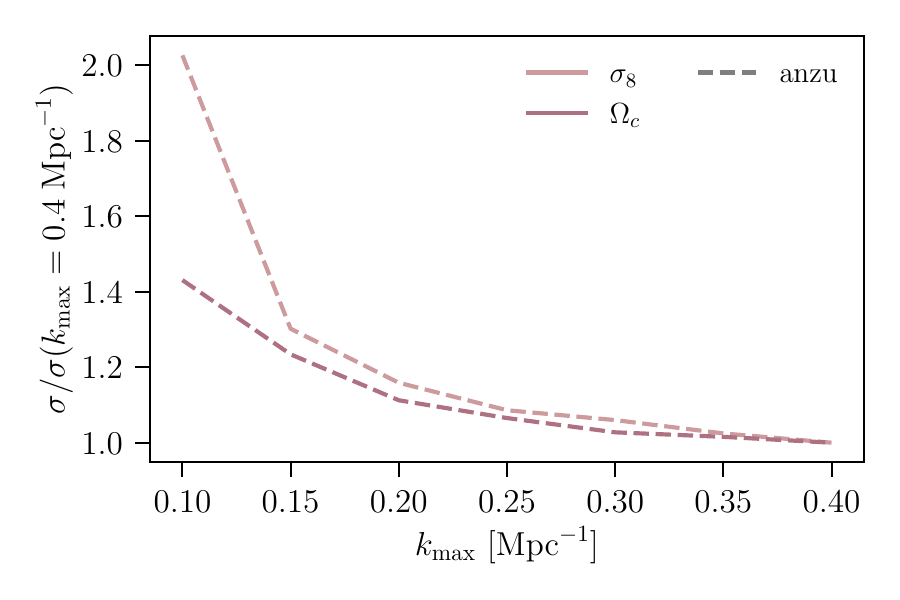}
 \caption{Relative uncertainties on $\sigma_{8}$ and $\Omega_{c}$ as a function of maximal wavenumber, $k_{\mathrm{max}}$, obtained when jointly fitting galaxy clustering and galaxy-galaxy lensing from our fiducial sample with \texttt{anzu}.}
\label{fig:rel-err-red-gg+ggl}
\end{center}
\end{figure*}

The lack of improvement in cosmological constraining power with increasing $k_{\mathrm{max}}$ appears to be driven by the cosmic shear data, as we find that the errors on $\sigma_{8}$ and $\Omega_{c}$ decrease by roughly a factor of $1.5-2$ when we increase $k_{\mathrm{max}}$ from $0.1 \; \mathrm{Mpc}^{-1}$ to $0.4 \; \mathrm{Mpc}^{-1}$ and only consider galaxy-galaxy and galaxy-shear correlations, as can be seen from Fig.~\ref{fig:rel-err-red-gg+ggl}\footnote{While the figure only shows \texttt{anzu}, we find similar results for the other bias models.}. To further test this hypothesis, we rerun our fiducial analysis, freeing up the Hubble parameter $h$ and the scalar spectral index $n_{s}$ in addition to $\sigma_{8}$ and $\Omega_{c}$. While the improvements in uncertainties for $\sigma_{8}$ and $\Omega_{c}$ are similar, we find larger gains for $h$ and $n_{s}$ of up to $30\%$. In particular, the gains do not seem to saturate at high $k_{\mathrm{max}}$ as they do in our fiducial analysis.

While the relatively modest gains in cosmological constraining power seem to be partly driven by our inclusion of weak lensing data, which yields tight constraints on $\sigma_{8}$ and $\Omega_{c}$, and thus limits the gains from small-scale clustering, we do see a general trend that small-scale clustering improves bias parameter constraints more significantly than constraints on their cosmological counterparts. However, we caution the reader that these conclusions might change when constraining a larger number of cosmological parameters, including systematic uncertainties, or considering data vectors different from the ones we are investigating in this work (e.g. including higher-order correlators, or redshift-space distortions in spectroscopic survey analyses). As an example, galaxy-galaxy lensing at small scales has been shown to help constrain intrinsic galaxy alignments and photometric redshift systematics, which would otherwise significantly degrade the constraining power of cosmic shear (see e.g. Ref.~\cite{Sanchez:2022}). 

\section{Conclusions}\label{sec:conclusions}

In this work, we compare the performance of a number of nonlinear galaxy bias models when applied to an LSST Y10-like tomographic joint analysis of galaxy clustering, galaxy-galaxy lensing and cosmic shear (\txtp analysis). Specifically, we compare two perturbative approaches, Lagrangian perturbation theory (LPT) \cite{Matsubara:2008} and Eulerian perturbation theory (EPT) to two implementations of Hybrid Effective Field Theory (HEFT), \texttt{anzu} and \texttt{BACCO}, which combines a perturbative bias expansion in Lagrangian space with an exact treatment of the gravitational evolution via  cosmological simulations \cite{Modi:2020, Kokron:2021, Zennaro:2021}. We test all the methods using simulated data vectors computed from the \textsc{AbacusSummit} \citep{Garrison:2019, Garrison:2021} cosmological simulation, considering several different galaxy samples: a DESI-like red sample, a magnitude-limited sample based loosely on HSC DR1, and a galaxy sample with assembly bias. We fit these simulated data using all bias models considered, keeping terms up to second order, and account for nonlocal bias as well as deviations from Poissonian stochasticity. In a final step, we compare their performance based on the accuracy and precision of the constraints obtained for the cosmological parameters $\sigma_{8}$ and $\Omega_{c}$ as well as the goodness-of-fit.

For our fiducial, red galaxy sample we find that the two HEFT implementations allow us to jointly model galaxy clustering, galaxy-galaxy lensing and cosmic shear with LSST Y10-like precision up to at least a maximal wavenumber of $k_{\mathrm{max}}=0.4 \; \mathrm{Mpc}^{-1}$. This is also true for LPT and EPT when we combine these methods with non-perturbative predictions for the matter power spectrum entering some of the terms in the expansion (\texttt{NLPk}). In contrast, when we use the predictions from perturbation theory for these terms (\texttt{PTPk}), the LPT and EPT implementations lead to biased constraints on cosmological parameters for $k\gtrsim0.2\,{\rm Mpc}^{-1}$. We find comparable results when analyzing the galaxy sample with assembly bias. For the magnitude-limited sample on the other hand, we find good performance for all bias models, including EPT and LPT with a perturbative prediction for the matter power spectrum. We further consider an extension of our fiducial galaxy sample with CMB lensing cross-correlations loosely matching the specifications for CMB S4, finding unbiased constraints on cosmological parameters as well as good minimal $\chi^{2}$-values. In all these analyses, we find significant detections of non-Poissonian stochasticity in the galaxy clustering auto-correlations.

We further investigate the effect of reducing bias model complexity by setting the tidal and nonlocal bias to zero respectively, finding sample- and model-dependent results. We find that the HEFT approaches are able to obtain unbiased constraints and provide a good fit to the data with this reduced models in most cases, the only exception being \texttt{BACCO} for vanishing nonlocal bias and $k_{\mathrm{max}}=0.4 \; \mathrm{Mpc}^{-1}$. In turn, while LPT and EPT perform well on the magnitude-limited sample, they lead to biases on cosmological parameters when applied to the more highly-biased red sample (with or without assembly bias) within these reduced parameterizations. This is the case regardless of the prescription used to model the matter power spectrum (\texttt{NLPk} or \texttt{PTPk}).

Investigating the constraints on cosmological and bias parameters obtained as a function of maximal wavenumber $k_{\mathrm{max}}$, we find the uncertainties on cosmological parameters to decrease only by around $15 \%$ as we increase $k_{\mathrm{max}}$ from $0.05 \; \mathrm{Mpc}^{-1}$ to $0.4 \; \mathrm{Mpc}^{-1}$. The bias parameter uncertainties on the other hand decrease significantly, in some cases by more than an order of magnitude. Removing the weak lensing auto-correlations from our data vector yields larger relative improvements on cosmological parameter uncertainties, of up to a factor of 2. This is a useful case to consider as the weak lensing auto-correlations can have separate systematic uncertainties (e.g. PSF induced additive shear contributions) while not being sensitive to galaxy bias, so separating out the cosmology inferred from those is a powerful consistency check. Nevertheless, the qualitative result is that pushing towards smaller scales seems to lead to significantly larger improvements in bias parameters than in cosmological parameters. These results are subject to a number of caveats: most importantly, we only consider constraints on galaxy bias as well as two cosmological parameters, $\sigma_{8}$ and $\Omega_{c}$, and we do not account for systematics such as photometric redshift uncertainties or calibration biases in cosmic shear. This might artificially increase the constraining power of weak lensing, thus reducing the gain obtained from small-scale clustering. We have also not considered uncertainties due to magnification, intrinsic alignments and their interplay with photo-z uncertainties -- these effects can modify our findings on the improvements from small scale information (likely in the direction of greater gain). We leave an investigation of these effects to future work.
 
In summary, our results confirm the performance of HEFT approaches found in previous work, while suggesting a potentially higher reach for EPT and LPT than previously found. This is likely in part due to our focus on projected clustering statistics (as opposed to three-dimensional clustering in redshift space), and to the specific metrics used to quantify goodness of fit (based on the expected performance of an LSST-like experiment, rather than on ad-hoc precision requirements).

With regards to LSST, the results of our analysis suggest that current nonlinear bias models appear promising for the analysis of tomographic galaxy clustering, galaxy-galaxy lensing and weak lensing from LSST Y10 data. Amongst the methods investigated, we find the recently developed HEFT methods to show particular promise. This bodes well for future tomographic galaxy clustering and \txtp analyses using small-scale information for both current and future photometric surveys such as LSST and Euclid. Nevertheless, more work is necessary in order to be able to fully exploit these methods in a robust and reliable manner. First, application of these methods to future data will require investigating the impact of observational effects such as photometric redshift uncertainties or large-scale galaxy clustering systematics and associated scale cuts. In addition, as we have shown, a good characterization of the theoretical uncertainties associated with the non-linear scheme used is vital to obtain unbiased constraints with adequate errors. A more precise treatment and modeling of these uncertainties than that used in our analysis will therefore be needed before the bias models studied here can be applied to future data sets. Furthermore our results have highlighted the susceptibility of perturbative bias prescriptions to modeling of the matter power spectrum, thus requiring higher-accuracy models. Although our analysis has covered the most likely target samples for galaxy clustering (LRG-like and magnitude-limited samples), including some amount of assembly bias, the results found here should be validated against a wider variety of galaxy samples, incorporating different physical effects such as satellite segregation and baryonic effects in the dark matter distribution \cite{2023MNRAS.tmp..417C}. Furthermore, as the internal consistency relations between different bias coefficients provide an avenue to reduce the freedom allowed to the bias model, studying the applicability of these relations to these samples in the context of LSST would be a useful exercise. Finally, higher-order statistics have the potential to unlock a significant amount of untapped non-Gaussian information in the galaxy distribution, and thus studying the ability of the bias models explored here to describe these observables is of high priority. 

\acknowledgments
This paper has undergone internal review by the LSST Dark Energy Science Collaboration. We kindly thank the internal reviewers Simone Ferraro, Andrew Hearin and Shivam Pandey for providing helpful comments, which helped us improve the quality and clarity of the paper.

\noindent We are very happy to thank Toshiya Namikawa and Colin Hill for sharing the CMB S4 lensing noise curves and for help with their usage. With pleasure we would also like to thank Martin White for many very helpful comments and suggestions. 

\noindent The contributions from the primary authors are as follows: DA: Co-designed the project, constructed simulated data vectors, contributed to likelihood software. NF: Performed initial fits with LPT and EPT models to determine the minimum scales to which unbiased results could be recovered. CGG: Implemented CMB lensing in likelihood, data and covariances. ZG: Implemented and tested Fisher matrix for uncertainty estimation. BH: Generated mocks from AbacusSummit and contributed to some iteration of the likelihood. BJ: Helped compare the results in this study with prior work on galaxy bias models and discussed/edited Sections 1, 2.4, 2.5, 4, 5.2, 6.7 and 7. NK: Implemented the hybrid EFT code anzu into the CCL-based likelihood used to compare models and data in the challenge. AN: Co-designed the project, lead the analysis and writing of the paper. AS: Contributed to the design of the experiment, help with interpretation of results, contributed to the paper text.

\noindent CGG acknowledges support from the European Research Council Grant No:  693024 and the Beecroft Trust. DA acknowledges support from the Beecroft Trust, and from the John O'Connor Research Fund, at St. Peter's College, Oxford. ZG and CW were supported by the Office of Science of the U.S. Department of Energy, grant DE-SC0010007.

\noindent The DESC acknowledges ongoing support from the Institut National de Physique Nucl\'eaire et de Physique des Particules in France; the Science \& Technology Facilities Council in the United Kingdom; and the Department of Energy, the National Science Foundation, and the LSST Corporation in the United States.  DESC uses resources of the IN2P3 Computing Center (CC-IN2P3--Lyon/Villeurbanne - France) funded by the Centre National de la Recherche Scientifique; the National Energy Research Scientific Computing Center, a DOE Office of Science User Facility supported by the Office of Science of the U.S.\ Department of Energy under Contract No.\ DE-AC02-05CH11231; STFC DiRAC HPC Facilities, funded by UK BEIS National E-infrastructure capital grants; and the UK particle physics grid, supported by the GridPP Collaboration.  This work was performed in part under DOE Contract DE-AC02-76SF00515.

\noindent The author(s) are pleased to acknowledge that the work reported on in this paper was substantially performed using the Princeton Research Computing resources at Princeton University which is consortium of groups led by the Princeton Institute for Computational Science and Engineering (PICSciE) and Office of Information Technology's Research Computing. 

\noindent This work made use of the following software packages: \texttt{astropy}\footnote{\url{https://www.astropy.org/}.}, \texttt{matplotlib}\footnote{\url{https://matplotlib.org/}.}, \texttt{numpy}\footnote{\url{https://numpy.org/}.} and \texttt{scipy}\footnote{\url{https://scipy.org/}.}.

\appendix

\section{Description of methodology to create smoothed power spectra}\label{sec:ap.smoothed_spectra}

\begin{figure*}
\begin{center}
  \includegraphics[width=0.49\textwidth]{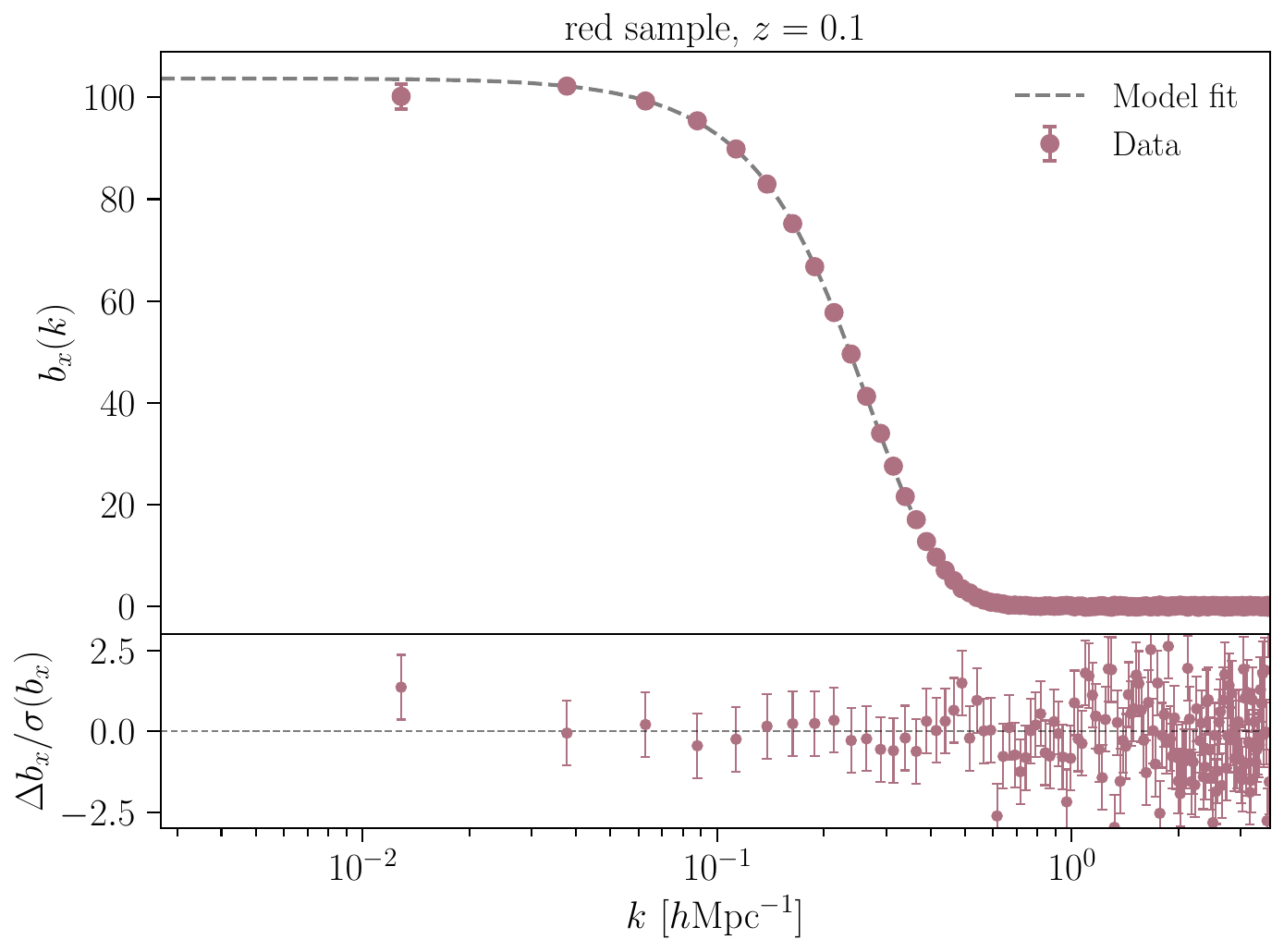}
  \includegraphics[width=0.49\textwidth]{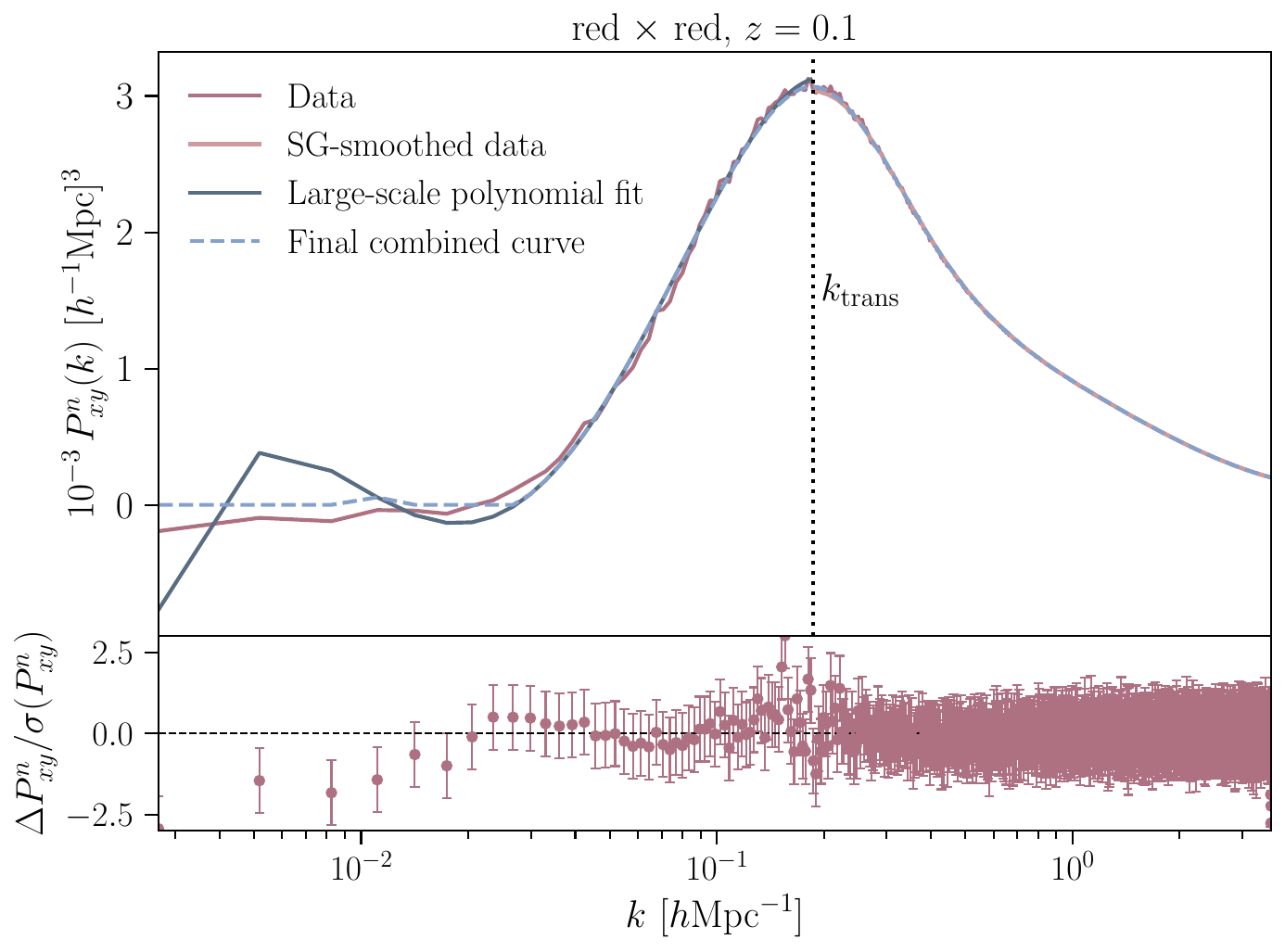}
  \includegraphics[width=0.7\textwidth]{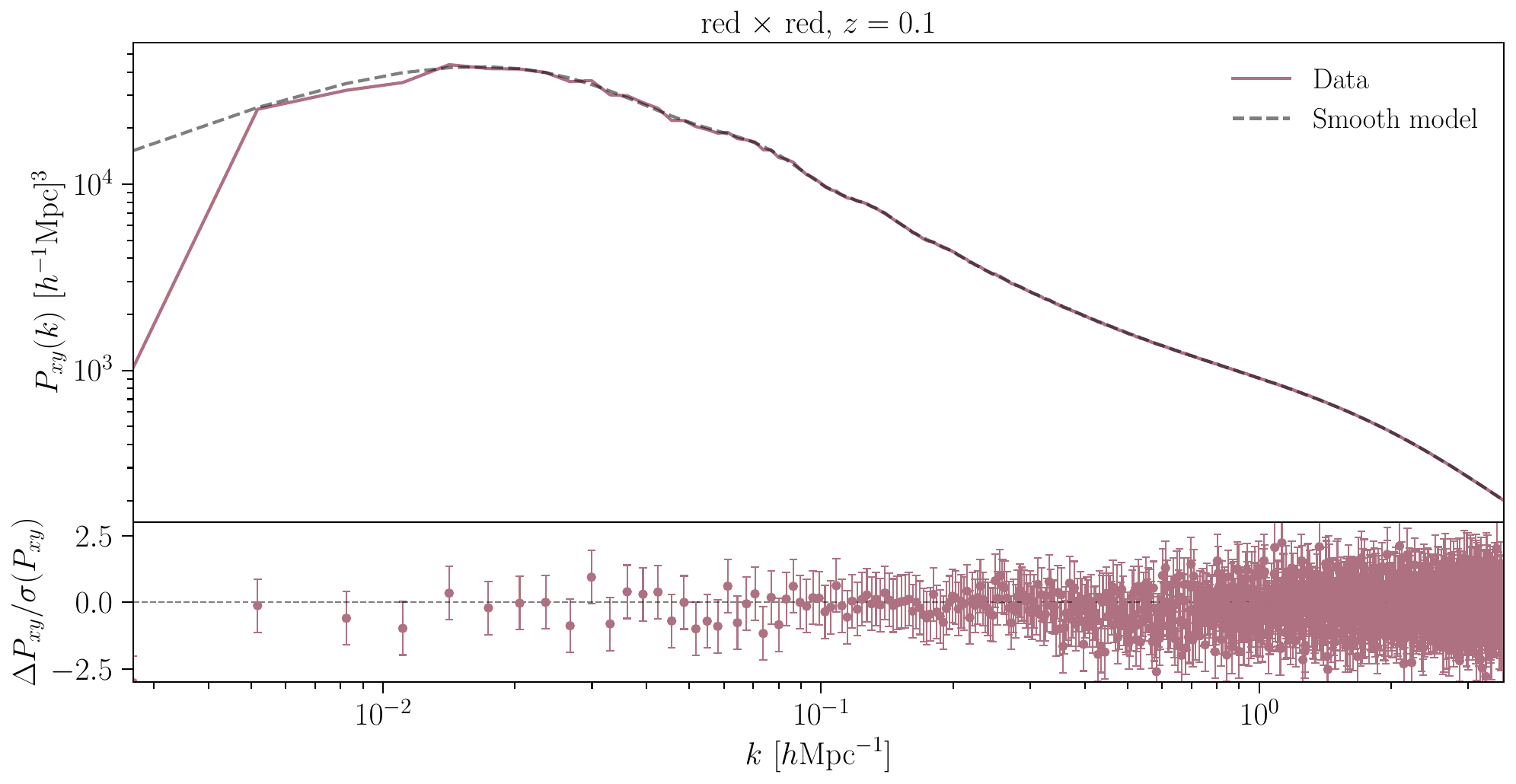}
  \caption{Steps involved in producing smooth power spectra from the measurements made on the \textsc{AbacusSummit} simulation. Results shown for the red galaxy sample in the $z=0.1$ snapshot. The top left panel shows the measurement of the large-scale bias function $b_x(k)$, together with its fit to Eq. \ref{eq:bk_model}. The top right panel shows the measured power spectrum of the small-scale component, together with the different contributions to the final smooth fit (described in the text). The bottom panel shows the global fit to the galaxy power spectrum constructed by combining the results in the two top panels. The bottom sub-panels in each plot show the relative deviation between data and model fits as a fraction of the statistical uncertainties. The error bars were computed from the scatter between each group of 8 adjacent data points.}\label{fig:pk_smooth}
\end{center}
\end{figure*}

To eliminate the noise in the measured power spectra due to the finite size of the \textsc{AbacusSummit} simulation boxes, we start by modeling the overdensity of a given tracer $x$ as
\begin{equation}\label{eq:ls_ss}
\delta_x({\bf k})=b_x(k)\,\delta_{\rm IC}({\bf k}) + n_x({\bf k}),
\end{equation}
where $\delta_{\rm IC}$ denotes the linear matter overdensity in the initial conditions, $b_x(k)$ is a deterministic function of $k$ and, by definition, $n({\bf k})$ is the small-scale component of the overdensity field, $\delta_x$, that does not correlate with $\delta_{\rm IC}$.

Then at each snapshot, we compute the following power spectra from the simulation:
$P_{gg}(k)$, $P_{gm}(k)$, $P_{mm}(k)$, $P_{g,{\rm IC}}(k)$, $P_{m,{\rm IC}}(k)$, and $P_{{\rm IC},{\rm IC}}(k)$, where $m$ and $g$ represent the overdensities of matter and of the target sample of galaxies. We also have a theoretical prediction for the linear power spectrum of the initial conditions, $\bar{P}_{{\rm IC},{\rm IC}}(k)$. From the model in Eq. \ref{eq:ls_ss}, we compute a first estimate of the bias functions $b_g(k)$ and $b_m(k)$ as
\begin{equation}\label{eq:bk_model}
\hat{b}_x(k)\equiv \frac{P_{x,{\rm IC}}(k)}{P_{{\rm IC},{\rm IC}}(k)}.
\end{equation}
Since both $P_{x,{\rm IC}}$ and $P_{{\rm IC},{\rm IC}}$ come from the same realization, the resulting bias function is reasonably smooth on large scales. To obtain a fully smooth function that is defined on the full continuum of $k$, we fit the resulting measured $\hat{b}_x$ to a smooth function of the form:
\begin{equation}
b_x(k)=b_0\,e^{-(k/k_0)^\alpha}\,\left[1+c\,e^{-\left(\frac{k-k_1}{0.1}\right)^2}\right],
\end{equation}
with $b_0$, $k_0$, $\alpha$, $c$ and $k_1$ as free parameters. We found this functional form to provide a good fit in all cases explored (see top left panel of Fig. \ref{fig:pk_smooth}).

After determining the $b_x$, we estimate the power spectrum between the small-scale components $n_x$ and $n_y$ as
\begin{equation}
P^n_{xy}(k)=P_{xy}(k)-\frac{P_{x,{\rm IC}}(k)P_{y,{\rm IC}}(k)}{P_{{\rm IC},{\rm IC}}(k)}.
\end{equation}
The resulting curve is fairly smooth on large $k$, but exhibits residual noise for small and intermediate wavenumbers ($k\lesssim0.2\,h\,{\rm Mpc}^{-1}$), which we correct for as follows (for illustration, see the top right panel of Fig.~\ref{fig:pk_smooth}):
\begin{enumerate}
\item $P^n_{xy}(k)$ reaches a maximum at a transition scale $k_{\rm trans}$ around $\sim 0.1\,h\,{\rm Mpc}^{-1}$, which provides a subdivision between the large-scale and small-scale components of $P^n_{xy}(k)$. In a first step, we determine this transition scale as the value at which the $P^n_{xy}(k)$ estimated from the simulation reaches its maximum value.
\item We fit the data on scales below $k_{\rm trans}$ using a 4th-order polynomial. Since this polynomial can take negative values, which are purely driven by large-scale noise, we apply a positivity prior on the resulting function. We note that on the largest scales, where this occurs, the final power spectrum is dominated by the large-scale correlated part from before, so these choices have a negligible impact on the final result.
\item On scales above $k_{\rm trans}$, we find the measured spectra to exhibit small noise-like oscillations, which we reduce by smoothing the spectra with a Savitzky-Golay (SG) filter of order 1 and window size 25.
\item Finally, we combine the large-scale polynomial $P^{n,{\rm L}}_{xy}(k)$ and the SG-smoothed small-scale component $P^{n,{\rm S}}_{xy}(k)$ by smoothing the transition between the two regimes such as:
\begin{equation}
  P^n_{xy}(k)=e^{-(k/k_{\rm trans})^5}P^{n,{\rm L}}_{xy}(k)+\left[1-e^{-(k/k_{\rm trans})^5}\right]P^{n,{\rm S}}_{xy}(k).
\end{equation}
\end{enumerate}

This procedure thus provides us with a smooth set of tabulated measurements of $P^n_{xy}$, which we then interpolate linearly in $\log(k)$. In a last step, we compute the final power spectrum from the theoretical linear power spectrum of the initial conditions combined with the models for $b_x(k)$ and $P^n_{xy}(k)$ as:
\begin{equation}
 P_{xy}(k)=b_x(k)b_y(k)\bar{P}_{{\rm IC},{\rm IC}}(k)+P^n_{xy}(k).
\end{equation}
To go beyond the smallest scale measured by \textsc{AbacusSummit} ($k_{\rm max}\simeq 2.7 \,{\rm Mpc}^{-1}$) we use a power law extrapolation with a logarithmic slope calculated from the last 50 points in the measurement. As our results will be mostly based on scales $k\lesssim0.4\,{\rm Mpc}^{-1}$, the effects of this extrapolation are largely irrelevant.

Note that other approaches have been recently proposed in the literature \cite{Kokron:2022a, DeRose:2023b} to reduce the impact of noise on simulation-based measurements of summary statistics. We verified that the scheme described above is able to provide a good description of the measured power spectra, well below the statistical uncertainties of the \textsc{AbacusSummit} simulation box (see bottom panel of Fig. \ref{fig:pk_smooth}), and leave a comparison to other approaches to future work.

\section{Detailed description of goodness-of-fit tests}\label{sec:ap:gof}

\subsection{HEFT}\label{ssec:ap.gof.heft}

In the following, we describe the goodness-of-fit tests employed to test the performance of the two HEFT implementations considered. 

\begin{figure*}
\begin{center}
\includegraphics[width=0.45\textwidth]{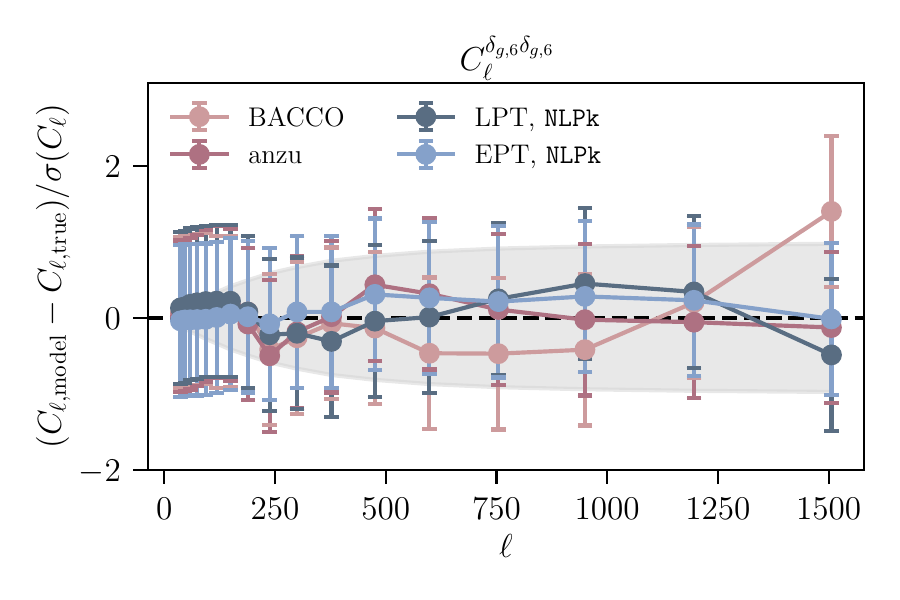}
\includegraphics[width=0.45\textwidth]{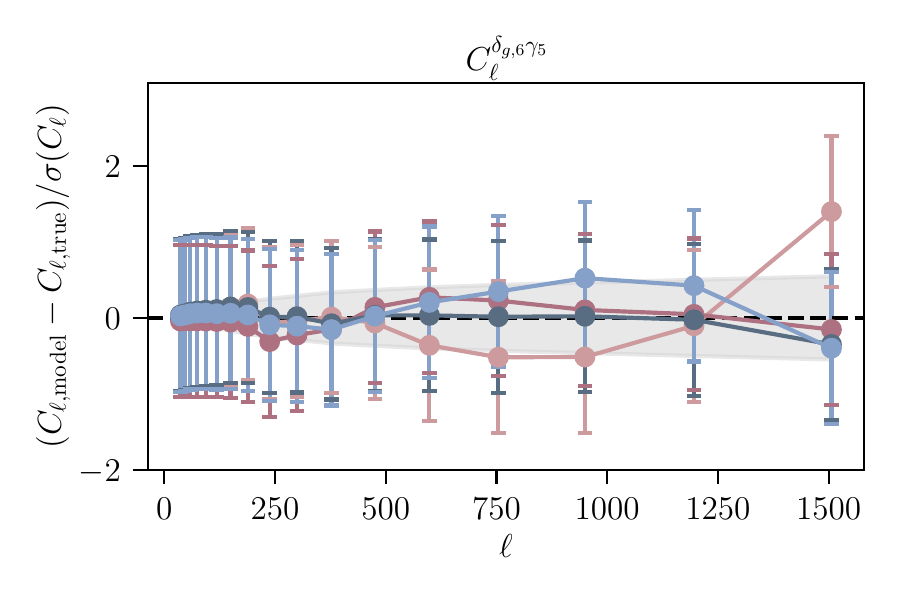}
 \caption{Normalized residuals between the \textsc{AbacusSummit} data and the predictions for the red galaxy sample obtained using the bias models considered in this work. \textit{Left panel:} Results for the auto-correlation of the highest clustering redshift bin. \textit{Right panel:} Results for the cross-correlation between the highest clustering bin and the highest weak lensing bin. In both figures, the gray area denotes relative deviation of 1\% with respect to the theoretical prediction.}
\label{fig:cl-residuals}
\end{center}
\end{figure*}

The number of degrees-of-freedom for our fiducial red sample is given by $\chi^{2}_{\mathrm{noise}}=478$, where we have used $n_{C_{\ell}}=786$, $n_{C_{\ell}^{\gamma \gamma}}=270$, and $n_{p}=38$. Therefore, the maximal $\chi^{2}_{\mathrm{theory}}$ allowed by our $p$-value criterion is given by $\chi^{2}_{\mathrm{theory, max}}=51$. The recovered $\chi^{2}$-values for both HEFT models pass this test, with $\chi^{2}_{\mathrm{theory}} \simeq 11$ for \texttt{anzu}, and $\chi^{2}_{\mathrm{theory}} \simeq 17$ for \texttt{BACCO}. The associated $p$-values are $p=0.36$ and $p=0.28$, respectively. In Fig.~\ref{fig:cl-residuals}, we also show the normalized fit residuals for the auto-correlation of the highest clustering redshift bin as well as its cross-correlation with the highest weak lensing bin\footnote{We choose these particular two combinations as they display some of the largest differences between the simulated data and the models, and thus serve as illustration of the worst model performance.}. As can be seen, we find most of the residuals to be within $1\sigma$, and the relative differences generally lie within $1\%$.

Additionally, we consider two noisy realizations of the synthetic data, finding $\chi^{2}$ $p$-values of $p=0.096$ ($p=0.63$) and $p=0.012$ ($p=0.11$) for \texttt{anzu} and \texttt{BACCO} respectively. The distributions of galaxy clustering and galaxy-galaxy lensing residuals are consistent with Gaussians, as can be seen in Fig.~\ref{fig:cl-residuals-heft} for one of the realizations, and a Kolmogorov-Smirnov (KS) test yields $p$-values of $p=0.38$ ($p=0.93$) for \texttt{BACCO} and $p=0.21$ ($p=0.89$) for \texttt{anzu}. Given these results, we conclude that the two HEFT implementations we consider in this work are both suited to the analysis of high-precision data as expected from LSST, and we regard the failed $p$-value test for one of the noisy realizations for \texttt{BACCO} as a statistical fluctuation.

\subsection{LPT/EPT}\label{ssec:ap.gof.pt}

For the \texttt{NLPk} implementations of LPT and EPT, we obtain minimal $\chi^{2}$-values of $\chi^{2}_{\mathrm{theory}}=8.6$ for LPT, and $\chi^{2}_{\mathrm{theory}}=13.8$ for EPT when using the noiseless data vector. Furthermore, as can be seen from Fig.~\ref{fig:cl-residuals}, we find the residuals between model and data to mostly lie within their $1\sigma$ uncertainties, with relative differences largely smaller than $1\%$ as we found for the HEFT models. For the \texttt{PTPk} models on the other hand, we find $\chi^{2}$-values of $\chi^{2}_{\mathrm{theory}}=28.7$ for LPT, and $\chi^{2}_{\mathrm{theory}}=59.0$ for EPT at our minimal cutoff scale ($k_{\rm max}=0.4\,{\rm Mpc}^{-1}$). 

\begin{figure*}
\begin{center}
\includegraphics[width=0.45\textwidth]{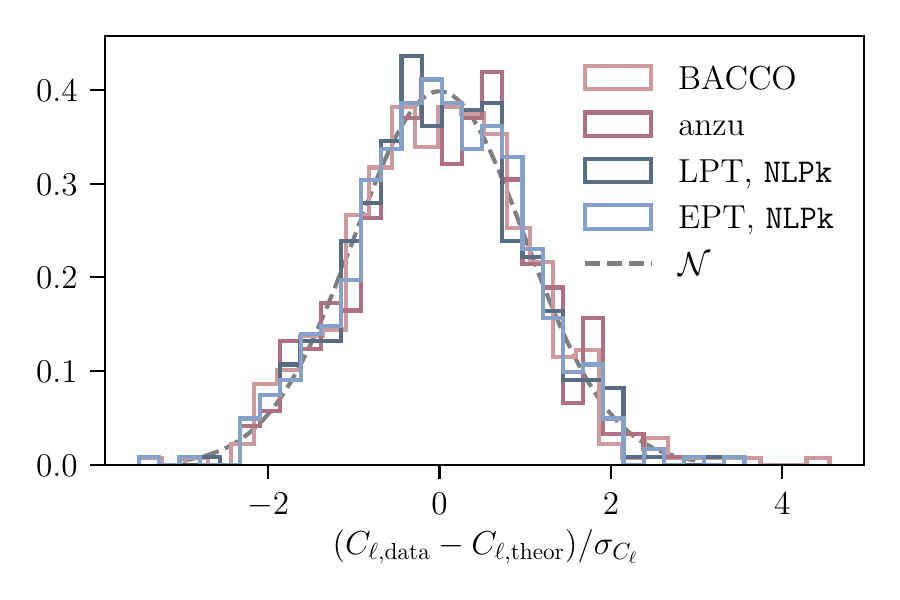}
 \caption{Distribution of galaxy clustering and galaxy-galaxy lensing $C_{\ell}$ residuals for a fit to noisified data, and using a maximal wavenumber $k_{\mathrm{max}}=0.4\;\mathrm{Mpc}^{-1}.$}
\label{fig:cl-residuals-heft}
\end{center}
\end{figure*}

Finally, we also analyze the same two noisy realizations of the data vector as used in the HEFT case with our fiducial \texttt{NLPk} implementation of LPT and EPT, finding $\chi^{2}$ $p$-values of $p=0.16$ ($p=0.77$) and $p=0.064$ ($p=0.83$) for LPT and EPT respectively. As shown in Fig.~\ref{fig:cl-residuals-heft} for one of the realizations, we also find the distribution of fit residuals to be consistent with a Gaussian (with corresponding $p$-values of $p=0.99$ for LPT and $p=0.98$ for EPT).

\section{Consistency of bias parameter values from different models}\label{sec:ap.bias_model_cons}

\begin{figure*}
\begin{center}
\includegraphics[width=0.32\textwidth]{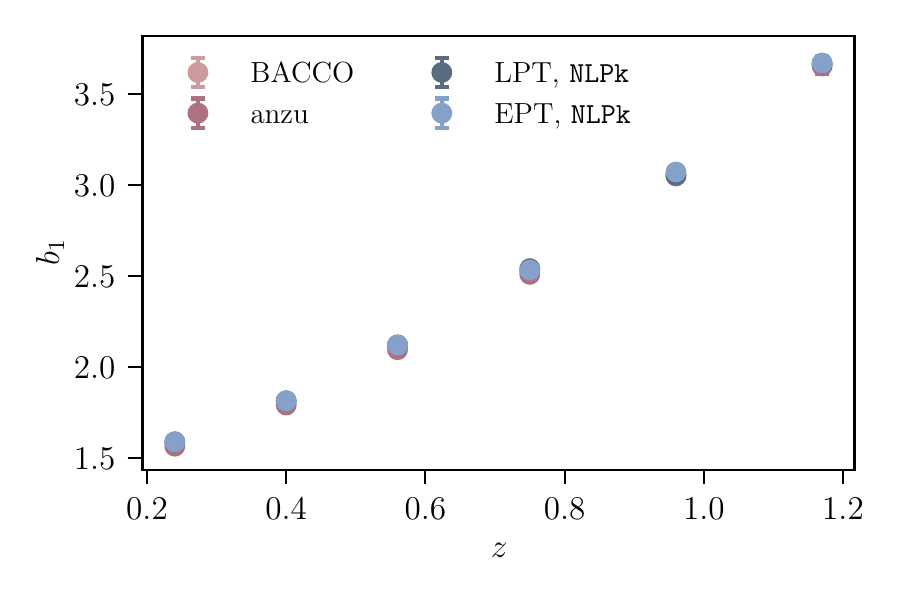}
\includegraphics[width=0.32\textwidth]{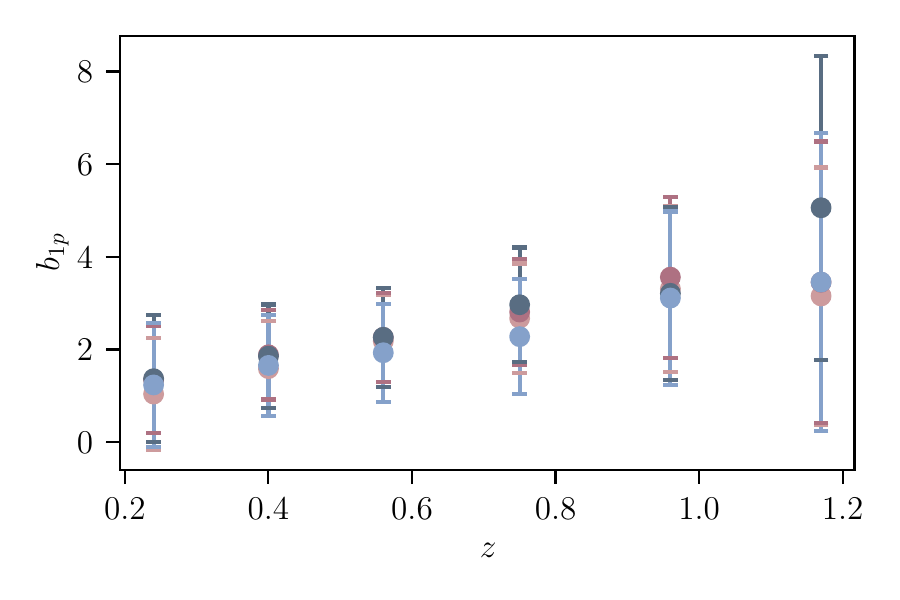}
\includegraphics[width=0.32\textwidth]{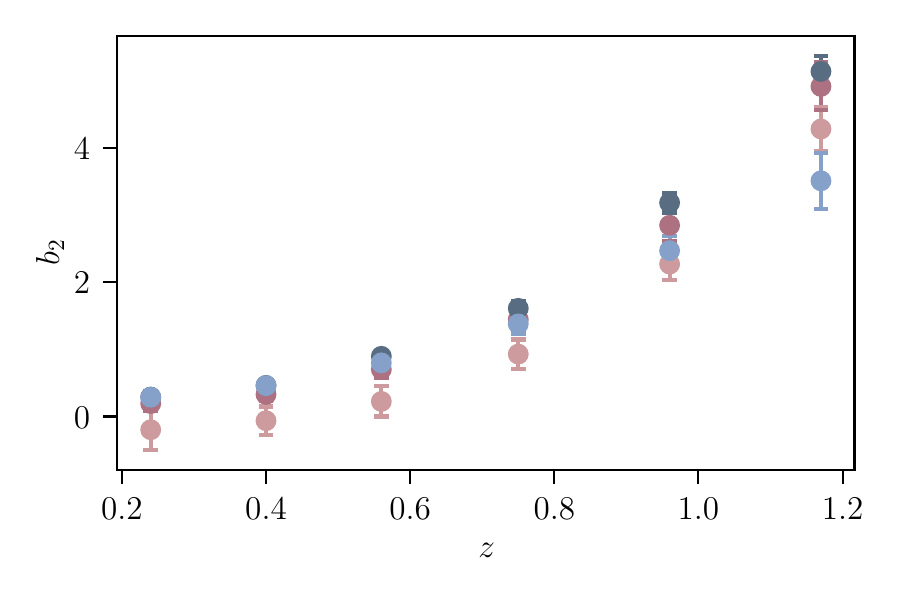}\\
\includegraphics[width=0.33\textwidth]{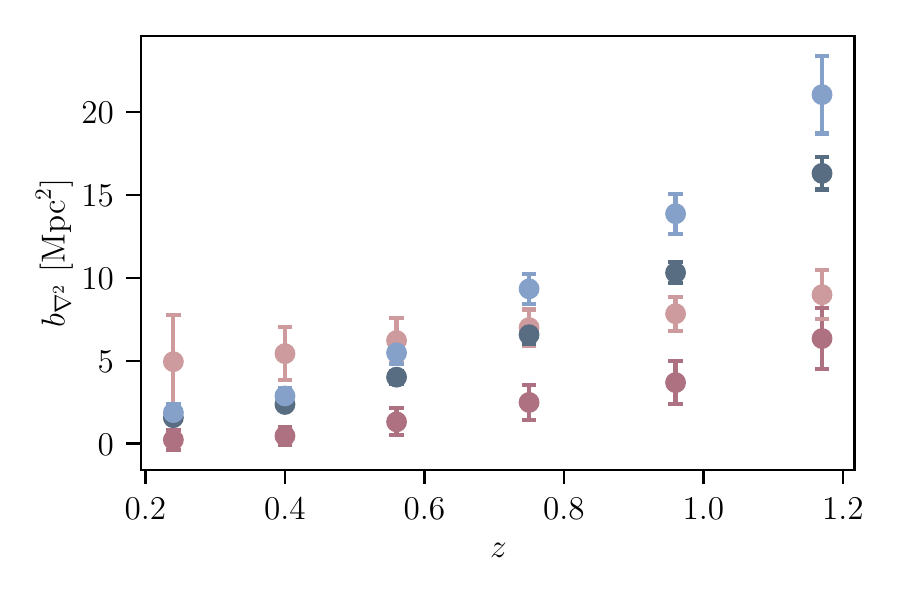}
\includegraphics[width=0.33\textwidth]{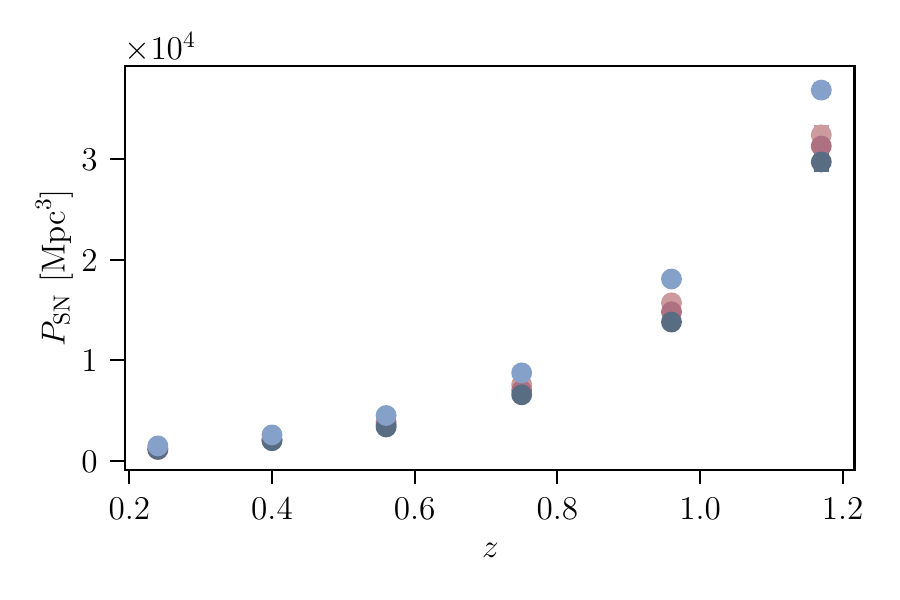}
 \caption{Comparison of bias parameter values obtained from fitting our fiducial red galaxy sample with all nonlinear bias models considered in this work.}
\label{fig:bias-params-all}
\end{center}
\end{figure*}

In order to test for possible systematic modeling uncertainties, we compare the consistency of the bias parameter values obtained for the two HEFT implementations. As discussed in more detail in Appendix \ref{sec:ap.MCMC_comp} below, we find the Fisher matrix uncertainties on the bias parameters to be rather unstable when varying both $b_{s^{2}}$ and $b_{\nabla^{2}}$. The constraints obtained for the cosmological parameters on the other hand, are stable in all cases considered. Fixing one of the two parameters, $b_{s^{2}}$ or $b_{\nabla^{2}}$, yields stable error bars in all cases, which are additionally consistent with their MCMC analogs (see Appendix \ref{sec:ap.MCMC_comp}). We therefore compare the recovered bias parameters for \texttt{anzu} and \texttt{BACCO} fixing $b_{s^{2}}$ to its recovered value, and the results are shown in Fig.~\ref{fig:bias-params-all} for all parameters except $b_{s^{2}}$\footnote{We find the recovered values for $b_{s^{2}}$ to be generally consistent between all four models considered here, and thus do not show the plots as we are unable to provide error bars for the measurements.}. As can be seen, these are generally consistent, the only exception being $b_{\nabla^{2}}$ for which we find consistently higher values for \texttt{BACCO} than we do for \texttt{anzu}. These differences might be due to the fact that the nonlocal bias parameter $b_{\nabla^{2}}$ depends on small-scale properties of the fields considered and is thus affected by implementation details such as different smoothing scales used when deriving template spectra (see e.g. Ref.~\cite{Desjacques:2018}). In addition, \texttt{anzu} and \texttt{BACCO} employ different methods to model power spectra involving higher derivative terms: while in \texttt{BACCO} these terms are determined from the simulations, \texttt{anzu} uses the approximation $\langle X, \nabla^{2}\delta_{L} \rangle = -k^{2}\langle X, 1 \rangle$, where $X$ denotes one of the fields described in Sec.~\ref{sec:pt}. 

We find similar results when comparing the bias parameter values obtained using our fiducial \texttt{NLPk} implementation of LPT and EPT. As can be seen from Fig.~\ref{fig:bias-params-all}, we generally find good agreement between both methods, although the EPT approach seems to prefer significantly larger values for $b_{\nabla^{2}}$ and $P_{\rm SN}$ compared to LPT. It is also interesting to note that the recovered bias parameters for the two PT-based models are generally consistent with those obtained from the HEFT methods. As above, the notable exception is $b_{\nabla^{2}}$, for which we obtain significantly lower values for the HEFT implementations than we do for the perturbation-theory-based models, particularly at high redshift. We see two possible explanations for these findings: (i) As discussed above, our results could be another consequence of different bias parameter normalizations due to implementation details such as smoothing. (ii) Nonlocal bias terms parameterized by $b_{\nabla^{2}}$ have the same functional form as EFT counter-terms, which absorb corrections to the bias model due to small-scale physics. These corrections are partly accounted for in the HEFT approach, while they are not present in PT-based models. Without providing rigorous confirmation, these results suggest that the larger nonlocal bias values obtained for LPT and EPT are due to larger contributions from EFT counterterms for these models (see Ref.~\cite{Pandey:2020} for similar results).

\section{Comparison to MCMC results}\label{sec:ap.MCMC_comp}

\begin{figure*}
\begin{center}
\includegraphics[width=0.45\textwidth]{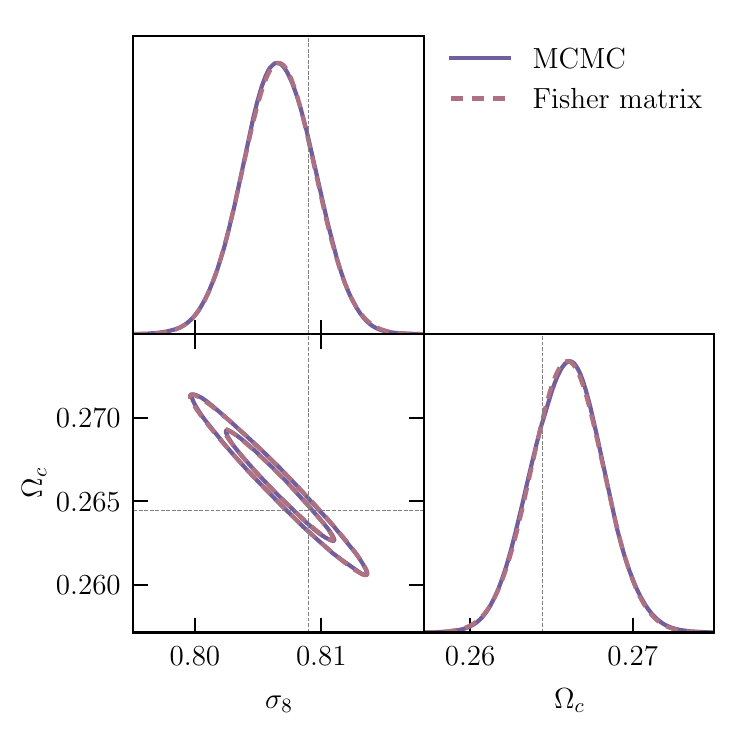}
\includegraphics[width=0.45\textwidth]{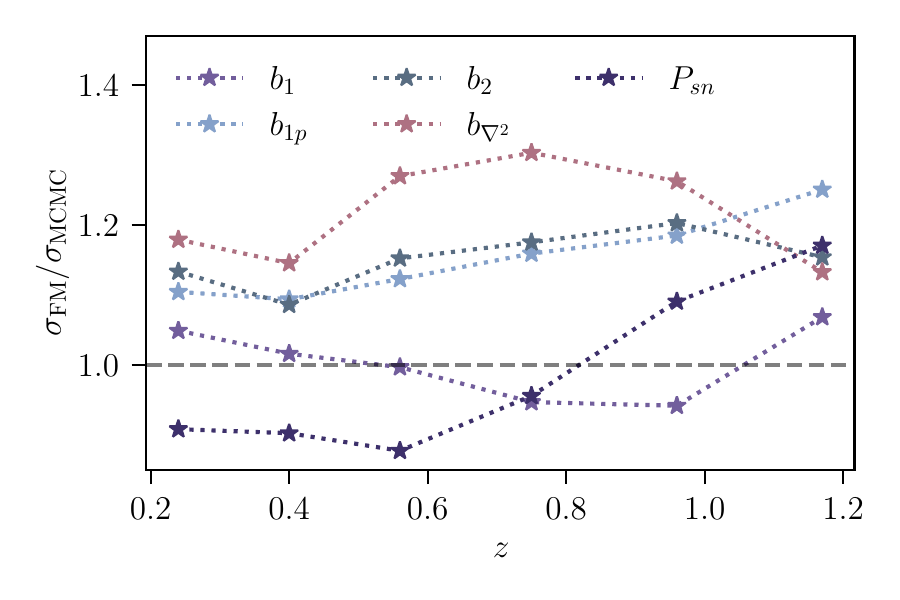}
 \caption{Comparison of parameter constraints obtained from \texttt{anzu} for $k_{\mathrm{max}}=0.4 \; \mathrm{Mpc}^{-1}$ using an MCMC or a Fisher matrix approximation. \textit{Left panel}: Constraints on $\Omega_{c}$ and $\sigma_{8}$ allowing all parameters to vary. \textit{Right panel}: Ratio of FM and MCMC 1-$\sigma$ uncertainties on bias parameters obtained when setting $b_{s^{2}}=0$.}
\label{fig:mcmc-vs-fish}
\end{center}
\end{figure*}

Fisher matrix analyses are prone to numerical instabilities (see e.g. Refs.~\cite{Euclid:2020, Bhandari:2021, Yahia:2021}), and it is therefore essential to validate our results by comparing our FM constraints to those derived using a Monte Carlo Markov Chain (MCMC). Here we focus on our fiducial case using \texttt{anzu} with $k_{\mathrm{max}}=0.4 \; \mathrm{Mpc}^{-1}$, and perform two separate MCMC analyses using the same specifications as used for the Fisher matrix computation: in our first analysis we allow for variations in all cosmological and bias parameters, while in the second case we fix $b_{s^{2}}=0$. The comparison of the MCMC constraints on cosmological parameters obtained in the first case and their FM counterparts are shown in the left hand panel of Fig.~\ref{fig:mcmc-vs-fish}. As can be seen, we find the two approaches to yield consistent constraints on the $\Omega_{c}-\sigma_{8}$ plane. This is not the case for the constraints on bias parameters as discussed above and we thus repeat this analysis setting $b_{s^{2}}=0$ in both cases. The ratio of the 1-$\sigma$ uncertainties obtained from the FM and MCMC analyses respectively are shown in the right hand panel of Fig.~\ref{fig:mcmc-vs-fish} for all bias parameters considered. Similarly to before, we find those to be consistent within roughly $30\%$. This is a reasonable level of agreement, given the approximations and numerical instabilities involved in Fisher matrix analyses.

\section{Fitting data from \texttt{UNIT}}\label{ap:sec:unit}

\begin{figure*}
\begin{center}
\includegraphics[width=0.45\textwidth]{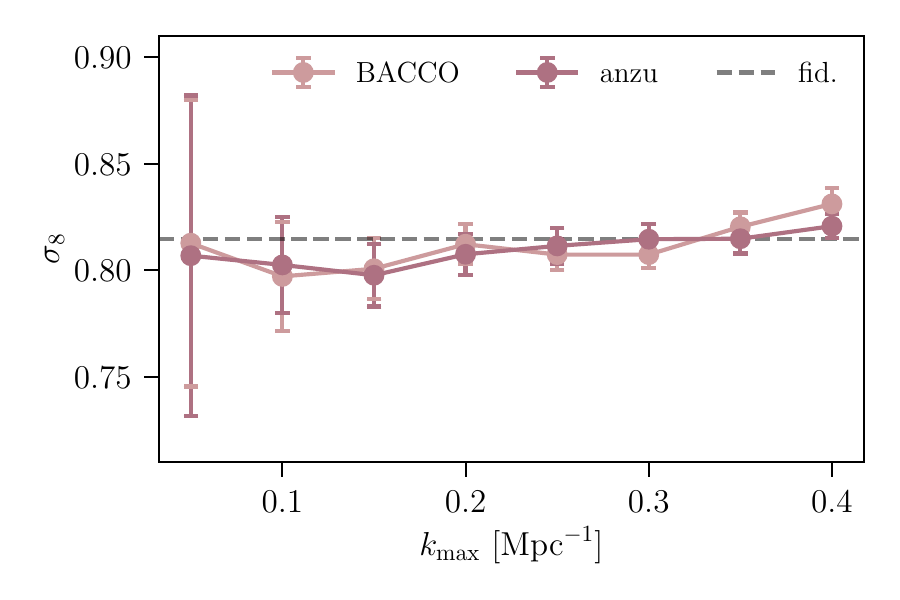}
\includegraphics[width=0.45\textwidth]{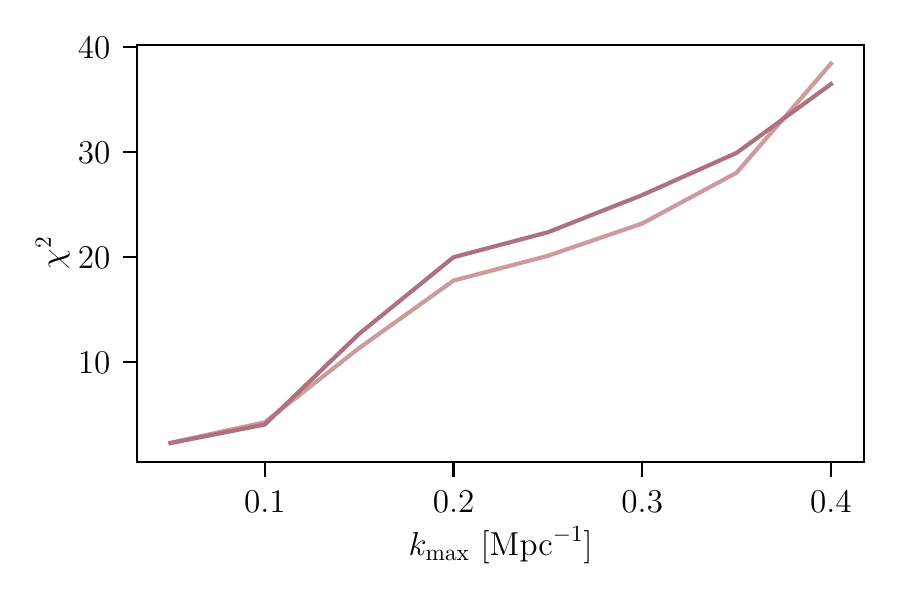}\\
\includegraphics[width=0.45\textwidth]{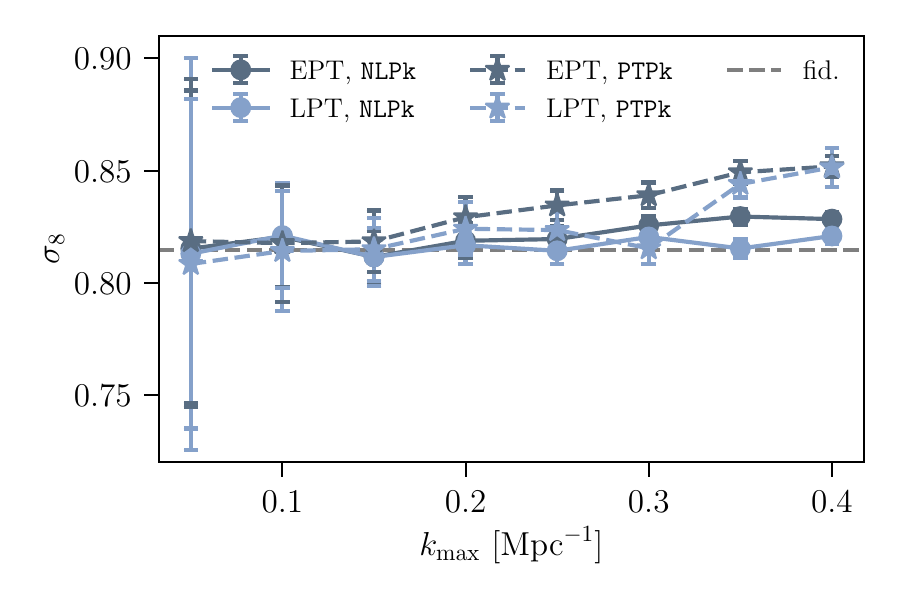}
\includegraphics[width=0.45\textwidth]{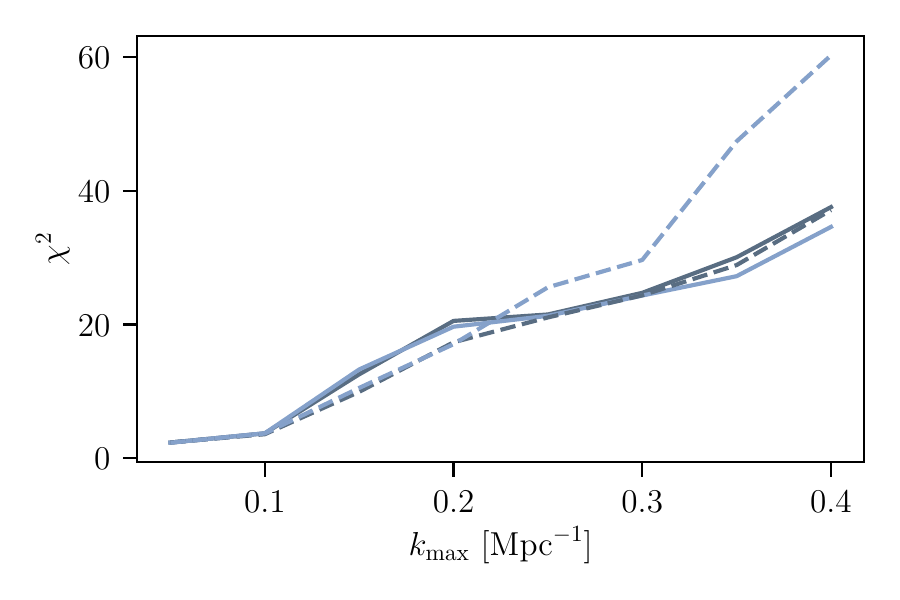}
 \caption{Results of fitting the \textsc{UNIT} data with the main nonlinear bias models considered in this work, LPT, EPT, \texttt{anzu} and \texttt{BACCO}. \textit{Top panels:} Recovered values of $\sigma_{8}$ and $\Omega_{c}$ as a function of maximal wavenumber, $k_{\mathrm{max}}$ for the two HEFT implementations. \textit{Bottom panels:} Recovered values of $\sigma_{8}$ and $\Omega_{c}$ as a function of maximal wavenumber, $k_{\mathrm{max}}$ for \texttt{NLPk} and \texttt{PTPk} implementations of LPT and EPT.}
\label{fig:unit-fits}
\end{center}
\end{figure*}

As an additional consistency test of the results presented in Sections \ref{ssec:results.heft} and \ref{ssec:pt}, we repeat our analysis using the \texttt{redmagic} sample from the \texttt{UNIT} simulations \cite{Chuang:2019} employed in Ref.~\cite{Kokron:2021}. Specifically, we work with the three-dimensional power spectrum and fit the combination of $\mathbf{d}=\{P_{gg}(z, k), P_{gm}(z, k)\}$ using the covariance matrix derived in Ref.~\cite{Kokron:2021}. The results for all bias models considered in this work are shown in Fig.~\ref{fig:unit-fits}. As can be seen, we find results very similar to those obtained for \textsc{AbacusSummit} data, i.e. both HEFT and \texttt{NLPk} implementations of EPT and LPT yield unbiased on $\sigma_{8}$ and $\Omega_{c}$, while the \texttt{PTPk} methods give rise to biases in the recovered cosmological parameters for large maximal wavenumbers $k_{\mathrm{max}}$. Compared to the results shown in Fig.~\ref{fig:parameter-fits-spt}, we find the \texttt{PTPk} models to break down at smaller wavenumbers. This is particularly true for EPT, which yields significantly biased constraints on cosmological parameters for $k_{\mathrm{max}} \gtrsim 0.2$ Mpc$^{-1}$. These results suggest that spherical harmonic power spectra might indeed be less susceptible to nonlinear bias modeling systematics than their three-dimensional counterparts, and provide confirmation that the findings reported in Sections \ref{ssec:results.heft} and \ref{ssec:pt} are not driven by our usage of data from the \textsc{AbacusSummit} simulations.

\section{Three-dimensional power spectrum analysis}\label{sec:ap.pk}

In order to investigate stochasticity in the red and maglim galaxy samples, we analyze three-dimensional power spectrum data. Specifically, we use power spectra at six discrete redshifts covered by the \textsc{AbacusSummit} simulations, i.e. $z=\{0.1, \ 0.3, \ 0.5, \ 0.8, \ 1.1, \ 1.4\}$. We consider a data vector $\mathbf{d}=\{P_{gg}(z, k), P_{gm}(z, k), P_{mm}(z, k)\}$ and assume a Gaussian likelihood with Gaussian covariance matrix given by (see e.g. \cite{Kokron:2021, Scoccimarro:1999})
\begin{equation}
    \mathrm{Cov}(P_{ij}(k), P_{ln}(k')) = \kappa \frac{2\pi^{2} \delta_{kk'}}{k^{2}\Delta k V}
    \begin{cases}
    2 P^{2}_{ii}(k), \; \mathrm{if} \; i=j=l=n \\
    2 P_{gg}(k)P_{gm}(k), \; \mathrm{if} \; i,j=g \; \mathrm{and} \; l=g, n=m \\
    [P_{gg}(k)P_{mm}(k) + P^{2}_{gm}(k)],  \; \mathrm{if} \; i,l=g,  \; \mathrm{and} \; j,n=m,
    \end{cases}
\end{equation}
where $V$ denotes the volume of the survey and $\Delta k$ is the width of each $k$-bin. For consistency with our previous analysis, we aim to ensure a similar signal-to-noise ratio for the $P(k)$ measurements. We therefore follow a rather crude approach and introduce a scaling factor $\kappa$ that scales the relative $P(k)$ errors to be equal to those for $C_{\ell}$ data obtained for bins with matching mean redshift\footnote{We note that we repeat this analysis without the scaling factor and recover very similar values for the stochasticity.}. We then use this likelihood and fit the data with the \texttt{anzu} implementation of HEFT, keeping the cosmological parameters fixed at their fiducial \textsc{AbacusSummit} values\footnote{We find this to be necessary, as the single bin fits exhibit a strong degeneracy between cosmological and bias parameters, and thus the minimization and error bars become unstable when fitting all parameters jointly.}. 

\section{Consistency of results for different data sets}\label{sec:ap.consistency}

\begin{figure*}
\begin{center}
\includegraphics[width=0.32\textwidth]{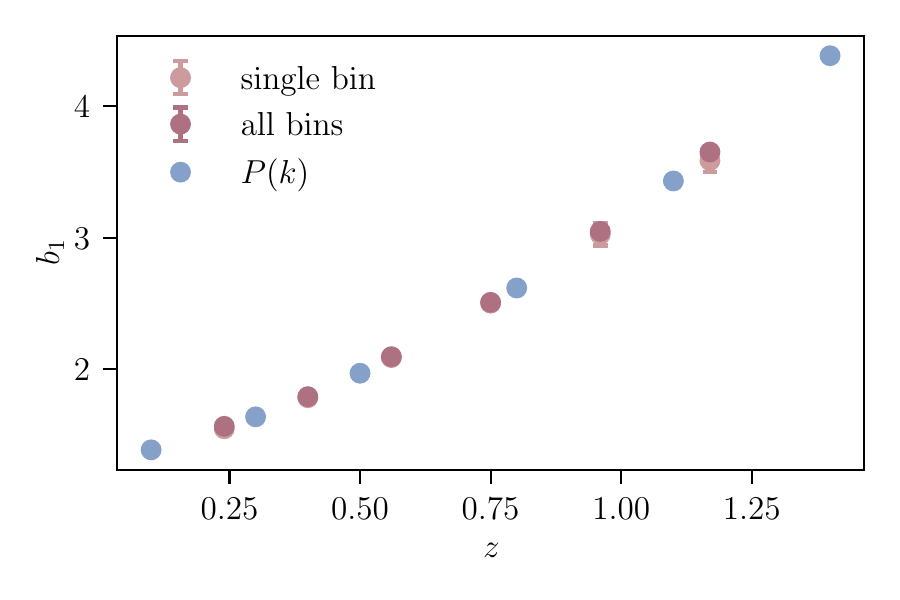}
\includegraphics[width=0.32\textwidth]{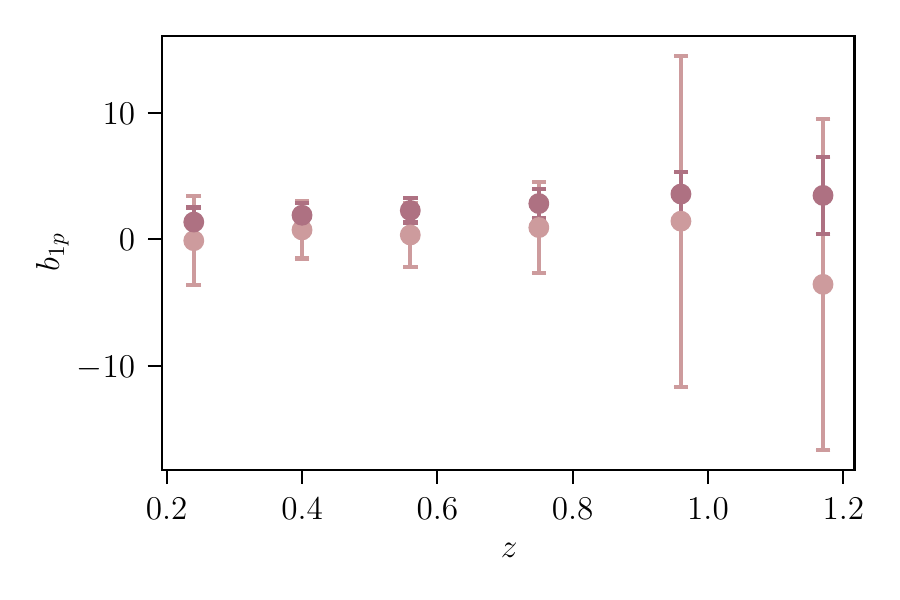}
\includegraphics[width=0.32\textwidth]{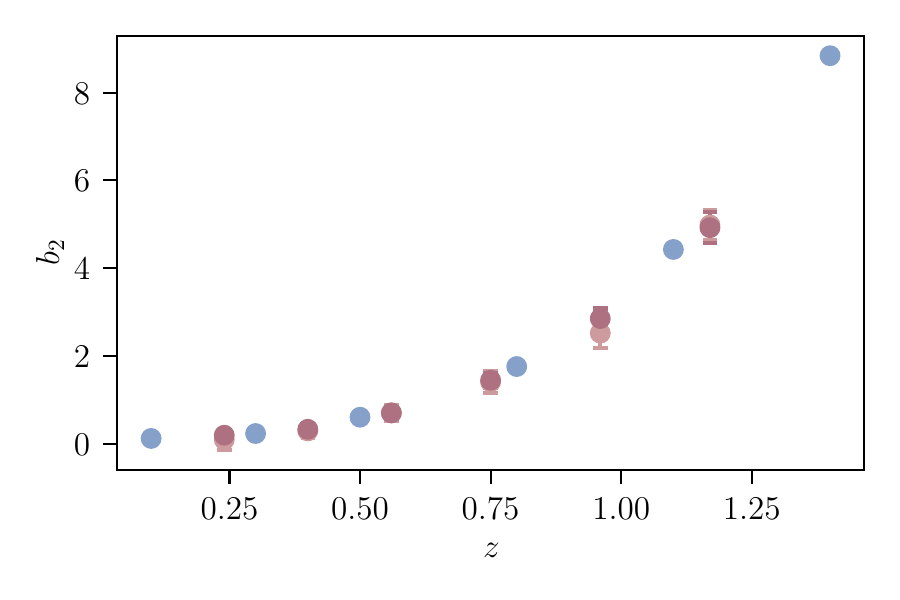}\\
\includegraphics[width=0.33\textwidth]{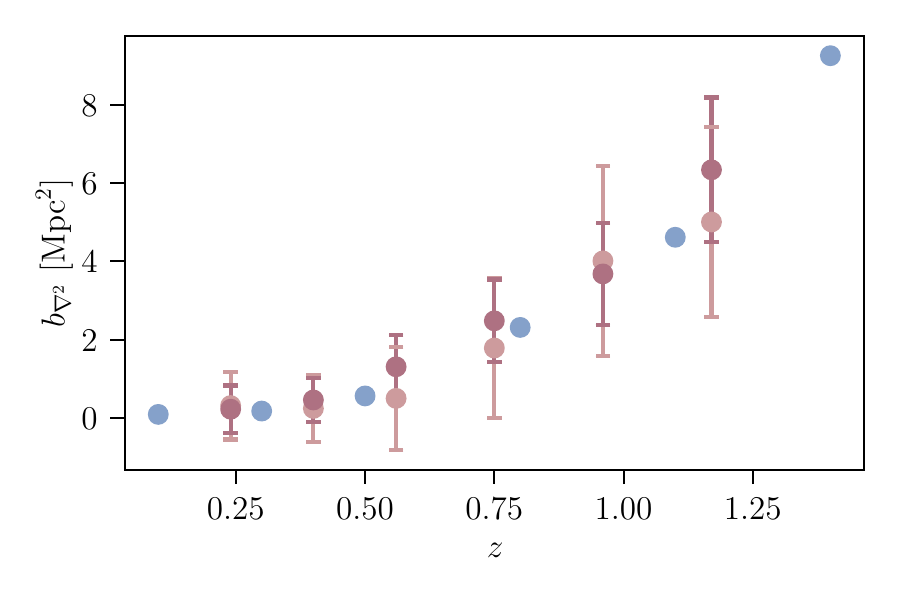}
\includegraphics[width=0.33\textwidth]{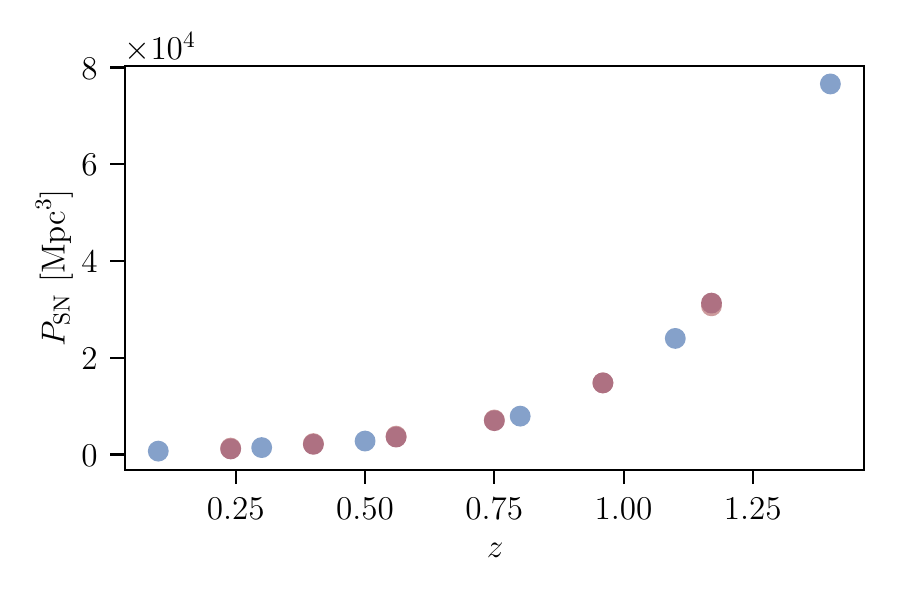}
 \caption{Comparison of bias parameter obtained by fitting three different data sets using \texttt{anzu}: (i) fit each redshift bin separately, (ii) fit all redshift bins simultaneously, (iii) fit three-dimensional power spectra instead of spherical harmonic power spectra.}
\label{fig:bias-params-anzu-single-all-pk}
\end{center}
\end{figure*}

Extending the analysis described in Sec.~\ref{ssec:results.stochasticity} and as a further consistency test of our results, we compare the bias values obtained using different combinations of our simulated data. Specifically, we compare our fiducial constraints obtained using spherical harmonics and fitting all redshift bins simultaneously to those obtained by fitting each bin separately both using spherical harmonics and three-dimensional power spectra. For the $P(k)$ data we proceed as described in Sec.~\ref{ssec:results.stochasticity}, while for the single-bin spherical harmonic fits, we choose to only fit the galaxy auto-correlation and the cross-correlation between the galaxies and DM while keeping the cosmological parameters fixed at their fiducial values as above. As discussed in Sec.~\ref{ssec:results.heft}, we find that while our fiducial analysis fits for all bias parameters simultaneously, the Fisher matrix constraints obtained in this setup are numerically unstable. In order to compare the constraints obtained from different data sets, we therefore follow the approach described in Sec.~\ref{ssec:results.heft}, and consider the FM errors obtained at the best-fit values when setting $b_{s^{2}}=0$. In addition, given the ad-hoc procedure to rescale the covariance matrix used for analyzing $P(k)$ data, we do not consider any error bars for these measurements. The results obtained for \texttt{anzu} are shown in Fig.~\ref{fig:bias-params-anzu-single-all-pk}, and as can be seen we find a very good agreement between the bias values derived in all cases. This suggests that the fits to the spherical harmonic power spectra show internal consistency, and that the redshift-averaged bias parameters obtained from these are consistent with the single-redshift fits obtained from the $P(k)$ analysis.

\bibliography{bibliography}{}

\end{document}